\documentclass[10pt,preprint]{aastex}

\shorttitle{Evolution of Quasar X-ray Emission}
\shortauthors{Kelly et al.}

\begin{document}

  \title{Evolution of the X-ray Emission of Radio-Quiet Quasars}

\author{Brandon C. Kelly, Jill Bechtold}
\affil{Steward Observatory, University of Arizona, 933 N Cherry Ave., 
  Tucson, AZ 85710}
\email{bkelly@as.arizona.edu, jbechtold@as.arizona.edu}

\author{Aneta Siemiginowska, Tom Aldcroft}
\affil{Harvard-Smithsonian Center for Astrophysics, 60 Garden Street, 
  Cambridge, MA 02138}
\email{asiemiginowska@cfa.harvard.edu, taldcroft@cfa.harvard.edu}

\and

\author{Ma\l{\normalsize g}orzata Sobolewska}
\affil{Department of Physics, University of Durham, 
  South Road, DH1 3LE, Durham, UK}
\email{m.a.sobolewska@durham.ac.uk}

\begin{abstract}

We report new {\it Chandra} observations of seven optically faint, $z
\sim 4$ radio-quiet quasars. We have combined these new observations
with previous \emph{Chandra} observations of radio-quiet quasars to
create a sample of 174 sources. These sources have $0.1 < z < 4.7$,
and $10^{44} {\rm\ ergs\ s^{-1}} < \nu L_{\nu} (2500 \AA) < 10^{48}
{\rm\ ergs\ s^{-1}}$.  The X-ray detection fraction is $90\%$.  We
find that the X-ray loudness of radio-quiet quasars decreases with UV
luminosity and increases with redshift.  The model that is best
supported by the data has a linear dependence of optical-to-X-ray
ratio, $\alpha_{\rm ox}$, on cosmic time, and a quadratic dependence
of $\alpha_{\rm ox}$ on $\log L_{UV}$, where $\alpha_{\rm ox}$ becomes
X-ray quiet more rapidly at higher $\log L_{UV}$. We find no
significant evidence for a relationship between the X-ray photon
index, $\Gamma_X$, and the UV luminosity, and we find marginally
significant evidence that the X-ray continuum flattens with increasing
$z$ ($2\sigma$). The $\Gamma_X$--$z$ anti-correlation may be the
result of X-ray spectral curvature, redshifting of a Compton
reflection component into the observed \emph{Chandra} band, and/or
redshifting of a soft excess out of the observed \emph{Chandra}
band. Using the results for $\Gamma_X$, we show that the $\alpha_{\rm
ox}$--$z$ relationship is unlikely to be a spurious result caused by
redshifting of the observable X-ray spectral region. A correlation
between $\alpha_{ox}$ and $z$ implies evolution of the accretion
process.  We present a qualitative comparison of these new results
with models for accretion disk emission.

\end{abstract}


\section{INTRODUCTION}

\label{s-intro}

It is widely accepted that the extraordinary activity associated with
quasars involves accretion onto a supermassive black hole, with the
UV/optical emission arising from a geometrically thin, optically thick
cold accretion disk, and the X-ray continuum arising from a hot,
optically thin corona that Compton upscatters the disk's UV
photons. The geometry of the X-ray emitting region is uncertain, but
possibilities include an accretion disk that evaporates into a hot
inner flow \citep[e.g.,][]{shap76,zdz99}, a hot ionized `skin' that
sandwiches the cold disk \citep[e.g.,][]{bis77,liang77,nayak00}, a
combination of a hot inner flow and a corona that sandwiches the disk
\citep[e.g.,][]{pout97,sob04a}, or a patchy corona, consisting of a
number of hot spots above the accretion disk
\citep[e.g.,][]{gal79,mal01,sob04b}. In addition, the UV and X-ray
producing processes may be coupled as a result of radiation pressure
from the UV photons driving a flow from the disk into the corona
\citep{proga05}. Investigating the relationships between the UV and
X-ray emission is an important step towards understanding the origin
of the X-ray emission. Furthermore, learning how the X-ray and UV
emission change with $z$ provides insight into evolution of the
accretion process, quasar black hole mass, and accretion rate.

Many studies have investigated whether $\alpha_{ox}$, the ratio of
X-ray to UV/optical flux, depends on redshift or UV luminosity,
$L_{UV}$
\citep[e.g.,][]{avni82,wilkes94,yuan98a,bech03,vig03b,strat05,steffen06}. The
parameter $\alpha_{\rm ox}$ is a simple measure of the amount of X-ray
radiation, dominated by non-thermal processes, in respect to the
amount of UV radtion, dominated by thermal processes.  Most studies
have concluded that there is no evidence for a redshift dependence of
$\alpha_{\rm ox}$ \citep[e.g.,][]{avni86,wilkes94,strat05}, although
\citet{bech03} argued that $\alpha_{ox}$ is significantly correlated
with both $z$ and UV luminosity, and \citet{yuan98a} found evidence
for a slight dependence of $\alpha_{\rm ox}$ on redshift at $z <
0.5$. \citet{vig03b}, \citet[][S05]{strat05}, and
\citet[][S06]{steffen06} used a partial correlation and regression
analysis to conclude that there is no evidence for a dependence of
$\alpha_{\rm ox}$ on $z$, after accounting for the $\alpha_{\rm
ox}$--$L_{UV}$ and $L_{UV}$--$z$ correlations. These authors also
found evidence that RQQs become more X-ray quiet with increasing UV
luminosity.

Previous investigations of the X-ray photon index, $\Gamma_X$, have
also produced mixed results. \citet{bech03} used a sample of
\emph{ROSAT} observations over from \citet{yuan98a} and \emph{Chandra}
observations of high redshift quasars to conclude that $\Gamma_X$ is
correlated with both luminosity and $z$. A $\Gamma_X$--$L_{UV}$
correlation was also seen by \citet{dai04}, using a sample of
gravitationally-lensed sources with \emph{XMM-Newton} and
\emph{Chandra} data. Some evidence for an anti-correlation between
$\Gamma_X$ and $z$ has also been found using \emph{ASCA} observations
\citep{reeves97, vig99} and \emph{XMM-Newton} observations
\citep{page03}. However, other investigations based on
\emph{XMM-Newton} data \citep[e.g.,][]{ris05} and spectral fitting of
composite spectra from \emph{Chandra} observations
\citep[e.g.,][]{vig03c,vig05,shem06a} have not revealed any evidence
for a relationship between $\Gamma_X, L_{UV},$ and $z$. Similarly,
other \emph{ASCA} observations have also not produced evidence for a
$\Gamma_X$--$z$ correlation \citep{reeves00}. In addition, there has
been evidence for a correlation between $\Gamma_X$ and the Eddington
ratio \citep[e.g.,][]{lu99,wang04,shem06b}, and \citet{gall05} found
evidence for an anti-correlation between $\Gamma_X$ and the UV
spectral slope.

All of these conclusions are by necessity based on flux-limited
samples.  Flux-limited samples typically suffer from an artificial
correlation between $z$ and $L_{UV}$, making it difficult to
disentangle which parameter is more important in determining X-ray
properties.  To help break the $L_{UV}$--$z$ degeneracy, we observed
seven optically faint $z \sim 4$ radio-quiet quasars with {\it
Chandra}.  We combined these sources with \emph{Chandra} data of
optically-selected radio-quiet quasars, drawn mostly from the Sloan
Digital Sky Survey \citep[SDSS,][]{york00}, to create a flux-limited
sample of 174 sources, $90\%$ of which have detections. Because the
X-ray emission in radio-loud sources can have an additional component
from the jet \citep[e.g.,][]{zam81,wilkes87}, we focus our analysis on
the radio-quiet majority. We use these sources to perform a
multivariate analysis of $\alpha_{\rm ox}, \Gamma_X, L_{UV},$ and $z$
in a manner that allows us to effectively separate the dependence of
the X-ray spectral properties on $L_{UV}$ and $z$.

We adopt a cosmology based on the the WMAP best-fit parameters
\citep[$h=0.71, \Omega_m=0.27, \Omega_{\Lambda}=0.73$,][]{wmap}. For
ease of notation, we define $l_{UV} \equiv \log \nu L_{\nu} (2500
\AA)$, and $l_X \equiv \log \nu L_{\nu} (2\ {\rm keV})$.

\section{OBSERVATIONS AND COMPARISON SAMPLE}

\label{s-sample}

The new observations targeted seven non-BAL RQQs with $z>4$ from the
literature.  These were known to be among the faintest ($\nu L_{\nu}
(2500 \AA) \lesssim 3 \times 10^{46}\ {\rm ergs\ s^{-1}}$) $z \sim 4$
optically-selected quasars to be observed thus far by {\it Chandra} or
{\it XMM-Newton}.  All seven were observed on-axis on the ACIS-S3 chip
with exposure times 10--23 ksec.  The exposure times were chosen in
order to ensure that the X-ray source would be detected if
$\alpha_{ox} < 1.9$.  All targets were, in fact, detected.  The new
observations are summarized in Table \ref{t-newobs}.

The other $z \gtrsim 4$ sources were selected from the literature
\citep{bech03,vig01,vig03a}, and had been observed as targeted
observations with Chandra.  The $z \lesssim 4$ sources were found by
cross-correlating the SDSS DR3 quasar catalogue \citep{dr3qsos} with
the {\it Chandra} public archive as of 2005 February 22. We selected
those SDSS DR3 quasars that were serendipitiously within $12'$ of a
{\it Chandra} target. The radio-quiet sources were selected to have
$R_i = 0.4(i - t_{\rm 1.4 GHz}) < 1.5$ \citep{ive04}, where, $t_{\rm
1.4 GHz}$ is the FIRST 1.4 GHz AB magnitude, and $i$ is the SDSS
$i$-band magnitude. The radio-loud sources were omitted because such
sources have an additional component of X-ray emission arising from
the jet \citep[e.g.,][]{wilkes87,worrall87}. Almost all of the $z < 4$
quasars have their \emph{Chandra} data reported here for the first
time.

The optical/UV spectra for each source were inspected by eye to
exclude the BALs or any sources that had significant absorption. It is
necessary to remove the BAL QSOs because their high column density
gives them the appearance of being X-ray weak \citep[e.g.,][]{green01,
gall02, gall06}, potentially biasing our analysis. We are unable to
remove the high-ionization BAL quasars for $z < 1.5$, as their
identification requires observations of the C IV line. We are able to
remove low-ionization BALs at $0.45 < z < 2.25$ based on Mg II
absorption. \citet{reich03} found the fraction of BALs in the SDSS to
be $\sim 14\%$, and therefore we expect there to be $13 \pm 3$ BALs in
our sample at $z < 1.5$. We did not include seven sources with an
obvious contribution in their spectra from the
host-galaxy. Host-galaxy contamination is likely negligible for all
included sources, except for possibly the lowest luminosity quasars,
since $\nu L_{\nu}^* \sim 10^{44} {\rm\ ergs\ s^{-1}}$ at $2500\AA$
for galaxies \citep{bud05}.

We visually inspected the {\it Chandra} events files to find those
sources that fell on an ACIS chip.  We did not include any sources
that were observed on chip S4 due to higher read-out noise (Data
Caveats on CIAO web pages
\footnote{http://cxc.harvard.edu/ciao/caveats}). All X-ray sources
reported are within 1''--2'' of the optical position.
The archival sources and their X-ray properties 
are listed in Table \ref{t-archival_samp}.

Altogether, the sample consists of 174 radio-quiet quasars, with a
broad range in redshift ($0.1 < z < 4.7$) and luminosity ($10^{44}
{\rm\ ergs\ s^{-1}} \lesssim \nu L_{\nu} (2500 \AA) \lesssim 10^{48}
{\rm\ ergs\ s^{-1}}$).  All sources have been observed with the {\it
Chandra X-ray Observatory} using the ACIS-S or ACIS-I detectors, and
157 ($90\%$) of them are detected.  The $(L_{UV},z)$ distribution of
the sample is shown in Figure 1.

\subsection{X-ray Spectra}

\label{s-xray}

Source extraction was done using {\it CIAO 3.2.2}\footnote{Chandra
  Interactive Analysis of Observations (CIAO),
  http://cxc.harvard.edu/ciao/} and {\it CALDB 3.1}. We extracted the
  PHA spectrum for all 174 sources using a circular aperture with
  radius chosen to include $95\%$ of 3.5 keV photons. Typical
  extraction regions range from 5'' for on-axis sources, to 10''--30''
  for most off-axis sources. The background was extracted from an
  annular region centered on the source, and any nearby sources were
  removed from the extraction regions. A correction for pileup was
  necessary for only one source in our sample, MRK 1014. The
  distribution of source counts is shown in Figure \ref{f-counts}.

Spectral fitting was done using the {\it CIAO} tool {\it SHERPA}
\citep{sherpa}. We estimated the parameters for a power law of the
form
\begin{equation}
N(E) = n_0 \left ( \frac{E}{1\ {\rm keV}} \right )^{-\Gamma_X},
\label{eq-xpow}
\end{equation}
where $\Gamma_X$ is the photon index and $n_0$ is the normalization at
1 keV, in units of photons keV$^{-1}$ cm$^{-2}$ s$^{-1}$. We
restricted our fits to energies 0.3--7.0 keV, and included Galactic
absorption with $N_H$ fixed to that inferred from 21 cm maps
\citep[COLDEN\footnote{For COLDEN, see
http://cxc.harvard.edu/toolkit/colden.jsp}][]{colden}. If the source
had $< 200$ counts, we fit the unbinned spectrum using the Cash
statistic \citep{cash}. For these sources, the background was fit
simultaneously with Equation (\ref{eq-xpow}) using an empirically
determined background for the ACIS-S and ACIS-I, respectively. If the
source had $< 50$ counts, we calculated $n_0$ by fixing $\Gamma_X =
1.9$, a typical value for RQQs \citep{reeves00, picon03}. We also
estimate $\Gamma_X$ for these sources, but fix $\Gamma_X = 1.9$ when
calculating $n_0$ to stabilize the estimates of $L_X$ and $\alpha_{\rm
ox}$.  We fit $n_0$ and $\Gamma_X$ simultaneously for sources with
counts between 50 and 200. If the source had $> 200$ counts, we fit
the binned spectrum by minimizing $\chi^2$ and included an intrinsic
neutral absorber if justified by the data; only two of the sources
with $> 200$ counts showed evidence for intrinsic absorption. These
were 1438+0335 and 0958+0734, with intrinsic $N_H = 5.34^{+ 0.745}_{-
0.203} \times 10^{21}\ {\rm cm^{-2}}$ and $N_H = 1.363^{+ 2.034}_{-
0.626} \times 10^{22}\ {\rm cm^{-2}}$, respectively; the errors are at
$95\%$ confidence. The background for the $> 200$ count sources was
binned and subtracted before spectral fitting. There were 86 sources
with $< 50$ counts, 44 sources with between 50 and 200 counts, and 44
sources with $>200$ counts.

We included an intrinsic neutral absorber for sources 0259+0048,
0918+5139, 1002+5542, and 1411+5205, despite the fact that they have
$< 200$ counts. These sources initially exhibited unusually hard X-ray
spectra ($\Gamma_X \lesssim 1$), so we fit $\Gamma_X$ and $N_H$
simultaneously to test if these low values of $\Gamma_X$ were caused
by unrecognized absorption. Even after including an absorber at the
quasar redshift, sources 0259+0048, 0918+5139, and 1002+5542 still
have rather hard X-ray continua, with photon induces of $1.02 \pm
0.24$, $0.92 \pm 0.33$, and $1.27 \pm 0.15$, respectively. Such hard
X-ray spectra could be the result of more complex absorption, such as
an ionized or partial covering absorber. However, the low number of
counts for these sources preclude obtaining meaningful spectral fits
for more complex models. In addition, the mean $\Gamma_X$ for our
sample is $\approx 2$ and the observed dispersion in $\Gamma_X$ for
our sample is 0.44. Thus, we might expect to observe a few sources
with $\Gamma_X \sim 1$ in a sample with 157 X-ray detected sources,
and therefore these three sources are probably not outliers but just
represent the tail of the RQQ $\Gamma_X$ distribution.

Source 1411+5205 shows evidence for considerable absorption, $N_H \sim
10^{23}\ {\rm cm^{-2}}$, and may be a BAL QSO. A $95\%$ confidence
interval on the column density for this source is $1.8 \times 10^{22}\
{\rm cm^{-2}} < N_H < 2.45 \times 10^{23}\ {\rm cm^{-2}}$.

We used the projection method in {\it Sherpa} to estimate a $3\sigma$
confidence interval on $n_0$. Those sources that did not contain $n_0
= 0$ in their $3\sigma$ confidence interval were considered detected,
otherwise we set a $3\sigma$ upper limit on $n_0$. For those sources
with $< 50$ counts, the projection method was used to calculate the
$68\%$ ($1\sigma$) individual confidence intervals on the power law
parameters. We calculated the covariance matrix of the parameters for
those sources with $> 50$ counts.

Fifteen sources were in multiple observations. For each of these
sources, all of the observations were fit simultaneously assuming the
same power-law spectrum.

Based on our X-ray spectral fits, we estimate the mean value of
$\Gamma_X$ for our sample to be $\bar{\Gamma}_X = 2.033 \pm 0.034$,
with an observed dispersion of 0.44. After accounting for the
additional scatter in $\Gamma_X$ caused by measurement error, we find
that the intrinsic dispersion of $\Gamma_X$ is $\sim 0.31$. This is
consistent with the dispersion found in other studies
\citep[e.g.,][]{picon05, brock06, grupe06}. The intrinsic dispersion
in $\Gamma_X$ is similar to the intrinsic dispersion in the optical/UV
spectral slope \citep{rich01}.

\subsection{Optical/UV Spectra}

\label{s-uv}

Optical spectra were obtained for most sources from the SDSS. We also
obtained spectra for some of the high redshift quasars from
\citet{and01}, \citet{peroux01}, and \citet{const02}.

We corrected the optical spectra for Galactic absorption using the
$E(B-V)$ values taken from \citet{schlegel}, as listed in the
NASA/IPAC Extragalactic Database (NED), and the extinction curve of
\citet{ccm89}, assuming a value of $A_V / E(B-V) = 3.1$. We model the
continuum as a power law of the form $f_{\nu} \propto \nu^{-\alpha}$,
and the Fe emission as a scaled and broadened iron template extracted
from I Zw I in the UV by \citet{uvfe}. The continuum and iron emission
were fit simultaneously using the Levenberg-Marquardt method for
nonlinear $\chi^2$-minimization. The continuum fitting windows are
listed in Table \ref{t-contwin}. The median value of $\alpha$ for our
sample is $0.602$, and the dispersion in $\alpha$ is $\approx 0.4$.

We were not able to obtain a spectrum for Q 0910+564. For this source,
we calculated the flux density at $2500\AA$ from the $AB$ magnitude at
$1450(1+z)\AA$ and the spectral index reported by \citep{schneid91}.

We were not able to use a power-law fit to calculate $l_{UV}$ for the
$z < 0.3$ sources, as the SDSS spectral range for these sources does
not contain the rest-frame UV continuum.  We therefore 
performed a linear regression of the dependence of 
$l_{UV}$ on $\log \nu L_{\nu} (5100\AA)$ and $\alpha_{opt}$ for 
higher redshift sources (0.3 $< z <$ 1.2) for which we had all
three quantities.  Here  
$\alpha_{opt}$ is the spectral index of the optical continuum. 
Using the regression results, $l_{UV}$ was then estimated for the $z<0.3$ sources 
based on their optical luminosity and spectral index.  
The optical continuum parameters
were found in the same manner as for the UV continuum, except that we
used the Fe emission template from \citet{optfe}. For sources within $0.3
< z < 1.2$ the scatter about the
regression fit resulted in a `measurement' error on $l_{UV}$ of $\approx
0.07$ dex. For comparison, typical measurement errors on $l_{UV}$ for
the $z > 0.3$ sources are $\approx 0.001$--$0.01$ dex, ignoring
variability.   

\subsection{$\alpha_{\rm ox}$}

We calculate the ratio of optical to X-ray flux \citep{tan79} as
\begin{equation}
  \alpha_{\rm ox} = -\frac{\log (f_{X} / f_{UV})}{\log
    (\nu_{X} / \nu_{UV})},
  \label{eq-alfox}
\end{equation}
where $f_{X}$ and $f_{UV}$ are the rest-frame flux densities at 2 keV
and $2500\AA$, respectively. If the flux density from $2500\AA$ to 2
keV is a simple power law, then $\alpha_{\rm ox}$ is the spectral
slope of this continuum, and thus $\alpha_{\rm ox}$ may be thought of
as a crude estimate of the shape of the ionizing continuum. The
parameter $\alpha_{ox}$ is an important parameter for model
comparison, as it summarizes the amount of energy emitted in the X-ray
region (most likely a Comptonized component), compared with that
emitted in the optical-UV (accretion disk component). The mean
$\alpha_{\rm ox}$ of our sample is $\bar{\alpha}_{\rm ox} = 1.49 \pm
0.01$, and the dispersion of $\alpha_{\rm ox}$ is estimated to be
$\sigma_{\rm ox} \approx 0.19$. Because some of the data points are
censored, these estimates of the mean and dispersion of $\alpha_{\rm
ox}$ were obtained by maximum-likelihood assuming a normal
density. The Kaplan-Meier estimate of the mean
\citep[e.g.,][]{feig85}, a non-parametric estimate, gives
$\bar{\alpha}_{\rm ox} = 1.49 \pm 0.06$.

In Figure \ref{f-xuvdist} we show the distributions of $l_X$ and
$\alpha_{\rm ox}$ as functions of $l_{UV}$ and $z$. We report the
X-ray and UV parameters in Table \ref{t-xuvpar_samp}.

\section{DEPENDENCE OF $\alpha_{\rm ox}$ ON $L_{UV}$ AND $z$}

\label{s-alfox}

\subsection{Regression Results}

\label{s-alfox_reg}

In order to study the relationship between $\alpha_{\rm ox}$,
optical/UV luminosity, and redshift, we performed a multivariate
regression of $\alpha_{\rm ox}$ on $z$ and $l_{UV}$. In our analysis,
we perform the regression using several different parameteric models
for the redshift and $l_{UV}$ dependencies.  We compare the different
models simultaneously using the Kullback-Leibler information (KLI;
Kullback \& Leibler 1951), a well-studied method for comparing data to
models from information theory.  In Appendix A, we describe KLI
minimization in detail, and compare this approach to classical
statistical methods for testing for significance.

Currently, there is no \emph{a priori} reason to assume a certain
parameteric form for a dependence $\alpha_{\rm ox}$ on $L_{UV}$ and
$z$.  Initially, we are interested in testing for the existence of a
dependence of $\alpha_{\rm ox}$ on redshift and on UV luminosity, and
we are not concerned with the particular parameteric forms of the
possible redshift and luminosity dependencies. As described in
Appendix A, the KLI is a particularly powerful tool for comparing and
testing several different parameteric models simultaneously, and is
valid even if the `correct' parameterization is not among those
considered.

We compare models with redshift dependencies of the form $L_X \propto
e^{-t(z)/t_0}, L_X \propto e^{z/z_0},$ and $L_X \propto
(1+z)^{\beta_{\zeta}}$, and $L_{UV}$ dependencies of the form $L_{X}
\propto L_{UV}^{\beta_l}$. Here, $t(z)$ is the age of the universe at
$z$ in units of Gyr. In addition, \citet{steffen06} found some
evidence for a nonlinear dependence of $\alpha_{\rm ox}$ on $l_{UV}$,
and to test this we include models that contain a quadratic term for
$l_{UV}$.

We also tested for including the UV spectral slope, $\alpha_{UV}$, as
one of the independent variables. However, we found no evidence that
$\alpha_{\rm ox}$ depended on $\alpha_{UV}$.

Each of the statistical models considered here may be expressed as a
normal density with variance $\sigma^2$ and mean $E(\alpha_{\rm ox}) =
\bar{\alpha}_{\rm ox}(\gamma)$. Here, $E(\alpha_{\rm ox})$ is the
expectation value of $\alpha_{\rm ox}$ at a given $l_{UV}$ and $z$,
and $\gamma$ denotes the regression coefficients.  The five models
considered differ in their description of $\bar{\alpha}_{\rm
ox}(\gamma)$:
\begin{eqnarray}
  {\cal M}_z : \bar{\alpha}_{\rm ox}(\gamma) & = & \gamma_0 + \gamma_l l_{UV} + \gamma_z z 
               \label{eq-zmean} \\
  {\cal M}_{\zeta} : \bar{\alpha}_{\rm ox}(\gamma) & = & \gamma_0 + \gamma_l l_{UV} + 
                     \gamma_{\zeta} \log(1+z) \label{eq-lzmean} \\
  {\cal M}_t : \bar{\alpha}_{\rm ox}(\gamma) & = & \gamma_0 + \gamma_l l_{UV} + \gamma_t t(z) 
               \label{eq-tmean} \\
  {\cal M}_l : \bar{\alpha}_{\rm ox}(\gamma) & = & \gamma_0 + \gamma_l l_{UV} + \gamma_{l^2} l_{UV}^2 
               \label{eq-lmean} \\
  {\cal M}_{l+t} : \bar{\alpha}_{\rm ox}(\gamma) & = & \gamma_0 + \gamma_l l_{UV} + 
               \gamma_{l^2} l_{UV}^2 + \gamma_t t(z) \label{eq-ltmean}.
\end{eqnarray}
Here, we have introduced the notation that ${\cal M}_z$ stands for the
model that parameterizes the average value of $\alpha_{\rm ox}$ as
depending linearly on redshift, and similarly for the remaining four
models. We do not include models with terms higher than quadratic in
$l_{UV}$ because such models were estimated to give a poorer fit to
the data (cf. \S~\ref{s-modcomp}). Model ${\cal M}_t$ is almost
identical to the paramerization used by several other authors (e.g.,
AT86, W94, S06), with the exception that other authors have used the
fractional cosmological look-back time, $\tau(z) = 1 - t(z) / t(0)$.

Because some of the values of $l_{X}$ are only upper limits, we employ
the Expectation-Maximization (EM) algorithm \citep{em, aitken} to
calculate the maximum-likelihood solution for the regression
parameters. These parameters are the regression coefficients,
$\gamma$, and the intrinsic variance about the linear relationship,
$\sigma^2$. The likelihood functions for these five models are normal
densities with means given by Equations
(\ref{eq-zmean})--(\ref{eq-ltmean}). The regression is carried out
directly on $\alpha_{\rm ox} = 0.384(l_{UV} - l_X + 2.605)$ using
computer routines coded by the authors.

The results of the regressions using the entire \emph{Chandra} sample are
\begin{eqnarray}
  \bar{\alpha}_{\rm ox} & = & (-5.148 \pm 1.293) + 
    (0.147 \pm 0.029) l_{UV} - (0.014 \pm 0.018) z, \ \sigma = 0.157 \label{eq-zreg1} \\
  \bar{\alpha}_{\rm ox} & = & (-7.048 \pm 1.443) + 
    (0.190 \pm 0.033) l_{UV} - (0.293 \pm 0.137) \log(1+z), \ \sigma = 0.155 \label{eq-lzreg1} \\
  \bar{\alpha}_{\rm ox} & = & (-8.816 \pm 1.473) + 
    (0.223 \pm 0.031) l_{UV} + (3.061 \pm 0.890) \times 10^{-2} t(z), \ \sigma = 0.151 
    \label{eq-treg1} \\
  \bar{\alpha}_{\rm ox} & = & (62.83 \pm 30.48) - 
    (2.816 \pm 1.336) l_{UV} + (3.226 \pm 1.464) \times 10^{-2} l_{UV}^2, \ 
    \sigma = 0.155 \label{eq-lreg1} \\
  \bar{\alpha}_{\rm ox} & = & (24.22 \pm 32.88) - 
    (1.212 \pm 1.428) l_{UV} + (1.560 \pm 1.552) \times 10^{-2} l_{UV}^2 + \\
     & & (2.684 \pm 0.965) \times 10^{-2} t(z), \ \sigma = 0.151 \label{eq-ltreg1}.
\end{eqnarray}
For all of these models, $\alpha_{ox}$ increases (becomes more X-ray
quiet) with increasing luminosity and decreases (becomes more X-ray
loud) with increasing redshift. The intrinsic scatter about the
relationships is estimated to be $\sigma \sim 0.15$.

\subsection{Evidence for low-redshift BAL QSOs in the Sample}

\label{s-bals}

The estimates for the regression parameters are derived via
maximum-likelihood. However, the likelihood functions assume that the
residuals are normally distributed. If this assumption is not true,
then it may bias our results.  To test the assumption of normality in
the residuals, in Figure \ref{f-cdf} we compare the cumulative
distribution function (CDF) of the standardized residuals for ${\cal
M}_t$ with the standard normal. The standardized residuals are the
residuals normalized by the intrinsic scatter. The CDFs for the other
models were very similar.  As can be seen, there is evidence for a
violation of the assumption of normality. A Kolmogorov-Smirnov (KS)
test confirmed this, finding a probability of $\approx 0.001$--$0.007$
that the maximum difference between the CDFs of the standardized
residuals for the models and the standard normal is greater than that
observed, assuming that the two distributions are the same. In
addition, inspection of the residuals reveals that there are several
sources that are significantly more X-ray quiet than would be expected
from Equations (\ref{eq-zreg1})--(\ref{eq-ltreg1}), and are therefore
outliers. These sources are all censored (i.e., not detected in the
X-rays) and at $z < 1.5$. As noted in \S~\ref{s-sample}, the high
column density of BAL QSOs gives them the appearance of being X-ray
weak. Furthermore, we are unable to remove BALs at $z < 1.5$. These
two facts suggests that the regression outliers are BALs, as they are
unusually X-ray weak and at $z < 1.5$.

To test the possibility that the $z < 1.5$ censored sources are
dominated by BALs, and thus affecting our regression analysis, we
removed these 10 sources and recalculated the regressions. Note that
this number is consistent with the expected number of BALs in the $z <
1.5$ sample (cf., \S~\ref{s-sample}). While this may remove some
non-BALs from the $z < 1.5$ sample, it is unlikely that removing a few
censored non-BALs from the fit will significantly affect the results,
since these sources are not expected to be outliers and the regression
is dominated by the detected sources. However, the BALs can have a
non-negligible effect on the regression even if they are censored
because they have an additional absorption component that contributes
to the observed X-ray luminosity, and therefore are not expected to
follow the functional form assumed by the regression and can be
outliers. The existence of outliers has the effect of biasing the
estimate of the intrinsic scatter upwards, inflating the uncertainties
on the regression coefficients, and therefore reducing the statistical
significance of the regression coefficients.

After removing the $z < 1.5$ censored data points, we were left with a
sample of 164 sources, 157 $(96\%)$ of which are detected. Performing
the regressions on this second sample, we find
\begin{eqnarray}
  \bar{\alpha}_{\rm ox} & = & (-5.340 \pm 1.097) + (0.150 \pm 0.025) l_{UV} - 
    (0.004 \pm 0.015) z, \ \sigma = 0.131 \label{eq-zreg2} \\
  \bar{\alpha}_{\rm ox} & = & (-7.152 \pm 1.229) + (0.192 \pm 0.028) l_{UV} - 
    (0.224 \pm 0.116) \log(1+z), \ \sigma = 0.129 \label{eq-lzreg2} \\
  \bar{\alpha}_{\rm ox} & = & (-9.273 \pm 1.241) + (0.233 \pm 0.026) l_{UV} + 
    (2.870 \pm 0.750) \times 10^{-2} t(z), \ \sigma = 0.125 \label{eq-treg2} \\
  \bar{\alpha}_{\rm ox} & = & (74.61 \pm 25.17) - (3.349 \pm 1.103) l_{UV} + 
    (3.827 \pm 1.208) \times 10^{-2} l_{UV}^2, \ \sigma = 0.127 \label{eq-lreg2} \\
  \bar{\alpha}_{\rm ox} & = &  (40.97 \pm 27.34) - (1.949 \pm 1.187) l_{UV} + 
    (2.370 \pm 1.288) \times 10^{-2} l_{UV}^2 + \\
    & & (2.263 \pm 0.813) \times 10^{-2} t(z), \ \sigma = 0.124 \label{eq-ltreg2}
\end{eqnarray}
The results are very similar to Equations
(\ref{eq-zreg1})--(\ref{eq-ltreg2}), but the intrinsic scatter has
decreased and the significance levels of the regression coefficients
are in general higher. Note that Equations (\ref{eq-treg2}) and
(\ref{eq-ltreg2}) are equivalent to the form $L_X \propto e^{-t(z) /
t_0}$, where the $e$-folding time is $t_0 = 5.75^{+4.98}_{-1.83}$
($95\%$ confidence) Gyr for ${\cal M}_t$ and $t_0 =
7.25^{+14.4}_{-2.96}$ ($95\%$ confidence) Gyr for ${\cal M}_{l+t}$.

The CDF of the standardized residuals for Equation (\ref{eq-treg2}) is
also shown in Figure \ref{f-cdf}. As can be seen, the residuals no
longer show any evidence for a significant divergence from normality,
suggesting that we have minimized BAL contamination by removing the $z
< 1.5$ censored sources. A KS test also found that the empirical
distribution of the standardized residuals for all parameterizations
considered are not significantly different than the standard normal,
having $p$-values of $p \sim 0.1$.

\subsection{$\alpha_{\rm ox}$ Depends on Both $L_{UV}$ and $z$.}

It is apparent from Equations (\ref{eq-zreg2})--(\ref{eq-ltreg2}) that
there is statistically significant evidence for a dependence of
$\alpha_{\rm ox}$ on $l_{UV}$ ($> 6\sigma$ significance). In addition,
there is evidence from model ${\cal M}_t$ that $\alpha_{\rm ox}$
depends on cosmic time ($3.8\sigma$ significance), and evidence from
model ${\cal M}_l$ that the $\alpha_{\rm ox}$--$l_{UV}$ relationship
is nonlinear ($3.2\sigma$ significance). While the coefficients for
both parameterizations imply that our data are inconsistent with the
simple form $\bar{\alpha}_{\rm ox} = \gamma_0 + \gamma_l l_{UV}$, it
is unclear which parameterization is the preferred one. In particular,
because there is a strong correlation between $l_{UV}$ and $z$, it is
possible that the $\alpha_{\rm ox}$--$t(z)$ relationship is simply
correcting for the nonlinearity in the $\alpha_{\rm ox}$--$l_{UV}$
relationship, and is thus a spurious result. Because the models are
not nested, (i.e. one is not merely a subset of the next), we cannot
use classical statistical methods, such as the likelihood ratio or
$F$-test, to compare their relative merits
\citep[e.g.,][]{efron84,free99,prot02}. Instead, we adopt an approach
that attempts to find the model that minimizes the `distance' to the
true probability density that gives rise to the observed data. We do
this by finding the model that minimizes the Kullback-Leibler
information (KLI; see Appendix A).

We use the Akaike Information Criterion \citep[$AIC$,][]{aic} to
estimate the difference in KLI between models. We estimate the
difference in KLI between two models by multiplying their difference
in $AIC$ by $1/2$. Terms of order higher than $l_{UV}^2$ increased
the $AIC$ for models ${\cal M}_l$ and ${\cal M}_{l+t}$, and were not
included in the analysis. Denoting the maximum likelihood estimate of
the model parameters as $\hat{\theta}$, and the estimated KLI as
$H(\hat{\theta})$, we find $H(\hat{\theta}_z) - H(\hat{\theta}_{l+t})
= 7.599$, $H(\hat{\theta}_{\zeta}) - H(\hat{\theta}_{l+t}) = 5.798$,
$H(\hat{\theta}_l) - H(\hat{\theta}_{l+t}) = 2.770$, and
$H(\hat{\theta}_t) - H(\hat{\theta}_{l+t}) = 0.677$. Here,
$H(\theta_z)$ denotes the KLI for model ${\cal M}_z$, and likewise for
$H(\theta_{\zeta}), H(\theta_l), H(\theta_t),$ and
$H(\theta_{l+t})$. Model ${\cal M}_{l+t}$ is best supported by the
empirical evidence.

Plots showing the residuals and partial residuals for the ${\cal
M}_{l+t}$ regression are shown in Figure \ref{f-alfoxres}.  To
visualize the result, Figure \ref{f-3dfit} shows a 3-dimensional plot
of the best fit for this model.  The partial residual plots display
the dependence of $\alpha_{\rm ox}$ on $L_{UV}$, after accounting for
the dependence on cosmic time, and the dependence of $\alpha_{\rm ox}$
on $z$, after accounting for the dependence on $L_{UV}$. Both of these
figures correspond to the regression results after removing the
suspected BALs. Also shown in Figure \ref{f-alfoxres} are
non-parametric fits to the residuals, calculated using a
locally-weighted average based on a Gaussian smoothing kernel; the
kernel width was chosen using generalized cross-validation
\citep[e.g., see][]{hastie01}. As can be seen, according to this model
the nonlinearity in the $\alpha_{\rm ox}$--$l_{UV}$ dependence, if
real, is such that $\alpha_{\rm ox}$ increases (becomes more X-ray
quiet) faster at higher $l_{UV}$. This trend is in agreement with the
results of \citet{steffen06}, who found evidence that the slope of the
$\alpha_{\rm ox}$--$l_{UV}$ correlation may be steeper at higher
$l_{UV}$.

Based on the $AIC$, model ${\cal M}_{l+t}$, which contains the $t(z)$
parameterization with a quadratic $l_{UV}$ term, appears to provide
the best description of our data, followed by the $t(z)$
parameterization with only a linear $l_{UV}$ term.

\subsection{Effects of Sampling and Nonlinear Dependence of $\alpha_{\rm ox}$ on Luminosity}

\label{s-modcomp}

To assess how our estimate of the KLI varies under sampling from the
underlying joint distribution of $(\alpha_{\rm ox}, l_{UV}, z)$, we
use the non-parameteric bootstrap \citep{efron79}. We drew $10^4$
bootstrap samples and performed the regression for each
parameterization on each bootstrap sample. We then estimated the KLI
in the same manner, with the exception that we now use the sample mean
of the difference in log-likelihoods of the \emph{original} sample,
evaluated at the maximum likelihood estimate of $\theta$ based on the
bootstrapped samples. The sampling distributions of the differences in
KLI between ${\cal M}_z, {\cal M}_{\zeta}, {\cal M}_l,$ and ${\cal
M}_t$, with respect to ${\cal M}_{l+t}$, are shown in Figure
\ref{f-klidist}. The $t(z)$ parameterization with quadratic $l_{UV}$
term, ${\cal M}_{l+t}$, had the smallest estimated KLI for $\approx
99.9\%$ of the bootstrap samples. Therefore, the preference for ${\cal
M}_{l+t}$ is unlikely to have resulted from fluctuations caused by
random sampling, and thus it appears that this parameterization
provides the best description of our data.

It is unlikely that the evidence for evolution is a result of
nonlinearity in $l_{UV}$. Assuming that there is no dependence of
$\alpha_{\rm ox}$ on $z$, we can use a suitably large enough
polynomial expansion of $l_{UV}$ to approximate any smooth nonlinear
dependence of $\alpha_{\rm ox}$ on $l_{UV}$. However, as mentioned
above, model ${\cal M}_l$ had the best $AIC$ among the set of
polynomial expansions in $l_{UV}$, and thus our data does not prefer
terms of order higher than $l_{UV}^2$. Therefore, ${\cal M}_l$ should
be viewed as the best approximation to the $\alpha_{\rm ox}$--$l_{UV}$
relationship that is supported by our data without overfitting, and
assuming that $\alpha_{\rm ox}$ is independent of $z$. However,
because the models that included $t(z)$ had an $AIC$ lower than ${\cal
M}_l$, and because ${\cal M}_l$ had an $AIC$ lower than models that
included terms of higher order, if follows that models ${\cal
M}_{l+t}$ and ${\cal M}_t$ are preferred by our data over any
polynomial expansion of $\alpha_{\rm ox}$ as a function of
$l_{UV}$. This is not to say that models ${\cal M}_{l+t}$ and ${\cal
M}_t$ are preferred over any smooth nonlinear function of $l_{UV}$,
but that if such a function exists, it is unlikely to differ
significantly from ${\cal M}_l$. The nonlinear effects are not
extreme, and in fact are not `statistically significant' in the
classical sense. However, while there is not enough evidence in the
data to reject a null hypothesis that $\alpha_{\rm ox}$ is linear in
$l_{UV}$ and $t(z)$ at, say, $> 2\sigma$ significance, the empirical
evidence supports a nonlinear dependence of $\alpha_{\rm ox}$ on
$l_{UV}$ at a given $z$ over a linear dependence.

\subsection{Effect of Variability and Measurement Error on the Results}

\label{s-var}

Measurement error or variability may induce false correlations between
parameters.  In this section, we consider two effects.  The first is
especially important when many sources are near the flux limit of the
sample, and when the number of sources increases strongly with
decreasing flux.  In the second case, a bias can result even if all
the sources are far above the flux limit.

For the first case, we argue that the expected tendancy would be for
$\alpha_{ox}$ to {\it increase} with increasing redshift, the opposite
of what we claim from the data.  Consider the possiblity that the
sources are variable.  Assume for simplicity of argument that the
sources vary around some mean flux in all spectral bands, and that a
particular source spends an equal amount of time brighter than the
mean and dimmer than the mean.  Then the sources that were discovered
by SDSS near the flux limit are preferentially observed in their
`bright' state.  By the time we observed them with Chandra, they will
likely no longer be in their `bright' state, and thus may be
systematically X-ray quieter. Thus we expect the sources near the flux
limit to appear fainter in X-rays on average than they really are.
Since most of the sources near the flux limit are at high redshift,
the tendency may be for $\alpha_{ox}$ to increase (quasars are less
X-ray bright) with redshift.  Since this is the opposite of what we
see, variability of sources near the flux limit is not producing the
result.

For measurement error, the qualitative argument is similar.  Near the
flux limit, random errors in photon counts result in more sources just
below the limit being randomly included in the sample than sources
above the limit being randomly excluded, provided that the number of
sources is an increasing function of decreasing flux limit (which is
the case here).  Thus, the SDSS selection would again be biased
towards sources with optical fluxes that appear brighter in optical
than they really are.  Chandra then measured X-ray fluxes for
essentially all the sources.  Thus, we expect the sources near the
flux limit to be systematically more X-ray faint than they really are.
Again, this is the opposite of what we see in the data, so this
Malmquist-type bias is not important.
   
For the second case, measurement errors or variability may induce
false correlations even if all sources are detected far above the flux
limit.  However in this case, Monte Carlo simulations can be carried
out to see how important the effect might be.  Measurement errors on
the independent variables can bias the estimates of the regression
coefficients \citep[e.g.,][]{bces,fox97}, and errors on the dependent
variable can bias the coefficient estimates for censored
regression. When the dependent variable is measured with error and/or
variability, the measurement error and variability inflate the
observed variance in the regression residuals, biasing the estimate of
the intrinsic scatter, $\sigma^2$. For ordinary least-squares this is
not a problem, since the estimate for the intrinsic scatter,
$\sigma^2$, and the estimates of the regression coefficients,
$\gamma$, are statistically independent. However, for the censored
regression model, the estimates of the intrinsic scatter and the
coefficients are no longer statistically independent, and the bias in
the intrinsic scatter estimate also carries over to the coefficient
estimates \citep{stap84}. These facts are confirmed by Monte Carlo
simulations, which have shown that the \emph{observed} relationship
between $l_X$ and $l_{UV}$ can differ from the intrinsic relationship
when the observed $l_X$ and $l_{UV}$ differ significantly from the
intrinsic $l_X$ and $l_{UV}$ \citep[e.g.,][]{yuan98b}. Since the UV
and X-ray data are measured with error, we are not fitting the
intrinsic distribution of $l_X$ given $l_{UV}$ and $z$, but rather the
distribution of $l_X + \epsilon_X$ at a given $l_{UV} + \epsilon_{UV}$
and $z$, where $\epsilon_X$ and $\epsilon_{UV}$ are random error
terms. The errors for $l_{UV}$ are the usual measurement errors from
the continuum fitting and are very small in our analysis, with typical
values of $\sigma_{UV} \approx 0.001$--$0.01$ dex. However, the errors
for $l_X$ include the contribution from measurement errors and from
variability. The errors from variability of the X-ray emission arise
from the fact that the X-ray and optical observations are not
simultaneous. We are interested in the distribution of $l_X$ at a
given $l_{UV}$ and $z$; however, because the X-ray observations are
not simultaneous with the optical, we do not observe the value of
$l_X$ given $l_{UV}$ for each source, but some value of $l_X$ which
has varied from the original X-ray luminosity at the time of the
optical observations.

Typical long-term X-ray variability for Seyfert 1s is $20\%$--$40\%$
with no obvious trend with luminosity \citep{grupe01, uttley02,
mark03}. The measurement errors in $l_X$ for our sample are typically
$\sim 0.07$ dex. Assuming X-ray variability amplitudes of $30\%$ for
the sources in our sample, this implies typical uncertainties in the
X-ray luminosity of $\sim 0.15$ dex. Correcting the scatter in
$\alpha_{\rm ox}$ for the contribution from X-ray variability and
measurement error, we find an implied intrinsic scatter in $l_X$ of
$\sigma \approx 0.29$ dex.

To assess whether the observed dependence of $\alpha_{\rm ox}$ on $z$
is the result of bias arising from variability, we performed Monte
Carlo simulations. Because we are interested in testing if a spurious
redshift dependence may occur due to this type of bias, we simulate
values of $l_X$, given $l_{UV}$, assuming $L_X \propto
L_{UV}^{0.65}$. Within the framework of model ${\cal M}_t$, this form
corresponds to assuming $\gamma_l = 0.134$ and $\gamma_t = 0$, and
therefore $\alpha_{\rm ox}$ depends only on UV luminosity for these
simulations. The value of $\beta_l = 0.65$ was chosen because a linear
regression of $l_X$ on $l_{UV}$ found $L_X \propto L_{UV}^{0.631 \pm
0.088}$, consistent with the work of \citet{avni86}, \citet{wilkes94},
\citet{vig03b}, \citet{strat05}, and \citet{steffen06}.

The simulations were performed as follows. We first drew 164 values of
$l_{UV}$ and $t(z)$ from a kernel estimate of their joint distribution
\citep{silv86}, after removing the censored $z < 1.5$ sources. Then,
we calculated values of $l_X$, assuming $L_X \propto
L_{UV}^{0.65}$. The random Gaussian scatter in $l_X$ about the
$l_{UV}$ dependence had a standard deviation of $\sigma = 0.30$ dex;
this value was motivated by the regression results. To simulate the
upper limits, we randomly censored 6 of the values of $l_X$, and
increased their censored values by a small random amount. We added
random Gaussian noise to the uncensored values of $l_{X}$ to simulate
the effects of variability and measurement error, where the standard
deviation of this noise was 0.15 dex. To simulate the effect of the
measurement errors on $l_{UV}$, we also added random Gaussian noise of
standard deviation 0.005 dex to the values of $l_{UV}$. We then
performed censored regression on the simulated values. We repeated
this procedure for $10^4$ simulations, and calculated the average
simulated regression coefficients for model ${\cal M}_t$,
$\bar{\gamma} = (\bar{\gamma}_l, \bar{\gamma}_{t})$, and their
covariance matrix, $\Sigma_{\gamma}$.

We calculate the $\chi^2$ of our regression coefficients for ${\cal
M}_t$ estimated from our sample, $\hat{\gamma} = (0.233, 0.029)$, as
$\chi_2^2 = (\hat{\gamma} - \bar{\gamma})^T \Sigma_{\gamma}^{-1}
(\hat{\gamma} - \bar{\gamma})$. Here, $x^T$ is the transpose of
$x$. We found a value of $\chi^2_2 = 11.62$; under the null hypothesis
that $\gamma_l = 0.134$ and $\gamma_{t} = 0$, the probability of
observing a $\chi^2_2$ this high or higher is $\lesssim 3 \times
10^{-3}$. Similar results were found by calculating the $\chi^2$ of
the regression coefficients for ${\cal M}_{l+t}$. Therefore, our
observed values of $\gamma_t$ are highly unlikely to be a spurious
correlation resulting from variability and measurement error.

In summary, we argue that Malmquist-type biases from measurement error
or variability will induce a false correlation of $\alpha_{ox}$ with
$z$ in the {\it opposite} sense of what is observed, and therefore are
not causing our finding.  We further showed through simulations that
measurement errors or variability for objects within the sample are
likewise not capable of inducing a false correlation between
variables.

\subsection{Rank Correlation Analysis}

\label{s-tau}

An alternative test for evolution of the X-ray emission for a given UV
luminosity is Kendall's generalized partial $\tau$
\citep{akritas96}. Kendall's partial $\tau$ has been used by
\citet{vig03b}, \citet{strat05}, and \citet{steffen06}, where they did
not find any evidence for a partial correlation between $\alpha_{\rm
ox}$ and $z$ based on it, consistent with their parametric
analysis. However, Kendall's partial $\tau$ has some undesirable
properties that make it difficult to assess the statistical
significance of the result \citep{nel88}. In particular, conditional
independence between variables 1 and 2, given a third variable, does
\emph{not} necessarily correspond to a value of $\tau = 0$.
Alternatively, conditional dependence of two variables given a third
does not necessarily correspond to a value of $\tau \neq 0$. Within
the context of this work, this implies that a value of $\tau = 0$ does
not necessarily correspond to the null hypothesis that $\alpha_{\rm
ox}$ (or $L_X$) is independent of $z$ at a given $L_{UV}$; i.e., an
expected value of $\tau = 0$ does not necessarily result when $p(L_X |
L_{UV}, z ) = p(L_X | L_{UV})$.

In order to test the reliability of Kendall's partial $\tau$, we use
Monte Carlo simulations to compare the distributions of $\tau$ under
the assumption that $\alpha_{\rm ox}$ only depends on $L_{UV}$, and
under the assumption that $\alpha_{\rm ox}$ depends on both $L_{UV}$
and $z$. For each of the simulations, we calculated values of
Kendall's partial $\tau$ for $\alpha_{\rm ox}$ with $z$, $\tau_{\alpha
z,l}$, and $L_X$ with $z$, $\tau_{xz,l}$, controlling for the
correlation between $L_{UV}$ and $z$. We did this for two
hypotheses. The `null' hypothesis, $H_0$, assumed $L_X \propto
L_{UV}^{0.65}$, and the alternative (i.e., `evolution') hypothesis,
$H_1$, assumed $L_X \propto L_{UV}^{0.40} e^{-t(z) / 5.5}$. The
simulations under both hypothesis were performed in the same manner as
described in \S~\ref{s-var}. The results are shown in Figure
\ref{f-taudist}.

Under the null hypothesis of no evolution, the expected values of
$\tau$ for both $L_X$ and $\alpha_{\rm ox}$ are indeed non-zero. Using
the sample average of the simulations as an estimate of the
expectation values, we find that the expected value of $\tau_{\alpha
z,l}$ under the assumption that $\alpha_{\rm ox}$ does not depend on
$z$ is 0.105.  However, under the assumption that $\alpha_{\rm ox}$
does depend on $z$, the average simulated value of $\tau$ is
-0.001. Surprisingly, the expected value of $\tau$ for the
$\alpha_{\rm ox}$--$z$ partial correlation is approximately equal to
zero when $\alpha_{\rm ox}$ depends on $z$, at least for the
simulation performed here. We investigate the behavior of Kendall's
partial $\tau$ further in Appendix B.

For our quasar sample, we find a value of Kendall's partial $\tau$ for
$L_X$ and $z$ of $\tau_{xz,l} = 0.212$, and for $\alpha_{\rm ox}$ and
$z$ of $\tau_{\alpha z,l} = 0.057$. Our sample has a value of
Kendall's regular $\tau$ between $L_{UV}$ and $z$ of $\tau_{lz} =
0.686$. As can be seen from the distributions of the simulated $\tau$,
both of our observed values of $\tau$ are about as equally consistent
with evolution of the X-ray emission at a given $L_{UV}$ as with no
evolution. In fact, there is considerable overlap between the
distributions of $\tau$ under both hypotheses, thus making it
difficult to distinguish between the two. Unfortunately, we are not
able to decide in favor of either hypothesis using Kendall's
generalized partial $\tau$.

Based on the simulations, the lack of evidence for a significant
correlation between $\alpha_{\rm ox}$ and $z$ based on Kendall's
generalized partial $\tau$ \citep{vig03b, strat05, steffen06} may be
the result of an incorrect assumption about the distribution of $\tau$
under the null hypothesis. However, it should be noted that the
parametric tests performed by \citet{vig03b}, \citet{strat05}, and
\citet{steffen06} also did not reveal any evidence for evolution of
$\alpha_{\rm ox}$. In addition, although we have shown that one can
both incorrectly reject and accept the null hypothesis based on the
partial $\tau$ statistic, there has never been a claimed rejection of
the null hypothesis of no evolution in $\alpha_{\rm ox}$ in previous
studies.

\section{RESULTS FOR $\Gamma_X$}

\label{s-gamx}

To investigate any dependence of the X-ray photon index, $\Gamma_X$,
on UV luminosity and redshift, we performed a weighted linear
regression of $\Gamma_X$ on $l_{UV}$ and $\log(1+z)$ using all 157
detected sources. The weights are made up of a combination of the
intrinsic scatter in $\Gamma_X$ and the measurement errors on
$\Gamma_X$. The results are
\begin{equation}
  \Gamma_X = -3.43(\pm 3.879) + 0.125(\pm 0.087) l_{UV} - 
  0.678(\pm 0.347) \log(1 + z).
  \label{eq-gamx_reg}
\end{equation}
Based on this regression, there is no significant evidence for a
dependence of $\Gamma_X$ on $L_{UV}$ or $z$, although the $z$
dependence is marginally significant at $\approx 2\sigma$.

Similar to $\alpha_{\rm ox}$, we experimented with parameterizing the
$z$ dependence using $t(z)$ and $z$. There was no noticeable
difference between the different parameterizations, although the
$\log(1+z)$ model gave slightly better results in the sense of
minimizing mean squared error. In addition, if any $z$ dependence of
$\Gamma_X$ is due to a systematic hardening of the X-ray spectra at
higher energies, then we might expect the $z$ dependence to be best
parameterized using $\log (1+z)$.

We show the joint confidence regions of the $l_{UV}$ and $\log (1+z)$
coefficients in Figure \ref{f-gamx_confreg}. While there is no
significant evidence that $\Gamma_X$ is related to either $L_{UV}$ or
$z$, we note that the measurement errors on $\Gamma_X$ are large and
contribute signficantly to enlarging the confidence region of the
regression coefficients. Because the confidence region of the
regression coefficients is large, the possibility that $\Gamma_X$ is
significantly correlated with both UV luminosity and redshift is also
consistent with our data.

Previous work by \citet{gall05} has found evidence for an
anti-correlation between $\Gamma_X$ and $\alpha_{UV}$. Motivated by
their work, we also perform a linear regression of $\Gamma_X$ on
$\alpha_{UV}$. We only included those $z > 0.5$ sources detected by
Chandra, leaving us with 136 sources. The redshift limit was imposed
to ensure that an adequate amount of the UV continuum was available
for estimating $\alpha_{UV}$. We used the \emph{FITEXY} procedure
\citep{numrec} with the \citet{trem02} modification to account for the
measurement errors in both $\Gamma_X$ and $\alpha_{UV}$, and the
intrinsic scatter about the regression. The result of the regression
is:
\begin{equation}
  \Gamma_X = 2.21(\pm 0.07) - 0.25(\pm 0.07) \alpha_{uv}.
  \label{eq-gamx_reg2}
\end{equation}
The $\alpha_{UV}$ coefficient is significant at $3.5\sigma$. The
regression results are consistent with the assumption that $\Gamma_X$
is linearly related to $\alpha_{UV}$, and the residuals are
approximately normally distributed. We also performed a Spearman and
Kendall rank correlation test between the two spectral slopes, and
found an anti-correlation of similar significance.

To test if including the UV red quasars affect our limits on the
luminosity and redshift dependencies, we also performed the regression
after removing all sources with $\alpha_{UV} > 1.2$. Removing these UV
red quasars did not significantly change the confidence regions shown
in Figure \ref{f-gamx_confreg}.

\section{COMPARISON WITH PREVIOUS STUDIES OF $\alpha_{\rm ox}$.}

\label{s-prevcomp}

The parametric dependence of $\alpha_{\rm ox}$ on $L_{UV}$ and $z$ has
been studied previously by several authors
\citep[e.g.,][]{avni82,avni86,wilkes94,bech03,vig03b,strat05,steffen06}. In
this analysis, we confirm the anti-correlation between $\alpha_{\rm
ox}$ and $L_{UV}$ seen previously, but also find evidence for a
correlation between $\alpha_{\rm ox}$ and redshift. Most previous
studies have not found any significant evidence that $\alpha_{\rm ox}$
is related to $z$, with the exception of
\citet{bech03}. \citet{yuan98a} found evidence for a slight dependence
of $\alpha_{\rm ox}$ with $z$ for $z < 0.5$, but with opposite sign as
that found here. Using high-quality \emph{Chandra} data, we find that
$\alpha_{\rm ox}$ is related to both $L_{UV}$ and $z$.

We perform a quantitative comparison between our results and those of
\citet{avni86}, \citet{wilkes94}, \citet{strat05}, and
\citet{steffen06}. These authors have presented their results using a
different parameterization for evolution, where they have fit a linear
relationship of the form\footnote{The term $\log \nu_{UV}$ arises
because we define $l_{UV}$ to be the logarithm of $\nu_{UV} L_{UV}$}
\begin{equation}
  \alpha_{\rm ox} = A_l (l_{UV} - 30.5 - \log \nu_{UV}) + 
  A_{\tau} (\tau(z) - 0.5) + A.
  \label{eq-avpar}
\end{equation}
Here, $\tau(z)$ is the cosmological look-back time in units of the
present age of the universe, and $\nu_{UV}$ is the frequency
corresponding to $2500\AA$. We fit a relationship of this form and
find $A_l = 0.233 \pm 0.026$ and $A_{\tau} = -0.392 \pm 0.103,$ with a
correlation of $Corr(A_l, A_{\tau}) = -0.878$. In Figure
\ref{f-confreg} we show the $95\%$ joint confidence region on our
estimate of $(A_l, A_{\tau})$, as well as the $95\%$ confidence
ellipses for AT86, W94, S05, and S06. These authors do not
report equations for their confidence regions, so we matched them by
eye. We compare with the results of AT86 obtained using their entire
sample, i.e., their BQS+BF+HET85 sample, and the results of W94
obtained using only the radio-quiet sources and assuming $\Gamma_X =
2$ as displayed in their Figure 14a. The results for the
S05 sample are presented by S06.

Statistically, our results differ from the analysis of AT86 at the
$\approx 4\sigma$ level, from W94 at the $\approx 2\sigma$ level, from
S05 at the $\approx 2.5\sigma$ level, and from S06 at the $\approx
3\sigma$ level. However, there are a number of systematic differences
between our analysis and those of AT86, W94, S05, and S06, that, when
taken into account, may introduce an additional systematic component
to the errors. The AT86 sample includes both radio-quiet and
radio-loud sources, and as noted in \S~\ref{s-sample}, the radio-loud
sources can have an additional component in their X-ray emission from
the jet. To avoid this type of contamination, we have only included
RQQs in our sample. In addition, AT86 assumed a value of $\Gamma_X =
1.5$ when calculating the 2 keV flux. As has been found here and in
many other studies, a value of $\Gamma_X \approx 2$ is more typical
for RQQs. \citet{wilkes94} calculated $A_{\tau}$ for both $\Gamma_X =
1.5$ and $\Gamma_X = 2$ and found that assuming $\Gamma_X = 2$ had the
effect of shifting $A_{\tau}$ towards more negative values. If the
assumption on $\Gamma_X$ affects the estimated $A_{\tau}$ for AT86 in
the same way as for W94, then one would expect assuming $\Gamma_X = 2$
would shift the AT86 confidence ellipse toward our estimate.

The largest systematic difference between our work and that of AT86,
W94, S05, and S06 is in the differing levels of heterogeneity of the
samples and the different instruments used to collect the X-ray
data. The AT86 and W94 X-ray observations were done using the
\emph{Einstein} Observatory Imaging Proportional Counter (IPC). About
two-thirds of the X-ray data for the S06 sample was observed using the
\emph{ROSAT} Position Sensitive Proportional Counter (PSPC), with the
remaining X-ray data from \emph{Chandra} or \emph{XMM-Newton}. S06
combined sources from the SDSS, COMBO-17 survey \citep{wolf04}, Bright
Quasar Survey \citep{schmidt83}, a heterogeneous low-$z$ Seyfert 1
sample, and a heterogenous high-$z$ sample similar to our high-$z$
sample. The S06 sample is more heterogeneous than ours, but probes a
wider range in luminosity; the ranges in cosmic age probed by S06 and
our sample are very similar. The S05 sample is a subset of the S06
sample, and does not contain the COMBO-17 and BQS sources. The main
SDSS sample of S05 and S06 consists of 155 radio-quiet quasars that
were selected from the SDSS and contained within the inner $19'$ of
\emph{ROSAT} PSPC pointings with exposure times $> 11$ ksec. Thus, the
S05 sample is very similar to ours in its heterogeneity, with the
exception of the additional low redshift Seyfert 1 sample; S05 perform
their analysis both with and without the Seyfert 1 sample. The AT86
and W94 samples are both more heterogeneous than the S06 sample. In
addition, the redshift ranges probed by the AT86 and W94 samples are
smaller ($z < 3.3$) than that of the S06 sample, the S05 sample, and
our sample, probing a slightly smaller range in cosmic time.

Our sample only consists of \emph{Chandra} ACIS
observations. Increased sensitivity gives \emph{Chandra} the ability
to detect sources with rest-frame 2--10 keV flux down to $f_{HB} \sim
2 \times 10^{-15}\ {\rm ergs\ s^{-1}}$ in a $\sim 20$ ksec
observation. In this work, we estimate the \emph{Chandra} 2 keV flux
densities using data over a broader spectral range (0.3--7 keV) than
that probed by the \emph{Einstein} IPC (0.4--4 keV) and the
\emph{ROSAT} PSPC (0.1--2.5 keV). Our sample is also more homogeneous
than those used previously, except for the S05 SDSS + high-$z$ sample,
being made up almost entirely of SDSS sources which had serendipitous
\emph{Chandra} observations; unfortunately this also results in our
sample probing a smaller range in luminosity. Similar to our work, the
high-$z$ samples of S06 and S05 both consist of sources with
\emph{Chandra} and \emph{XMM} data, and the COMBO-17 sample of S06
also consists of \emph{Chandra} data; both the S05 and S06 samples
have slightly lower X-ray detection fractions than our sample.

\citet{avni86} and W94 estimate the $2500\AA$ flux density from
published $B$- and $V$-band magnitudes assuming a constant spectral
slope of $\alpha = 0.5$. The $2500\AA$ flux density for many of the
S05 and S06 sources were measured directly from the SDSS
spectra. However, the $2500\AA$ flux densities for the 52 COMBO-17
sources of S06 were estimated by interpolation and extrapolation. The
$2500\AA$ flux densities for the 46 BQS sources in S06 were estimated
from the $3000\AA$ flux assuming a constant spectral slope of $\alpha
= 0.5$. The dispersion in quasar spectral slopes is large
\citep[$\approx 0.3$,][]{rich01}, and this large dispersion can result
in a non-negligible error on $l_{UV}$ if one assumes a constant
spectra slope, especially if one is extrapolating over a large range
in wavelength. These issues are exacerbated when one fits $\alpha_{\rm
ox}$ instead of $l_X$, as the errors on $l_{UV}$ contribute to the
errors on $\alpha_{\rm ox}$, thus not only increasing the scatter
about the regression, but also correlating the errors on $l_{UV}$ and
$\alpha_{\rm ox}$. In addition, as noted in \S~\ref{s-var}, the X-ray
variability can also bias the coefficients for censored regression. In
particular, these issues will affect the AT84 and W94 results because
of the larger wavelength difference between $2500\AA$ and the $B$- and
$V$-bands for many of the sources, and the lower detection fraction
($\sim 60\%$). The analysis of S05 and S06 is unlikely to
be significantly affected by these issues due to the high detection
fraction and large number of sources with directly measured $2500\AA$
flux densities. Furthermore, these authors found consistent results
when analyzing different subsamples of their data.

\citet{strat05} report values of Kendall's generalized partial $\tau$
for a partial correlation between $\alpha_{\rm ox}$ and $z$, given
$L_{UV}$, for their main SDSS sample combined with their high-$z$
sample. Because the main + high-$z$ sample of S05 is very similar to
ours in distribution of $L_{UV}$ and $z$ and the number of sources, we
expect that the distribution of Kendall's partial $\tau$ under the
no-evolution and evolution hypotheses should also be
similar. \citet{strat05} find a value of $\tau = 0.03$, where we have
corrected for the sign difference between our definition of
$\alpha_{\rm ox}$ and theirs. This value of $\tau$ is consistent with
our value of $\tau = 0.057$. Because their value of $\tau$ is not
significantly different than $\tau = 0$, and because their parametric
analysis gave similar results, S05 concluded that there is no evidence
that $\alpha_{\rm ox}$ changes with redshift. However, as per the
discussion in \S~\ref{s-tau}, the expected value of $\tau$ under the
null hypothesis of no evolution in $\alpha_{\rm ox}$ is in general not
$\tau = 0$, and therefore it is inappropriate to calculate signficance
levels with respect to $\tau = 0$. Comparison with Figure
\ref{f-taudist} implies that the value of $\tau = 0.03$ found by S05
is about as equally consistent with the evolution hypothesis as with
the no-evolution hypothesis. However, the parametric analysis by S05
still differs from ours at the $\approx 2.5\sigma$ level.

Similar to \citet{bech03}, we find that $\alpha_{\rm ox}$ is
correlated with both UV luminosity and redshift. However, in contrast
to \citet{bech03}, we find that $\alpha_{\rm ox}$ depends more
strongly on $L_{UV}$ than on $z$. In addition, we find that RQQs are
systematically more X-ray loud at higher redshift. \citet{bech03}
found that $\alpha_{\rm ox}$ is larger (more X-ray quiet) for high-$z$
sources, but found an overall trend where $\alpha_{\rm ox}$ becomes
more X-ray loud as $z$ increases. \citet{bech03} did not perform a
regression or partial correlation analysis, but we note that their
observed marginal distribution of $\alpha_{\rm ox}$--$z$ is such that
$\alpha_{\rm ox}$ becomes more X-ray loud as $z$ increases for $z
\lesssim 2$. This is opposite the trend seen in our data, where
inspection of Figure \ref{f-xuvdist} reveals that $\alpha_{\rm ox}$ is
observed to become more X-ray quiet with increasing $z$, if one does
not correct for the $L_{UV}$--$z$ correlation.

The source of this discrepancy is likely the values of $\Gamma_X$ used
by \citet{bech03} for their \emph{ROSAT} sources. \citet{bech03} used
$\Gamma_X$ values taken from \citet{yuan98a}, which were calculated
using the two hardness ratios given by the Standard Analysis Software
System (SASS), to estimate the flux density at 2 keV. These values of
$\Gamma_X$ steadily decrease from $\Gamma_X \sim 2.6$ at $z \sim 0$,
to $\Gamma_X \sim 2$ at $z \sim 2$. However, values of $\Gamma_X \sim
2.6$ are steeper than is commonly seen in RQQs, as has been found in
this work and in other recent studies
\citep[e.g.,][]{reeves00,picon03}. Therefore, assuming a power-law and
the values of $\Gamma_X$ obtained by \citet{yuan98a} may not provide
an accurate estimate of the 2 keV flux density, and thus $\alpha_{\rm
ox}$. If $\Gamma_X \sim 2$, then assuming values of $\Gamma_X \sim
2.6$ will systematically under-predict the 2 keV flux density for a
given 0.1--2.4 keV flux, and consequently provide estimates of
$\alpha_{\rm ox}$ that are too large. Furthermore, a steady decrease
from $\Gamma_X \sim 2.6$ at $z \sim 0$ to $\Gamma_X \sim 2$ at $z \sim
2$ would produce a similar observed decrease in $\alpha_{\rm ox}$ from
$z \sim 0$ to $z \sim 2$, thus increasing the magnitude of any
$\alpha_{\rm ox}$--$z$ anti-correlation. Considering that $L_{UV}$
also increases with increasing $z$ due to flux limits, this would also
weaken any observed correlation between $\alpha_{\rm ox}$ and
$l_{UV}$, and thus lead \citet{bech03} to conclude that $\alpha_{\rm
ox}$ is a stronger function of redshift. This is what is observed in
the \citet{bech03} data, where $\alpha_{\rm ox}$ is observed to
decrease from $z \sim 0$ to $z \sim 2$. At $z \sim 2$, the
\citet{yuan98a} sources have values of $\Gamma_X$ that are more
typical of RQQs, $\Gamma_X \sim 2$. In addition, after $z \sim 2$, the
trend in $\alpha_{\rm ox}$ is observed to change sign, increasing with
increasing $z$, consistent with the data presented here. This is also
the redshift where the \citet{bech03} sample becomes dominated by
\emph{Chandra} sources, and have 2 keV flux densities calculated
assuming $\Gamma_X = 2.2$.

\section{DISCUSSION}

\label{s-discussion}

In this paper we were able to separate the dependence of $\alpha_{ox}$
on the quasar luminosity, $L_{UV}$, from that of cosmic epoch, $z$,
and we find that both dependencies are present, though with opposite
sign. From this, it follows that RQQs become more X-ray quiet
(increasing $\alpha_{\rm ox}$) with increasing UV luminosity, and
become more X-ray loud with increasing redshift. An analysis based on
the Kullback-Leibler information finds evidence that $\alpha_{\rm ox}$
may depend nonlinearly on $l_{UV}$ at a given $t(z)$, with
$\alpha_{\rm ox}$ increasing more rapidly as $l_{UV}$ increases.

One may be able to find a better parameterization for the redshift
dependence than the one adopted here, but that would only strengthen
our claims of evidence for a dependence of $\alpha_{\rm ox}$ on
$z$. In addition, as argued in \S~\ref{s-modcomp}, it is unlikely
that the observed redshift dependence can be explained by nonlinearity
in the $\alpha_{\rm ox}$--$l_{UV}$ relationship. However, expanding
the model space to include other parameters such as black hole mass or
accretion rate may provide a better fit and be preferred over models
which contain only a redshift and $L_{UV}$ dependence. For example, a
dependence of $\alpha_{\rm ox}$ on black hole mass, $M_{BH}$, is
predicted by some models for accretion disk and hot corona
\citep[e.g.,][]{jan00, bech03}. If such a correlation exists, than it
may be that the the $\alpha_{\rm ox}$--$t(z)$ relationship is simply
tracing the underlying evolution of the active black hole mass
function, which is then projected onto the $\alpha_{\rm ox}$-$t(z)$
plane via an $\alpha_{\rm ox}$--$M_{BH}$ relationship. In this case
the statistical model that contains $M_{BH}$ would provide the best
fit, and there would be no need for an additional redshift
dependence. However, in the absence of such information, we find that
the model that best describes our data is given in terms of a
quadratic dependence on $l_{UV}$ and a linear dependence on $t(z)$.

\subsection{$K$-Corrections and the $\alpha_{\rm ox}$--$z$ Relationship}

It may be suggested that the $\alpha_{\rm ox}$ dependence on $z$ is
caused by a varying $\Gamma_X$ as the observed \emph{Chandra} spectral
range shifts to harder rest-frame energies. If $\Gamma_X$ were to
steepen at higher energies, creating a softer X-ray continuum at these
energies, then we would be systematically over-estimating the 2 keV
flux densities, thus explaining the $\alpha_{\rm ox}$--$z$
relationship. While we do not find any strong evidence for a change in
$\Gamma_X$ with $z$, the results from \S~\ref{s-gamx} shown in Figure
\ref{f-gamx_confreg} suggest that if there is spectral curvature, then
$\Gamma_X$ likely flattens with increasing energy. A flattening of the
X-ray continua at higher energies is opposite the trend needed to
explain the $\alpha_{\rm ox}$--$z$ relationship, and thus our result
cannot be explained by a systematic steepening of the intrinsic X-ray
continuum at harder energies.

Because we do not fit an intrinsic absorber to most of our sources, it
may also be suggested that the observed redshift dependence of
$\alpha_{\rm ox}$ is caused by redshifting of soft X-ray absorption
out of the observed spectral region. An intrinsic absorber will more
strongly absorb the softer X-rays, and therefore will more
significantly affect the observed X-ray continuum of lower redshift
sources. This could then cause a spurious anti-correlation between
$L_X$ and $t(z)$.

To test if the observed dependence of $\alpha_{\rm ox}$ on $t(z)$ is
the result of soft X-ray absorption shifting out of the observed band,
we used \emph{SHERPA}'s FAKEIT routine to simulate observed X-ray
spectra as a function of $z$. We assumed a power-law continuum with
$\Gamma_X = 2$, and an intrinsic neutral absorber with column density
$N_H = 10^{21}\ {\rm cm^{-2}}$. We argue in the next paragraph that a
column density of $N_H = 10^{21}\ {\rm cm^{-2}}$ is greater than the
maximum $N_H$ allowed by the $\Gamma_X$--$z$ regression, and thus we
use $N_H = 10^{21}$ as an upper limit on the effect of unrecognized
neutral intrinsic absorption on the $\alpha_{\rm ox}$--$z$
relationship. The observed X-ray continuum was simulated for a source
at $z = 0, 1, 2, 3,$ and $4$, and the intrinsic luminosity of the
source was kept constant. We then fit each simulated spectrum with
only a power-law. This resulted in the inferred X-ray luminosity of $z
= 0$ sources being a factor of $\sim 2$ lower than the $z = 4$
sources. Therefore, based on these simulations, ignoring intrinsic
absorption can result in a spurious decline in $L_X$ from $z = 4$ to
$z = 0$ by a factor of $\sim 2$ when $N_H \sim 10^{21}\ {\rm
cm^{-2}}$. However, the results of our $\alpha_{\rm ox}$ regression
imply that the X-ray luminosity drops by a factor of $\sim 8$ from $z
= 4$ to $z = 0$, and therefore the observed $\alpha_{\rm ox}$--$z$
relationship cannot be explained as a spurious correlation resulting
from unidentified intrinsic neutral absorption.

We can use our simulated spectra to constrain a typical value of $N_H$
for our sources, assuming that $N_H$ remains roughly constant with
redshift. The observed photon index of the simulated spectra dropped
from $\Gamma_X = 2$ at $z = 4$ to $\Gamma_X \approx 1.4$ at $z =
0$. From Figure \ref{f-gamx_confreg}, we note that the $3\sigma$ limit
on the maximal drop in observed $\Gamma_X$ between $z = 4$ and $z = 0$
is $\Delta \Gamma_X \approx 0.35$. This is considerable less than the
observed drop in $\Gamma_X$ from the simulations, and thus represents
more than the maximal amount of change in $\Gamma_X$ with redshift
that is allowed by our data. Therefore, assuming only neutral
absorption, values of $N_H \gtrsim 10^{21}\ {\rm cm^{-2}}$ would
produce observed values of $\Gamma_X$ at $z = 0$ that are too flat,
and thus $N_H \lesssim 10^{21}\ {\rm cm^{-2}}$ for most of our
sources.

Because the redshift dependence of $\alpha_{\rm ox}$ cannot be
explained by a systematic steepening of $\Gamma_X$ at higher energies,
or by an unidentified intrinsic neutral absorber shifting out of the
observed \emph{Chandra} bandpass, we conclude that the $\alpha_{\rm
ox}$--$z$ relationship is likely the result of evolution of the
accretion mechanism and environment. However, more complex absorption
models, such as an ionized or partial covering absorber cannot be
ruled out as causing the observed $\alpha_{\rm ox}$--$z$ dependence,
but investigation of such models is beyond the scope of this work.

\subsection{$\Gamma_X$ Relationships}

We do not find significant evidence for a correlation between the
radio-quiet quasar X-ray spectral photon index, $\Gamma_X$, and UV
luminosity or redshift. This is consistent with results obtained using
\emph{XMM} observations of SDSS RQQs \citep{ris05}, \emph{ASCA}
observations of RQQs \citep{reeves00}, and fitting of composite
spectra of $z > 4$ RQQs \citep[e.g.,][]{vig03c, vig05,
shem06a}. However, this is in contrast with the work of \citet{dai04}
and \citet{bech03}. \citet{dai04}, found evidence for a correlation
between $\Gamma_X$ and $L_X$ using a small sample of
gravitationally-lensed RQQs. \citet{bech03} used Kendall's generalized
$\tau$ to assess the 2-dimensional correlations between $\Gamma_X$ and
both luminosity and $z$, and found evidence that $\Gamma_X$ is
correlated with luminosity and anti-correlated with $z$. In this work
we have used linear regression to control for the artifical
correlation between luminosity and redshift, and find that there is no
significant evidence that $\Gamma_X$ varies with $l_{UV}$ and
$z$. However, inspection of Figure \ref{f-gamx_confreg} reveals that
if $\Gamma_X$ does depend on $l_{UV}$ and $z$, then the directions of
these trends are likely in agreement with the correlations seen by
\citet{bech03}. In addition, a systematic flattening of $\Gamma_X$
with increasing $z$ has also been seen in \emph{XMM} data by
\citet{page03}, and \citet{vig99} found some evidence that $\Gamma_X$
is flatter on average for $z \sim 2$ RQQs than for lower $z$ RQQs..

It is interesting to note that at $z \sim 2$ the observed \emph{ROSAT}
band has shifted to $\sim 0.3$--$7.2$ keV in the quasar rest-frame,
overlapping with our rest-frame Chandra band at $z \sim
0$. Considering that the $z \sim 2$ \emph{ROSAT} sources of
\citet{yuan98a} have values of $\Gamma_X$ similar to those observed
here with \emph{Chandra} at $z \sim 0$, and noting that the source
rest-frame energies are approximately the same in these two observed
spectra regions, this implies that the soft X-ray spectra of RQQs may
be more complex than a simple power-law. In particular, the
\citet{yuan98a} sources may exhibit a soft excess at $\lesssim 0.3$
keV, causing steeper hardness ratios. Soft excesses have been seen in
good-quality spectra of other low-$z$ RQQs \citep[e.g.,][]{gier04},
but the origin of this component is still unclear. If such additional
complexity exists, it may be the cause of the steeper values of
$\Gamma_X$ seen in the low-$z$ \citet{yuan98a} \emph{ROSAT} sources,
and thus the strong anti-correlation between $\Gamma_X$ and $z$, and
consequently $\alpha_{\rm ox}$ and $z$, seen by \citet{bech03}.

Although, there is no significant evidence for a $\Gamma_X$--$z$
relationship, there is marginally significant evidence ($2\sigma$)
that $\Gamma_X$ flattens as $z$ increases. While this may be caused by
evolution in $\Gamma_X$, it may also represent a systematic flattening
of the X-ray continuum at harder energies or the result of a Compton
reflection component redshifting into the observed 0.3--7 keV band at
higher $z$. Similarly, a soft excess redshifting out of the observed
band may also contribute.

An observed anti-correlation between $\Gamma_X$ and $\alpha_{UV}$ may
result from not fitting an intrinsic absorber to most of the X-ray
spectra. The sources with redder $\alpha_{UV}$ may have higher $N_H$,
which would result in a lower value of $\Gamma_X$ inferred from the
power-law spectral fit. To test this, we estimated the increase in
$N_H$ between $\alpha_{UV} \sim 0.1$ and $\alpha_{UV} \sim 1.2$ needed
to produce the observed decrease in $\Gamma_X$. We used
\emph{Sherpa}'s FAKEIT command to simulate spectra assuming a value of
$\Gamma_X = 2.2$ and negligible intrinsic neutral absorption, $N_H =
10^{20}\ {\rm cm^{-2}}$. A value of $\Gamma_X \sim 2.2$ is typical for
the bluer sources, $\alpha_{UV} \sim 0.1$. Based on the simulations,
$N_H$ must increase to $\sim 10^{21}\ {\rm cm^{-2}}$ at $\alpha_{UV}
\sim 1.2$ to produce an observed decrease of $\Gamma_X$ from $\Gamma_X
\sim 2.2$ to $\Gamma_X \sim 1.8$. We fit an absorbed power-law for the
ten reddest sources with $> 50$ counts, and found that these sources
typically had $3\sigma$ upper limits of $N_H \lesssim 5 \times
10^{21}\ {\rm cm^{-2}}$. Values of $N_H \sim 10^{21}\ {\rm cm^{-2}}$
are well within the limits on $N_H$ at $\alpha_{UV} \sim 1.2$, and
therefore we cannot rule out the observed $\Gamma_X$--$\alpha_{UV}$
anti-correlation as resulting from unidentified intrinsic absorption.

\subsection{Expectation of Accretion Models}

\citet{sob04a} and \citet{sob04b} explored the general parameter space
available for accreting compact sources, and quasars in particular, in
respect to geometry of the disk and X-ray emitting region. They show
that $\alpha_{ox}$ is most sensitive to (1) the amount of energy
dissipated in the corona or (2) the size of the inner flow or a
structure and outflow velocity of the coronal flares. \citet{sob04b}
suggest that the $\alpha_{ox}$--$L_{UV}$ anti-correlation can be
explained by differences in the structure of the X-ray emitting
region. They point out that in the framework of the truncated disk and
hot inner flow geometry, $L_{UV}$ increases when the disk extends
further towards the last stable orbit, while $\Gamma_X$ steepens,
reducing the 2 keV emission.  In the patchy corona geometry, the
$\alpha_{ox}$--$L_{UV}$ relationship can be explained by changes in
the fraction of gravitational energy dissipated in the corona, where a
lower fraction results in a weaker and softer (higher $\Gamma_X$)
X-ray continuum.

The redshift dependence may similarly be explained as resulting from
evolution in the accretion geometry. This would imply that for a given
$L_{UV}$, the high redshift sources have larger radii of the inner hot
flow sphere or they generate more flares with higher outflow
velocities. Both of these explanations imply that $\Gamma_X$ is also
correlated with UV luminosity and anti-correlated with redshift, where
$\Gamma_X$ steepens with increasing $L_{UV}$ and flattens with
increasing $z$. We find no statistically significant evidence for a
correlation between $\Gamma_X$ and $L_{UV}$, and only marginally
significant evidence for an anti-correlation between $\Gamma_X$ and
$z$. However, it should be noted that there is considerable
uncertainty in the regression coefficients, and their joint confidence
region is large. While values of zero for the regression coefficients
cannot be ruled out, it is interesting to note that the trends of
$\Gamma_X$ with $L_{UV}$ and $z$ implied by the regression are
consistent with the model predictions. So long as the accretion disk
models do not predict too strong of a relationship between $\Gamma_X$
and $L_{UV}$, they may still be consistent with our results.

A dual dependency of $\alpha_{ox}$ on both $L_{UV}$ and $z$ must
relate, in current scenarios, to variations in $M_{BH}, \dot{m}$, and
chemical abundances, and their effects on the accretion disk and
corona. Unfortunately, the models do not yet predict a specific
relationship between $\alpha_{ox}$ and the model parameters, and thus
quantitative comparison of our results with the models is difficult;
this will be the subject of future research. In addition, the
discussion in this section has been model-dependent, and hopefully
magneto-hydrodynamic simulations will provide further insight
\citep[e.g.,][]{devill03,krolik05}.

\subsection{Improving the $\alpha_{\rm ox}$ Analysis}

The main source of statistical uncertainty in $\gamma_l$ and
$\gamma_t$ is the strong degree of correlation between $l_{UV}$ and
$t(z)$. For ordinary least-squares regression, the standard errors in
the regression coefficients are inflated upwards by a factor of $1 /
\sqrt{1 - r^2}$, where $r$ is the correlation between $l_{UV}$ and
$t(z)$ \citep{fox97}. For our sample, $r = -0.878$, and therefore the
standard errors on the regression coefficients are a factor of
$\approx 2$ higher than if $l_{UV}$ and $t(z)$ were
uncorrelated. Because the anti-correlation between $l_{UV}$ and $t(z)$
is so strong, even a small reduction in this correlation can give a
large reduction in the standard errors of $\gamma_l$ and
$\gamma_t$. For example, selecting a sample to have $r = -0.7$ will
result in a reduction in the standard deviations of $\gamma$ by about
$30\%$. Future $\alpha_{\rm ox}$ studies should try to select samples
that minimize $r$, as has been done by S06. This, along with the
larger sample size and range in luminosity probed by S06, is likely
the reason why their confidence regions are smaller (cf.,
Fig. \ref{f-confreg}).

Using the fact that the coefficient uncertainties are proportional to
$1 / \sqrt{1 - r^2}$, we investigated whether targeting more faint $z
\sim 4$ RQQs with \emph{Chandra} will significantly reduce the
standard errors in the regression coefficients. Unfortunately,
targeting a reasonable number of additional faint $z \sim 4$ RQQs will
not significantly improve the estimates of $\gamma_l$ and
$\gamma_t$. Targeting 10 additional RQQs uniformly distribution
between $45.5 < \log \nu L_{\nu} (2500\AA) < 46.5$ and $3.7 < z < 5$
will only result in a reduction in the standard errors of $\sim
8\%$. Including 30 additional faint, high-$z$ RQQs reduces the
standard errors by $\sim 18\%$, but about half of this reduction is
the result of the increased sample size.

\section{SUMMARY}

\begin{itemize}

\item 
  There is a significant relationship between $\alpha_{\rm ox}$,
  $L_{UV}$, and $t(z)$, and we did not find any evidence that
  $\alpha_{\rm ox}$ depends on the UV spectral slope. If we remove the
  10 suspected BALs ($z < 1.5$ non-detections), the two best
  $\alpha_{\rm ox}$ regressions are
  \begin{eqnarray}
  \bar{\alpha}_{\rm ox} & = & (-9.273 \pm 1.241) + (0.233 \pm 0.026) l_{UV} + 
    (2.870 \pm 0.750) \times 10^{-2} t(z), \ \sigma = 0.125 \nonumber \\
  \bar{\alpha}_{\rm ox} & = &  (40.97 \pm 27.34) - (1.949 \pm 1.187) l_{UV} + 
    (2.370 \pm 1.288) \times 10^{-2} l_{UV}^2 + \\
    & & (2.263 \pm 0.813) \times 10^{-2} t(z), \ \sigma = 0.124 \nonumber
  \end{eqnarray}
  Here, the notation $\bar{\alpha}_{\rm ox}$ denotes the average
  $\alpha_{\rm ox}$ at a given $l_{UV}$ and $t(z)$, and the intrinsic
  scatter in $\alpha_{\rm ox}$ at a given $l_{UV}$ and $t(z)$ has a
  dispersion of $\sigma \approx 0.125$ about $\bar{\alpha}_{\rm ox}$.
  Although the $l_{UV}^2$ term is not `statistically significant' in
  the classical sense, an analysis based on the Kullback-Leibler
  information found that this model is best supported by the evidence
  in our data. The KLI analysis found that both models are preferred
  over a purely quadratic dependence of $\alpha_{\rm ox}$ on $l_{UV}$,
  and over models that parameterized the redshift dependence as linear
  in $z$ or $\log(1+z)$.
\item 
  We used Monte Carlo simulations to show that the $\alpha_{\rm
  ox}$--$z$ relationship is not a spurious result caused by
  variability and measurement error. Based on the simulations, we
  calculate the $\chi^2_2$ of our regression coefficients for model
  ${\cal M}_t$ and find that the probability of observing a $\chi^2_2$
  this high or higher is $\lesssim 3 \times 10^{-3}$, under the
  assumption of no evolution in $\alpha_{\rm ox}$.
\item 
  We used Monte Carlo simulations to show that interpretation of
  Kendall's generalized partial $\tau$ is problematic. In particular,
  Kendall's partial $\tau$ for the $\alpha_{\rm ox}$--$z$ correlation
  is not necessarily expected to be zero when $\alpha_{\rm ox}$ is
  unrelated to $z$, given $L_{UV}$. Moreover, Kendall's partial $\tau$
  is not necessarily expected to be non-zero when $\alpha_{\rm ox}$ is
  correlated with $z$, given $L_{UV}$. This can have a significant
  effect on the power of Kendall's partial $\tau$, and therefore care
  must be taken when using $\tau$ to investigate whether $\alpha_{\rm
  ox}$ evolves or not. Based on our simulations, we are not able to
  decide for either evolution or no evolution in $\alpha_{\rm ox}$
  using $\tau$.
\item
  The $\alpha_{\rm ox}$--$z$ correlation cannot be explained as a
  result of a systematic steepening of the X-ray continuum at higher
  energies, as this is inconsistent with the regression of $\Gamma_X$
  on $z$. Furthermore, the $\alpha_{\rm ox}$--$z$ relationship is not
  the result of soft X-ray neutral absorption shifting out of the
  observed band. The observed factor of $\sim 8$ drop in $L_X$ from $z
  = 4$ to $z = 0$ is higher than the factor of $\sim 2$ drop in $L_X$
  that would result for an unidentified intrinsic neutral absorber
  with $N_H \sim 10^{21}\ {\rm cm^{-2}}$. Higher values of $N_H$ are
  inconsistent with the $\Gamma_X$--$z$ results.
\item 
  We do not find any evidence for a dependence of $\Gamma_X$ on UV
  luminosity, and only marginally significant evidence ($2\sigma$) for
  a dependence of $\Gamma_X$ on redshift. The $\Gamma_X$--$z$
  relationship may be caused by a systematic flattening of the X-ray
  continuum at higher energies, by redshifting of a Compton reflection
  component into the observed 0.3--7 keV band, and/or by redshifting
  of a soft excess out of the observable band. 
\item
  We find evidence for an anti-correlation ($3.5\sigma$) between
  $\Gamma_X$ and the UV spectral slope, where the X-ray continuum
  hardens as the UV continuum softens. This may be the result of
  unidentified intrinsic absorption, with the UV redder sources having
  higher intrinsic $N_H$, thus causing a flatter inferred X-ray
  continuum.

\end{itemize}

\acknowledgements

This research was supported by Chandra Observatory Guest Observer
award GO4-5112X, and in part by NASA contract NAS8-39073 (AS, TLA). We
are grateful to Martin Elvis for many helpful comments and discussions
on this work, Marianne Vestergaard for providing the UV Fe emission
template used in this study, and to the anonymous referee for a
careful reading and comments that contributed to the significant
improvement of this paper.

This research has made use of the NASA/IPAC Extragalactic Database
(NED) which is operated by the Jet Propulsion Laboratory,
California Institute of Technology, under contract with the
National Aeronautics and Space Administration.
   
Funding for the SDSS and SDSS-II has been provided by the Alfred
P. Sloan Foundation, the Participating Institutions, the National
Science Foundation, the U.S. Department of Energy, the National
Aeronautics and Space Administration, the Japanese Monbukagakusho,
the Max Planck Society, and the Higher Education Funding Council for
England. The SDSS Web Site is http://www.sdss.org/.

The SDSS is managed by the Astrophysical Research Consortium for the
Participating Institutions. The Participating Institutions are the
American Museum of Natural History, Astrophysical Institute Potsdam,
University of Basel, Cambridge University, Case Western Reserve
University, University of Chicago, Drexel University, Fermilab, the
Institute for Advanced Study, the Japan Participation Group, Johns
Hopkins University, the Joint Institute for Nuclear Astrophysics, the
Kavli Institute for Particle Astrophysics and Cosmology, the Korean
Scientist Group, the Chinese Academy of Sciences (LAMOST), Los Alamos
National Laboratory, the Max-Planck-Institute for Astronomy (MPIA),
the Max-Planck-Institute for Astrophysics (MPA), New Mexico State
University, Ohio State University, University of Pittsburgh,
University of Portsmouth, Princeton University, the United States
Naval Observatory, and the University of Washington.

\appendix

\section{THE KULLBACK-LEIBLER INFORMATION}

\label{s-appendix}

The Kullback-Leibler Information \citep[KLI,][]{kli} is a common
method for comparing models and their representation of data. The KLI
may be thought of as representing the information lost when a
parametric model is used to approximate the true distribution that
gave rise to the \citep{ander00}. This approach is appealing because
it attempts to find the model that is `closest' to the true
distribution. This approach differs from the classical method of
statistical hypothesis testing, in that the KLI looks for the model
that best describes the data, without assuming that the true model is
among the set of models considered. Furtermore, the KLI allows for the
comparison of both nested and nonnested models. This is important,
because the parametric models considered in Equations
(\ref{eq-zmean})--(\ref{eq-ltmean}) are idealizations that are
unlikely to be completely true, and do not form a set of nested
parametric forms. However, among these idealizations, we can attempt
to find the model that best describes the observed data, while fully
admitting that such a model is unlikely to be true exactly. In
contrast, the classical approach assumes that some null hypothesis is
correct, and then tests whether the model parameters are compatible
with this null hypothesis at some set significance level. The
significance level is usually set such that, if the null hypothesis is
true, then one would incorrectly reject it with low probability. In
other words, the classical approach assumes that a `null' model is
correct, and then looks for overwhelming evidence to the
contrary. However, a comparison of models based on the KLI does not
make any \emph{a priori} assumptions about which model is correct, but
rather assesses which (flawed) model best describes the observed data;
i.e., is `closest' to the true probability density that generated the
data. In this sense, the KLI evaluates the evidence \emph{for} the
models considered, whereas the classical approach only evaluates the
evidence \emph{against} some assumed null model. In addition,
comparing models based on the KLI has the advantage that all models
may be compared simultaneously, whereas the classical approach can
only compare two models at a time.

Problems with the classical approach to statistical hypothesis testing
and model selection have been known for some time, and many authors
have proposed using the KLI as an alternative
\citep[e.g.,][]{aic}. There is a large literature on these issues; see
\citet{ander00} and references therein for a more thorough discussion
of the problems with classical hypothesis testing and the advantages
of the information-theoretic approach. 

The KLI measures the discrepancy between the model distribution for
the data, $p(y|\theta)$, parameterized by $\theta$, and the true
distribution of the data $f(y)$; note that $p(y|\theta)$ is the
likelihood function for $y$. The KLI may be thought of as the relative
entropy of a statistical model, and is given by
\begin{equation}
H(\theta) = \int \log \left( \frac{f(y)}{p(y|\theta)} \right) f(y) dy.
\label{eq-kli}
\end{equation}
The difference in KLI between two statistical models,
$p_j(y|\theta_j)$ and $p_k(y|\theta_k)$, is then
\begin{equation}
  H(\theta_j) - H(\theta_k) = \int [ \log p_k(y|\theta_k) - \log
    p_j(y|\theta_j) ] f(y) dy.
  \label{eq-klidiff}
\end{equation}
As is apparent from Equation (\ref{eq-klidiff}), the difference in KLI
between two statistical models is the expectation of the difference in
their log-likelihoods, multiplied by -1. The Kullback-Leibler
information describes the information lost when a statistical model,
$p(y|\theta)$, is used to approximate the true distribution, $f(y)$.

For this work $y = \alpha_{\rm ox}$, and the true sampling
distribution of $\alpha_{\rm ox}$ is denoted as $f(\alpha_{\rm ox} |
l_{UV}, z)$. The distribution is made conditional on $(l_{UV}, z)$
because we are interested in how $\alpha_{\rm ox}$ is distributed at a
given $l_{UV}$ and $z$.

In order to choose the model that minimizes the KLI, it is necessary
to find the values of the model parameters, $\theta$, that minimize
Equation (\ref{eq-kli}). Unfortunately, this requires knowledge of the
unknown sampling density, $f(\alpha_{\rm ox}|l_{UV}, z)$. However, the
maximum likelihood estimate of $\theta$, $\hat{\theta}$, provides a
good estimate of the $\theta$ that minimizes the KLI, and in fact
converges to it as the sample size becomes large
\citep[e.g.,][]{shib97}. Furthermore, for non-censored Gaussian data,
$\hat{\theta}$ corresponds to the $\theta$ that minimizes the squared
error between the data and the model predictions, and thus finding the
model that minimizes the KLI is asymptotically equivalent to finding
the model that minimizes the expected squared error. In this work the
amount of censoring is small, and the censored data only contribute to
the log-likelihood at the $\approx 5\%$ level. Because the residuals
from the $\alpha_{\rm ox}$ regressions are approximately Gaussian, and
because the amount of censoring is small, we expect our statistical
models to behave similarly to the usual uncensored case for Gaussian
data. Therefore, finding the parameterization that minimizes the KLI
has the straight-forward interpretation of finding the
parameterization that approximately minimizes the expected squared
error.

Because the difference in KLI between two models is the expected
difference in their log-likelihoods, one can estimate
$H(\hat{\theta}_j) - H(\hat{\theta}_k)$ for a single data point using
the sample mean of the difference in log-likelihoods, evaluated at the
maximum likelihood estimates of $\theta$. However, this produces a
biased estimate of $H(\theta_j) - H(\theta_k)$, as we use the same
data to fit the model as to estimate its KLI. \citet{aic} showed that
this bias is on the order of the number of free parameters in the
statistical model. This led him to define the Akaike Information
Criterion (AIC):
\begin{equation}
  AIC = -2 \log p(y|\hat{\theta}) + 2 d.
  \label{eq-aic}
\end{equation}
Here, $\log p(y|\hat{\theta})$ is the log-likelihood of the data
evaluated at the maximum-likelihood estimate of $\theta$, and $d$ is
the number of free parameters in the model. The differences in $AIC$
between models may then be used as estimates of the differences in KLI
between the models. The model with the best estimated KLI is the model
that minimizes the $AIC$.

\section{Kendall's Partial $\tau$}

To further assess the behavior of Kendall's partial $\tau$ we
performed additional simulations, varying the degree of correlation
between $L_{UV}$ and $z$. We drew 165 values of $l_{UV}$ and $\log z$
from a multivariate normal density for 1000 simulations, with
correlations between $l_{UV}$ and $\log z$ of $\rho = 0.0, 0.3, 0.6,$
and $0.9$. The remainder of the simulations were performed in an
identical manner to those described in \S~\ref{s-var} and
\S~\ref{s-tau}, with the exception that we did not include the effects
of variability and measurement error. For each of these cases, we
calculated the average values of $\tau_{xz,l}$ and $\tau_{\alpha z,l}$
under both the null and alternative hypotheses. From the notation of
\S~\ref{s-tau}, $\tau_{xz,l}$ denotes the value of Kendall's partial
$\tau$ between $L_X$ and $z$, controlling for $L_{UV}$, and
$\tau_{\alpha z,l}$ denotes the value of Kendall's partial $\tau$
between $\alpha_{\rm ox}$ and $z$, controlling for $L_{UV}$.

In addition, we also calculate the power of the test when using either
$\tau_{xz,l}$ and $\tau_{\alpha z,l}$ under both hypotheses; the power
of a statistical test is the probability of choosing for the
alternative hypothesis. We have chosen a significance level of 0.05,
meaning that we reject the null hypothesis, $H_0$, when $|\tau -
E(\tau|H_0)| / \sigma_{\tau} > 1.96$, where $\sigma_{\tau}$ is the
standard deviation in $\tau$. Here, the notation $E(x|H)$ denotes the
conditional expectation value of $x$, given hypothesis $H$. Using this
significance level, we would expect to incorrectly reject the null
hypothesis $\approx 5\%$ of the time. To illustrate the effect of
incorrectly assuming that $E(\tau|H_0) = 0$, we calculate the power
assuming $E(\tau|H_0) = 0$. Therefore, the null hypothesis is rejected
when the observed value of $\tau$ falls within the region, $|\tau| >
1.96\sigma_{\tau} $.

The results of our simulations are shown in Table \ref{t-tau}. As can
be seen, incorrectly assuming $E(\tau|H_0) = 0$ has a significant
effect on the conclusions drawn from Kendall's generalized partial
$\tau$. This is particularly notable when the correlation between the
two `independent' variables is high, in this case between $L_{UV}$ and
$z$. For high correlations between $L_{UV}$ and $z$, one incorrectly
rejects the null hypothesis with probability $p \approx 0.988$ when
using $\tau_{xz,l}$, and $p \approx 0.845$ when using $\tau_{\alpha
z,l}$. Furthermore, when $L_{UV}$ and $z$ are highly correlated and
one is using $\tau_{\alpha z,l}$, one rarely ($p \approx 0.061$)
rejects the null hypothesis if the alternative hypothesis is
true. However, one does reject the null hypothesis when the
alternative is true with near certainty when using
$\tau_{xz,l}$. These problems are ameliorated when the degree of
correlation between the independent variables is reduced; however,
even when $\rho = 0.3$ one still incorrectly rejects the null
hypothesis with probability a factor of $\sim 2$--$4$ times higher
than the expected $5\%$.

We also performed our simulations experimenting with different sample
sizes. We did not notice any dependence of $E(\tau | H_0)$ on the
sample size. This results in the seemingly paradoxical behavior that,
as the sample size increases, the null hypothesis is rejected with
increasing probability even if it is true, if one assumes $E(\tau |
H_0) = 0$. While the expected value of $\tau$ under the null
hypothesis does not appear to depend on sample size, the variance in
$\tau$, $\sigma^2_{\tau}$, does, decreasing as the sample size
increases. Thus, as the sample size increases, $\sigma_{\tau}$
decreases, and the observed values of $\tau$ become more concentrated
around $E(\tau | H_0)$. As a result, if one incorrectly assumes
$E(\tau|H_0) = 0$, it becomes more likely that $\tau$ falls inside of
the region where $|\tau| > 1.96 \sigma_{\tau}$, and thus one is more
likely to incorrectly reject the null hypothesis as the sample size
increases.

Similar results using Monte Carlo Simulations were found by
\citet{steffen06}. S06 concluded that their results implied that
spurious but false correlations between $L_X$ and $z$ result when
there is a high degree of correlation between $L_{UV}$ and
$z$. However, the results do not imply that spurious but false
correlation arise when there is a high degree of correlation between
$L_{UV}$ and $z$, but rather the simulations show that the expected
value of $\tau$ under the null hypothesis varies as the degree of
correlation between $L_{UV}$ and $z$ varies. This may then result in
apparently significant correlations because one is incorrectly
assuming $E(\tau | H_0) = 0$. This incorrect assumption can also
result in apparently insignificant correlations, even if such
correlations are significant.  Our simulations clearly show this,
since they were constructed to ensure that under the null hypothesis
$\alpha_{\rm ox}$ is independent of $z$, given $L_{UV}$,,
\emph{independent} of the distribution of $L_{UV}$ and $z$.

\citet{akritas96} also performed simulations to assess the behavior of
Kendall's partial $\tau$. They investigated the power of the test and
found that the test becomes more powerful as the departure from the
null hypothesis increases. They also found that when the null
hypothesis is true, and when the two independent variable are
statistically independent, the probability of rejecting $H_0$ is equal
to the chosen significance level. Similar to us, they also use a
significance level of 0.05 ($1.96\sigma$). However, when investigating
the probability of incorrectly rejecting the null hypothesis, they
only used simulations where the two independent variables ($L_{UV}$
and $z$ in this work) are statistically independent. We have confirmed
this result here, but have also expanded upon it, showing that the
null hypothesis can be incorrectly rejected with high probability when
the two independent variables have a moderate to high
correlation. Furthermore, \citet{akritas96} concluded that the
Kendall's partial $\tau$ test can become more powerful when the
alternative hypothesis is far from the null hypothesis, rejecting the
null hypothesis with high probability. However, we have shown here
that this is not always true, and that the power of the test depends
on the degree of correlation between the two independent
variables. Indeed, when $L_{UV}$ and $z$ are highly correlated, and
when one is investigating the partial correlation between $\alpha_{\rm
ox}$ and $z$, one almost always incorrectly claims that the data are
consistent with a null hypothesis of no evolution in $\alpha_{\rm
ox}$.

It should be noted that these results only apply to Kendall's
\emph{partial} $\tau$, and not to the usual Kendall's $\tau$. Both
Kendall's $\tau$ and the partial linear correlation have values of
zero under the null hypothesis of statistical independence between the
variables of interest \citep{nel88}.

Our simulations show that the expected value of Kendall's partial
$\tau$ under the null hypothesis depends on the distributional
properties of the sample, and that this can significantly effect the
power of the test. Unfortunately, we know of no way in which to
analytically calculate the expected value of $\tau$ under the null
hypothesis, and it must likely be calculated using
simulation. However, one must likely employ parametric methods in
order to simulate data, and this undermines the nonparametric nature
of Kendall's partial $\tau$. We also note that these results are not
meant to be a complete dismissal of the use of Kendall's generalized
partial $\tau$, but rather to point out the problems that can arise
when using Kendall's partial $\tau$. If one does not know
$E(\tau|H_0)$, then one is not able to calibrate the partial $\tau$
statistic against a physically meaningful null hypothesis, therefore
making statistical hypothesis testing based on it suspect.

\clearpage

\begin{figure}
\begin{center}
\scalebox{0.6}{\rotatebox{90}{\plotone{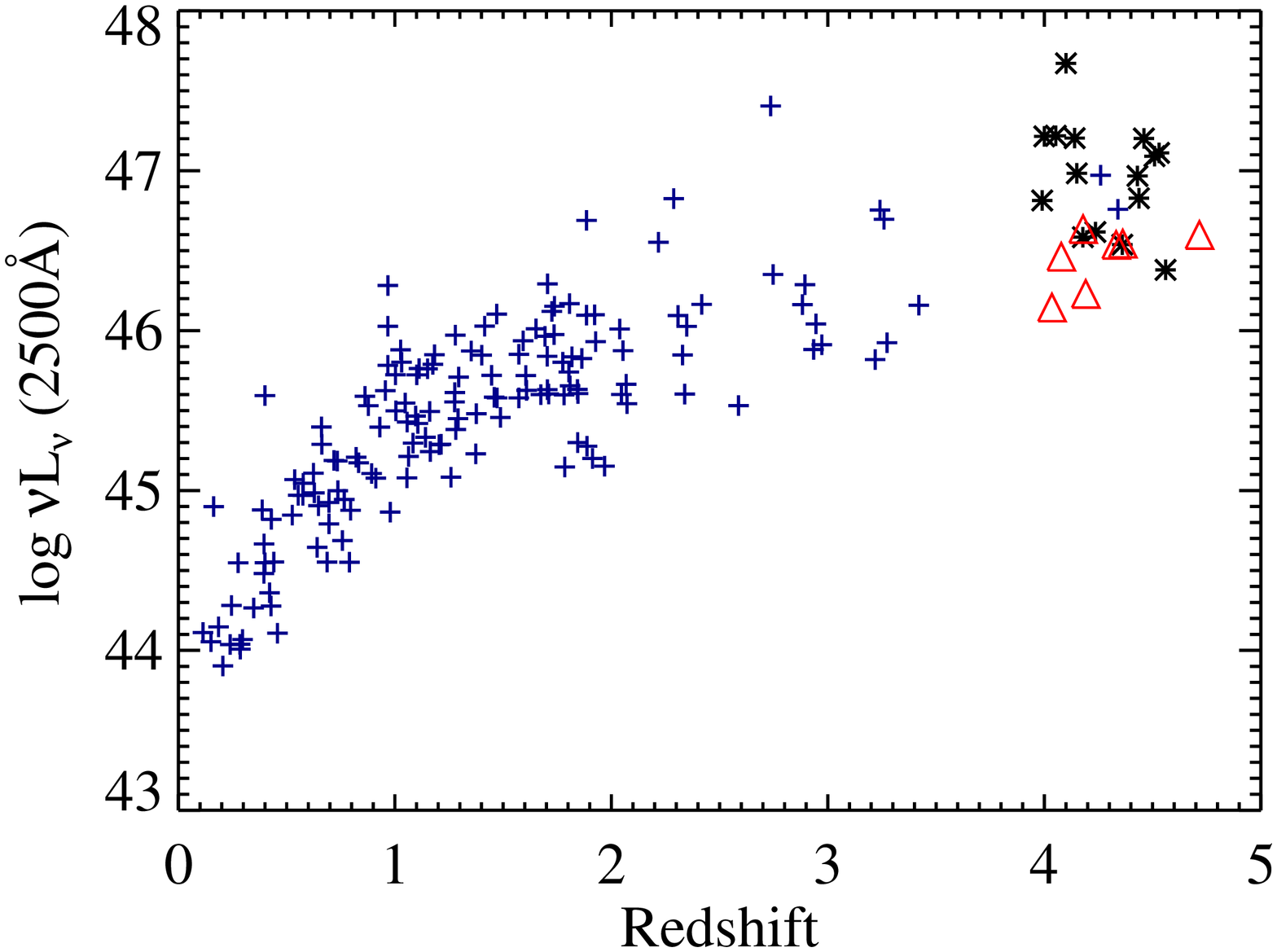}}}
\caption{The $(L_{UV},z)$ distribution of our sample. The seven
      new observations are denoted by red triangles, the SDSS sources
      are denoted by blue crosses, and the non-SDSS $z \gtrsim 4$
      sources are denoted by black asterisks. There appears to be data
      points for only six RQQs with new \emph{Chandra} observations
      because sources 0050-0053 and 2357+0053 have almost the same
      $L_{UV}$ and $z$, causing their symbols to
      overlap. \label{f-lzdist}}
\end{center}
\end{figure}

\clearpage

\begin{figure}
\begin{center}
\scalebox{0.6}{\rotatebox{90}{\plotone{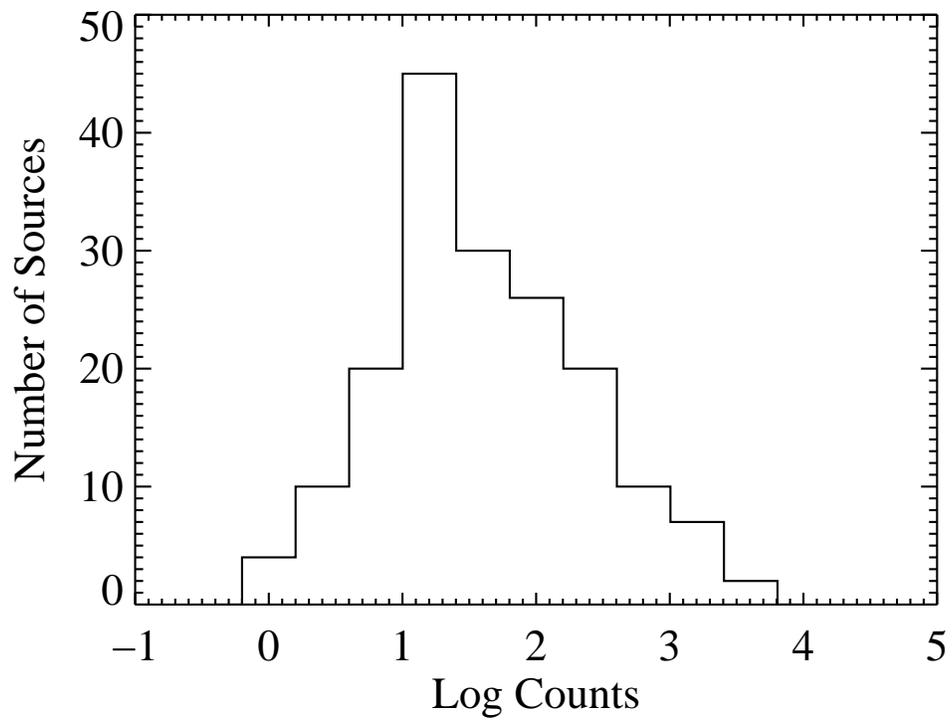}}}
    \caption{The distribution of observed, background-subtracted,
      $0.3$--$7.0$ keV photon counts for our sample.\label{f-counts}}
  \end{center}
\end{figure}

\clearpage

\begin{figure}
\begin{center}
\scalebox{0.6}{\rotatebox{90}{\plotone{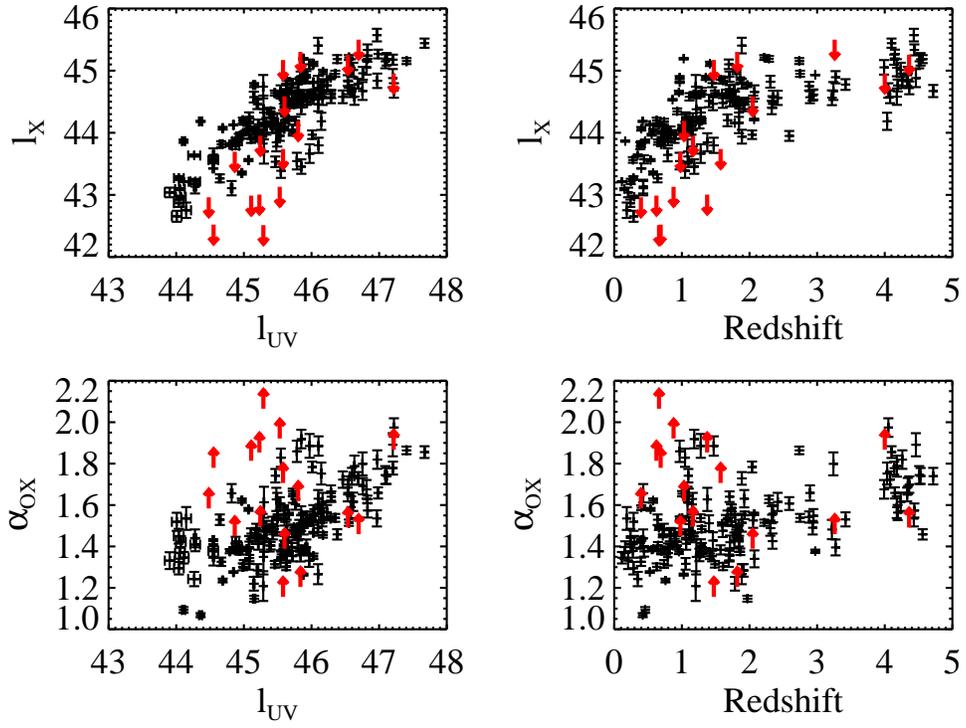}}}
\caption{The distribution of $l_X$ and $\alpha_{\rm ox}$ as
      functions of $l_{UV}$ and $z$. Red arrows denote upper limits
      for $l_X$ and lower limits of $\alpha_{\rm
      ox}$. \label{f-xuvdist}}
\end{center}
\end{figure}

\clearpage

\begin{figure}
\begin{center}
\includegraphics[scale=0.33,angle=90]{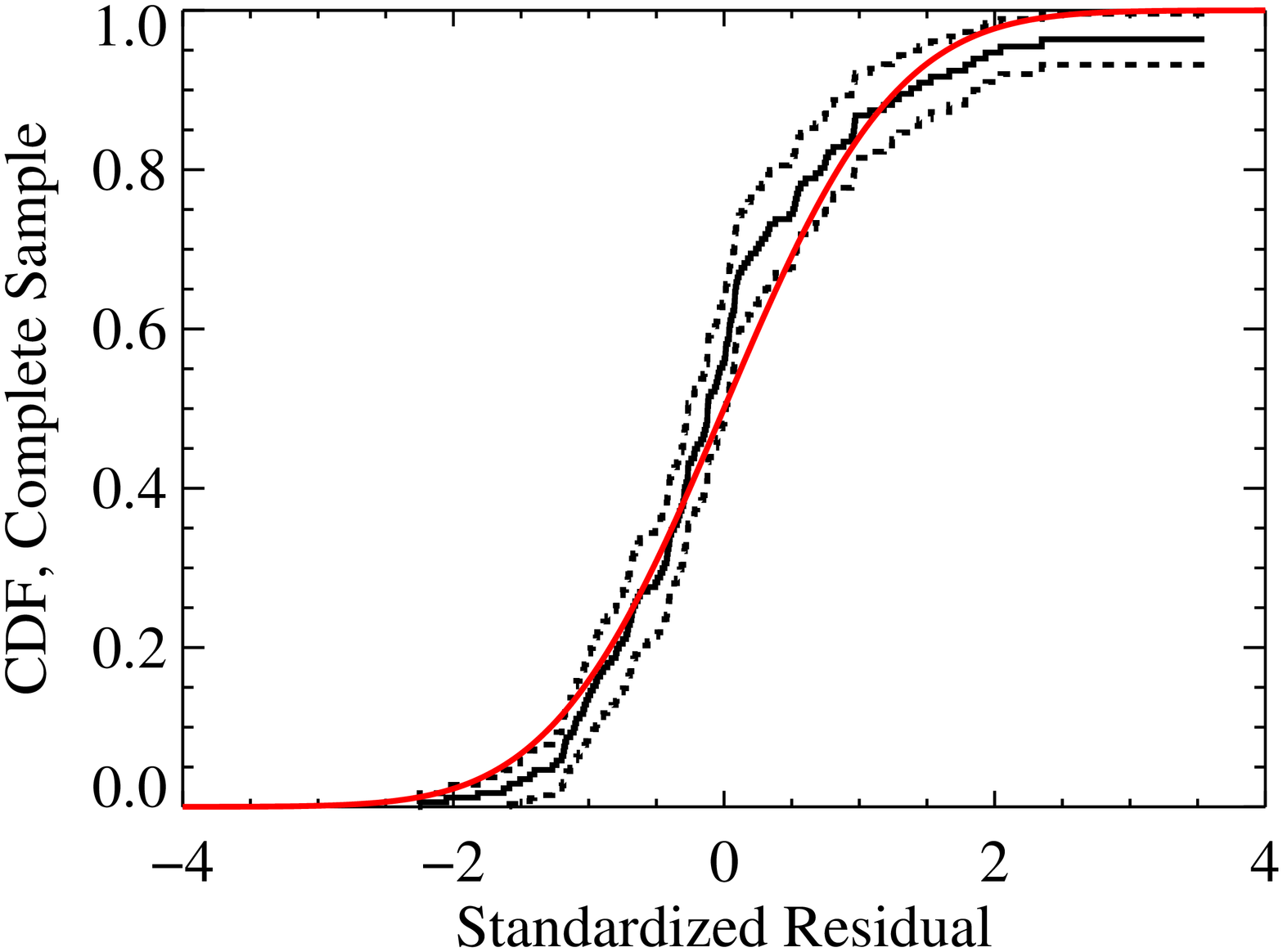}
\includegraphics[scale=0.33,angle=90]{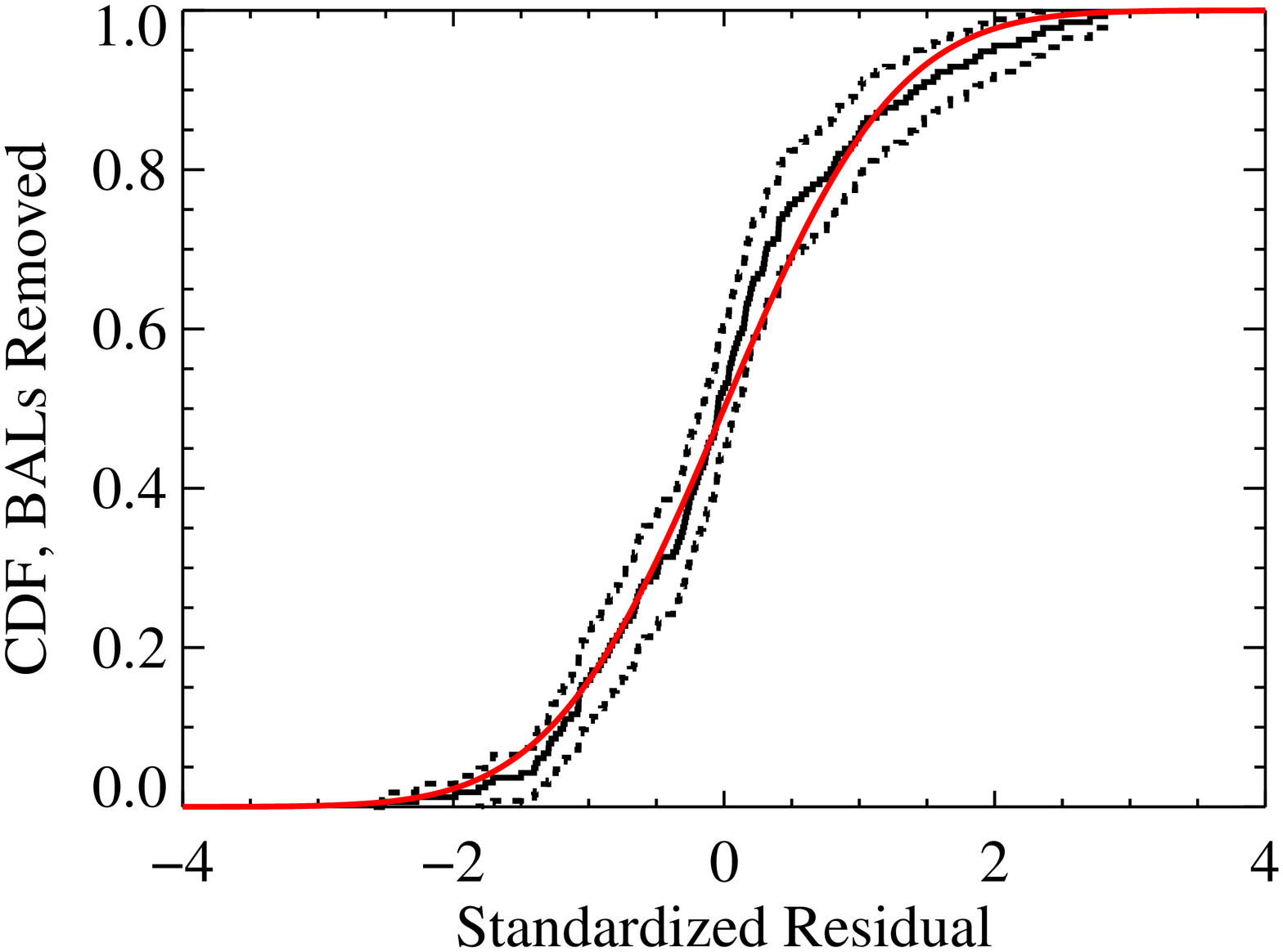}
\caption{Empirical cumulative distribution functions (CDF) of the
      standardized residuals for the regression of $\alpha_{\rm ox}$
      on $l_{UV}$ and $t(z)$ using the complete sample (left), and
      after removing suspected BALs (all $z < 1.5$ censored data
      points), right. The red line is the standard normal CDF, and the
      dashed lines denote the $95\%$ pointwise confidence interval on
      the empirical CDF. The CDF of the standardized residuals for the
      full sample shows evidence of diverging from normality, while
      the CDF of the sample with suspected BALs removed is consistent
      with the assumption of normality. \label{f-cdf}}
\end{center}
\end{figure}

\clearpage

\begin{figure}
\begin{center}
\scalebox{0.7}{\rotatebox{90}{\plotone{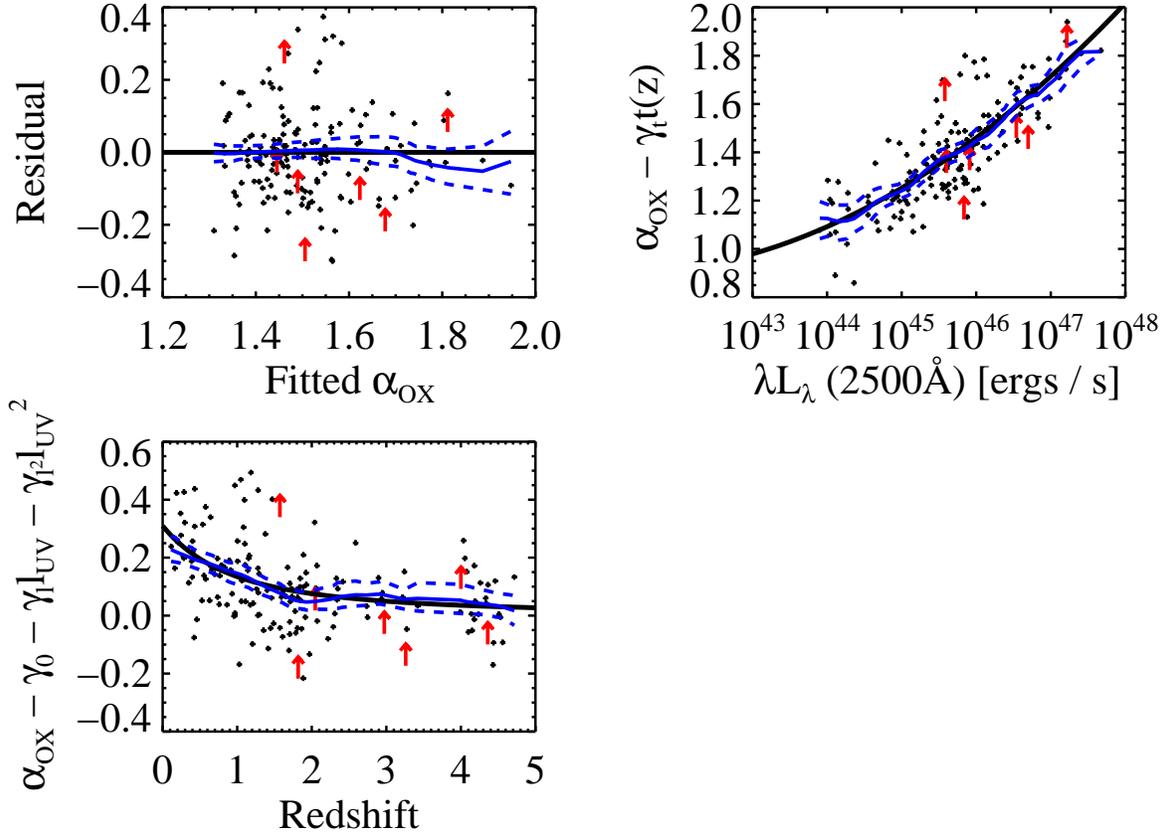}}}
\caption{The residuals of the $\alpha_{\rm ox}$ regression for model
      ${\cal M}_{l+t}$, shown as a function of the fitted $\alpha_{\rm
      ox}$, and partial residuals shown as functions of $L_{UV}$ and
      $z$. The partial residual plots show the dependence of
      $\alpha_{\rm ox}$ on $L_{UV}$ or $z$, after accounting for the
      dependence of $\alpha_{\rm ox}$ on $z$ or $L_{UV}$. Also shown
      are kernel-smoother fits to the residuals (solid blue lines), as
      well as approximate $95\%$ pointwise confidence intervals on the
      kernel-smoother fits (dashed blue lines). \label{f-alfoxres}}
\end{center}
\end{figure}

\clearpage

\begin{figure}
\begin{center}
\scalebox{0.7}{\rotatebox{90}{\plotone{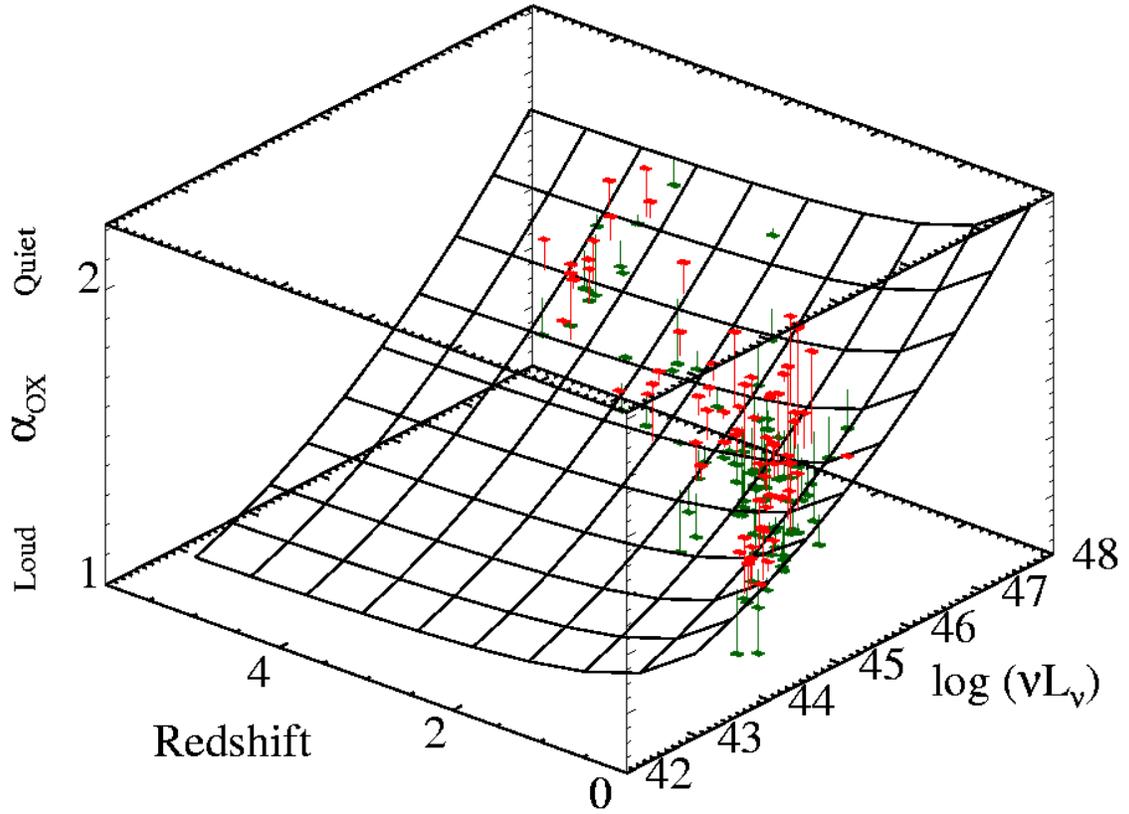}}}
\caption{The 3-dimensional distribution of $\alpha_{\rm ox}, l_{UV},$
      and $z$. The surface is the best fit to the data, obtained with
      model ${\cal M}_{l+t}$. Red denotes data points that fall above
      the fit, green denotes data points that fall below the
      fit.\label{f-3dfit}}
\end{center}
\end{figure}

\clearpage

\begin{figure}
\begin{center}
\scalebox{0.7}{\rotatebox{90}{\plotone{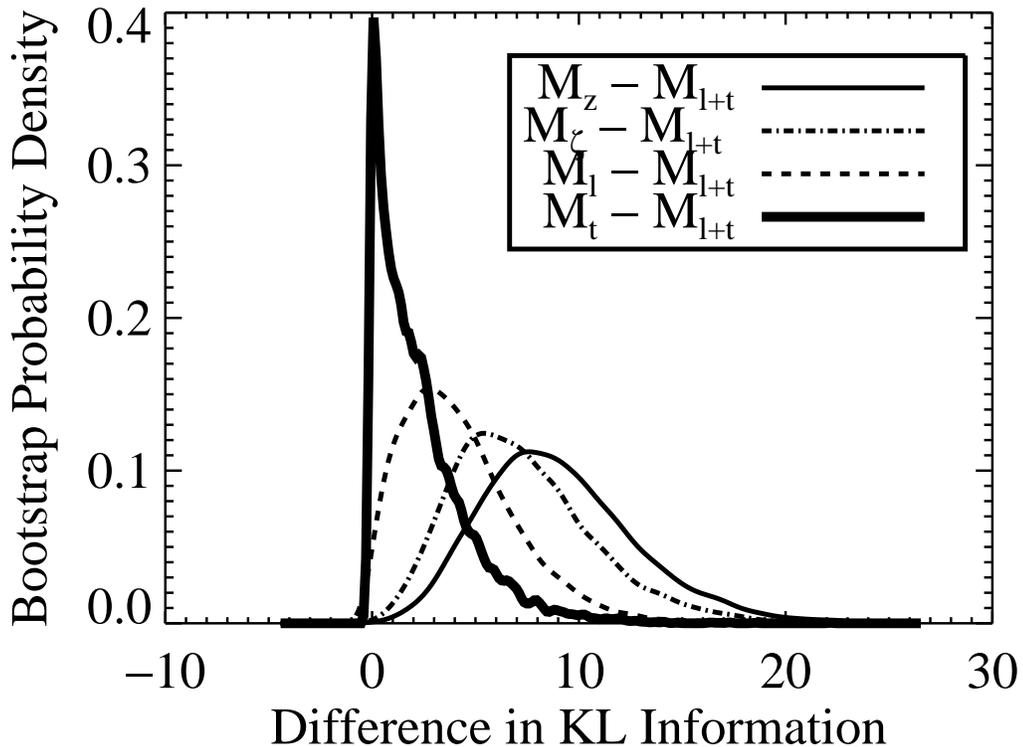}}}
\caption{Sampling distributions of the difference in
      Kullback-Leibler information relative to the $t(z)$
      parameterization with quadratic $l_{UV}$ term, ${\cal M}_{l+t}$,
      as determined from bootstrapping. Shown are the estimated
      distributions of the difference in KLI between ${\cal M}_z$ and
      ${\cal M}_{l+t}$ (thin solid line), ${\cal M}_{\zeta}$ and
      ${\cal M}_{l+t}$ (dot-dashed line), ${\cal M}_l$ and ${\cal
      M}_{l+t}$ (dashed line), and ${\cal M}_{t}$ and ${\cal M}_{l+t}$
      (thick solid line). As can be seen, the $t(z)$ parameterization
      with quadratic $l_{UV}$ term is almost always
      preferred.\label{f-klidist}}
\end{center}
\end{figure}

\begin{figure}
\begin{center}
\includegraphics[scale=0.33,angle=90]{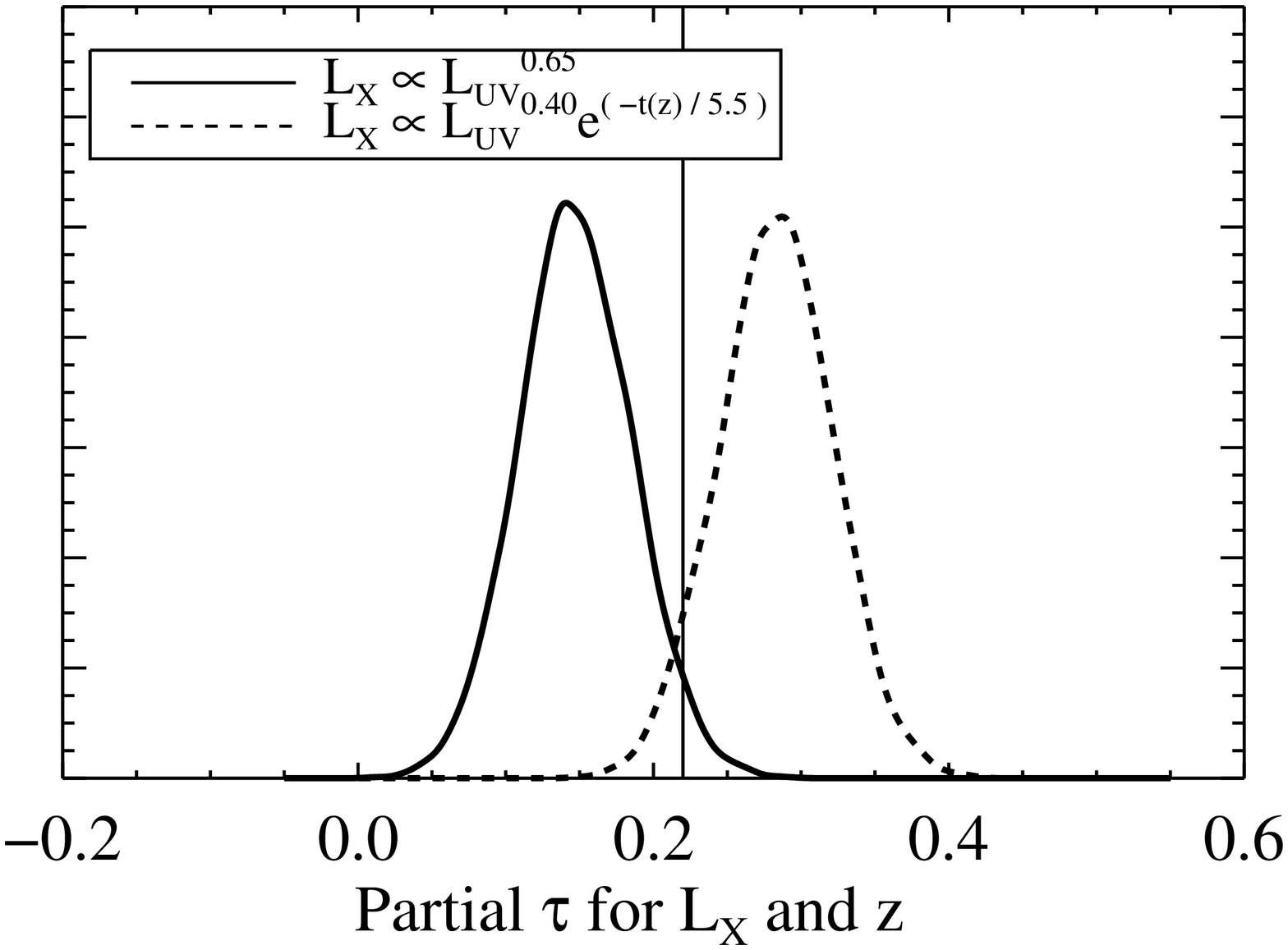}
\includegraphics[scale=0.33,angle=90]{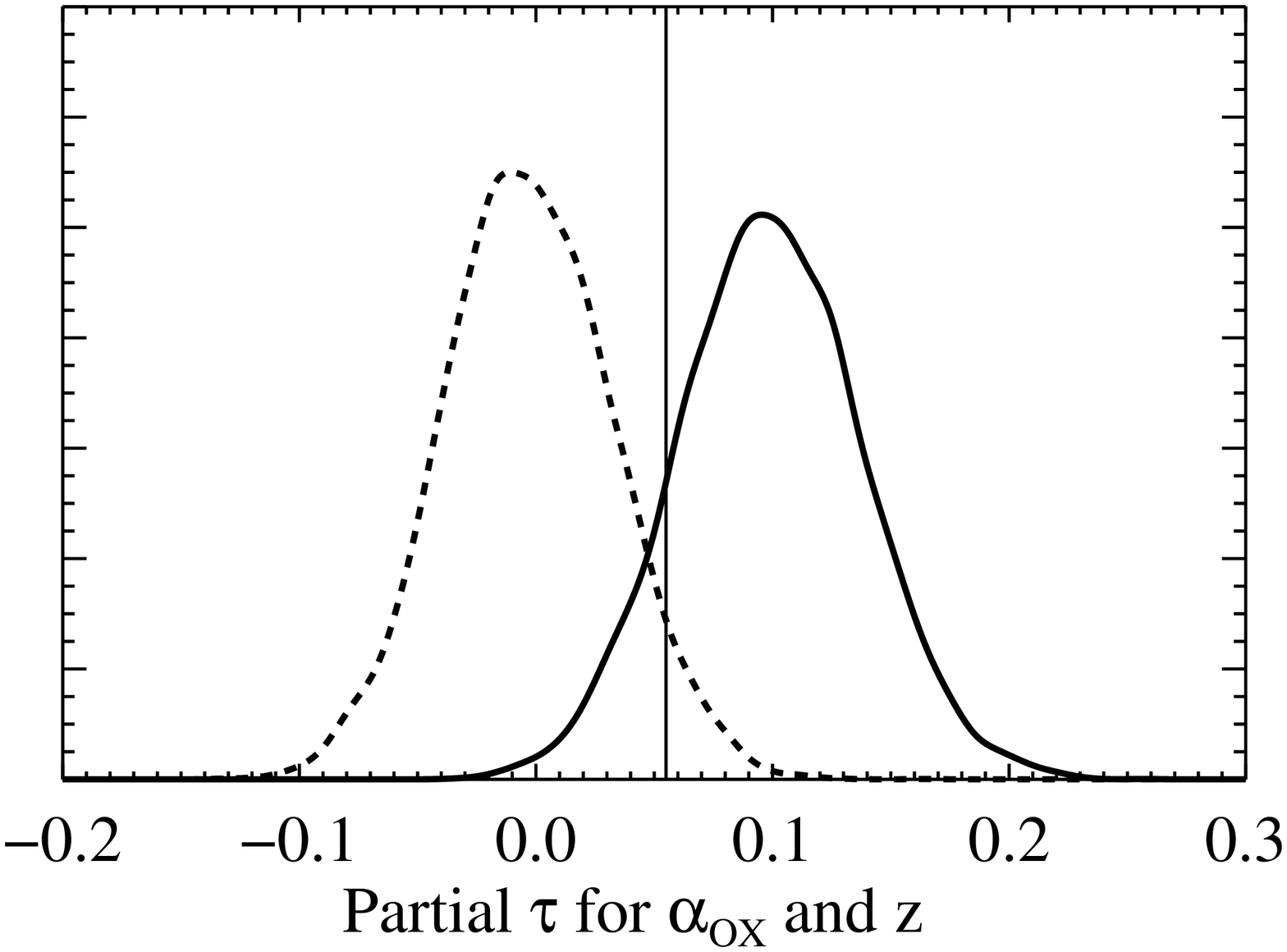}
\caption{Distribution of Kendall's generalized partial $\tau$ for
      $L_X$ and $z$ (left) and $\alpha_{\rm ox}$ and $z$ (right) under
      the no-evolution hypothesis (solid line) and the evolution
      hypothesis (dashed line). The vertical lines show the observed
      values of $\tau$ for our sample. As can be seen, our value of
      $\tau$ is about as consistent with the evolution model as with
      the no-evolution model. Also, note that $\tau \neq 0$ under the
      null hypothesis of no evolution (i.e., statistical independence
      of $\alpha_{\rm ox}$ and $z$ given $L_{UV}$), and therefore it
      is incorrect to calculate significance levels with respect to
      $\tau = 0$.
      \label{f-taudist}}
\end{center}
\end{figure}

\clearpage

\begin{figure}
\begin{center}
\scalebox{0.7}{\rotatebox{90}{\plotone{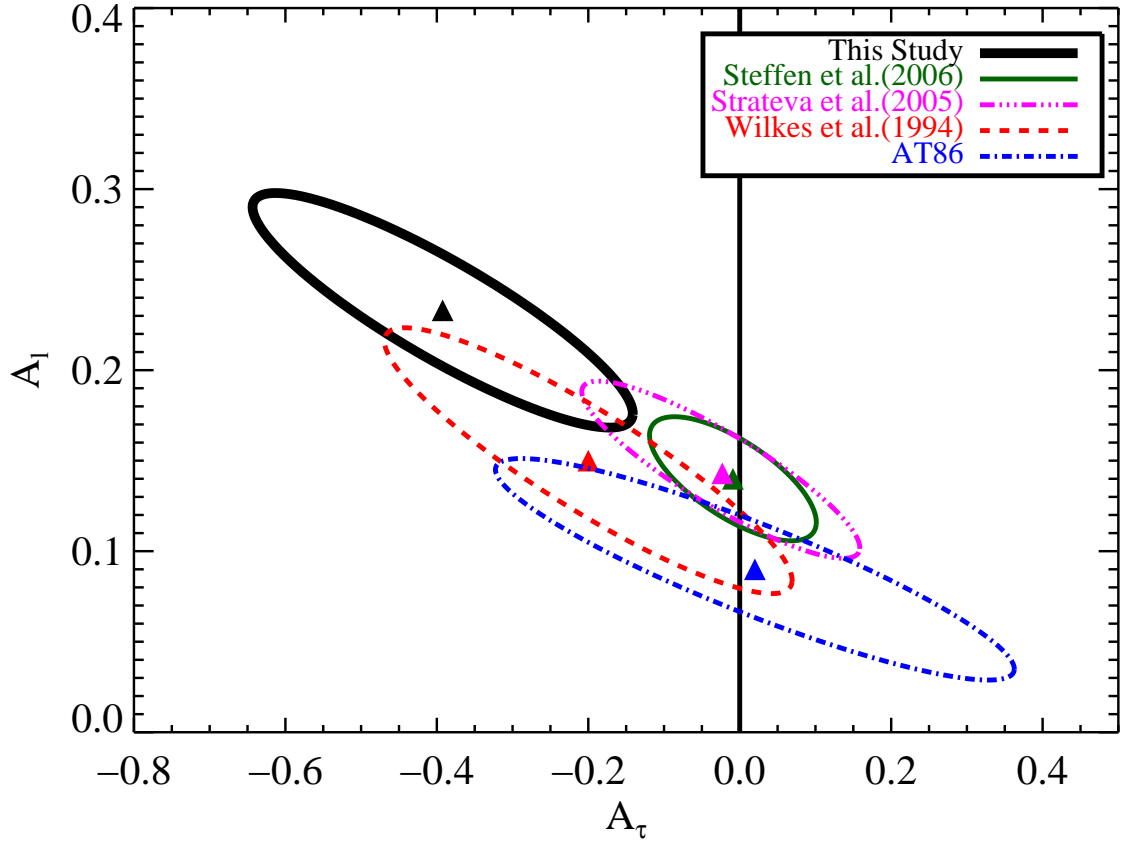}}}
      \caption{Comparison of our values of $A_l$ and $A_{\tau}$ with
      AT86 (blue, dashed-dotted contour), W94 (red, dashed contour),
      S05 (magenta, dash-dot-dot-dot contour), and S06
      (green, solid thin contour). The estimates of $(A_l, A_{\tau})$
      are denoted by triangles and the ellipses are approximate $95\%$
      ($2\sigma)$ confidence regions.\label{f-confreg}}
\end{center}
\end{figure}

\clearpage

\begin{figure}
\begin{center}
\scalebox{0.7}{\rotatebox{90}{\plotone{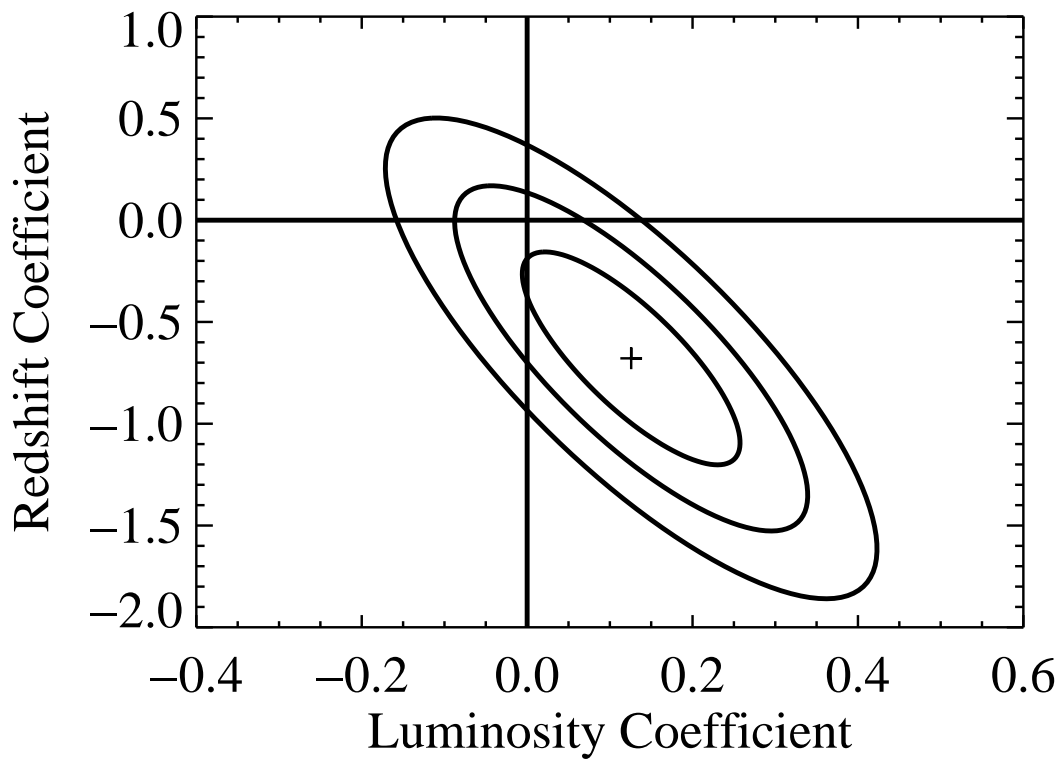}}}
\caption{Confidence regions for the $l_{UV}$ and $\log(1+z)$
    coefficients in the regression of $\Gamma_X$ on $l_{UV}$ and $\log
    (1+z)$. The cross denotes the best-fit value, and the contours are
    the $68\%, 95\%,$ and $99.7\%$ joint confidence
    regions. \label{f-gamx_confreg}}
\end{center}
\end{figure}

\clearpage

\begin{deluxetable}{lcccccccccc}
\tabletypesize{\scriptsize}
\rotate
\tablecaption{List of New Observations\label{t-newobs}}
\tablewidth{0pt}
\tablehead{
    \colhead{Source}
    & \colhead{RA}
    & \colhead{DEC}
    & \colhead{$z$}
    & \colhead{$r$\tablenotemark{a}}
    & \colhead{OBSID}
    & \colhead{Counts\tablenotemark{b}}
    & \colhead{$f_{X}$\tablenotemark{c}}
    & \colhead{Exp. Time}
    & \colhead{Spec. Ref.\tablenotemark{d}}
    & \colhead{Phot. Ref.\tablenotemark{e}} \\
    & J2000 & J2000 & & & &
    & $10^{-15}\ {\rm ergs\ cm^{-2}\ s^{-1}}$
    & ksec
    & &
  }
  \startdata
  SDSS 0050-0053 & 00 50 06.3 & -00 53 19.0 & 4.331 & 20.13 & 4825 & 26.3 & $6.91 \pm 1.38$ & 13.0 & 1 & 1 \\
  Q 0910+564     & 09 14 39.3 & +56 13 21.0 & 4.035 & 20.87 & 4821 & 13.4 & $1.79 \pm 0.57$ & 23.0 & 2 & 2 \\
  SDSS 1321+0038 & 13 21 10.8 & +00 38 22.0 & 4.716 & 21.30 & 4824 & 19.8 & $3.88 \pm 0.87$ & 17.8 & 1 & 1 \\
  SDSS 1413+0000 & 14 13 15.3 & +00 00 32.0 & 4.078 & 19.75 & 4823 & 23.0 & $4.33 \pm 1.02$ & 12.5 & 1 & 1 \\
  SDSS 1444-0123 & 14 44 28.7 & -01 23 44.0 & 4.179 & 19.64 & 4826 & 12.1 & $4.12 \pm 1.26$ & 10.0 & 1 & 1 \\
  PC 1450+3404   & 14 53 00.6 & +33 52 06.0 & 4.191 & 20.81 & 4822 & 20.2 & $6.20 \pm 1.36$ & 14.8 & 3 & 4 \\
  SDSS 2357+0043 & 23 57 18.3 & +00 43 50.0 & 4.362 & 19.92 & 4827 & 19.7 & $5.20 \pm 1.23$ & 12.7 & 1 & 1
  \enddata
  
  \tablenotetext{a}{$r$-band apparent magnitude.}
  \tablenotetext{b}{Number of observed background-subtracted source
    counts in the range 0.3--7.0 keV.}
  \tablenotetext{c}{Unabsorbed 2--10 keV flux, assuming a power law 
    with $\Gamma_X = 1.9$.}
  \tablenotetext{d}{Reference for the observed frame optical spectrum.}
  \tablenotetext{e}{Reference for the $r$-band magnitude.}
  \tablerefs{(1) SDSS; (2) \citet{schneid91}; (3) \citet{const02}; (4) \citet{schneid97}}

\end{deluxetable}

\clearpage
\begin{deluxetable}{lccccccccccc}
  \tabletypesize{\tiny}
  \rotate
  \tablecaption{List of Archival Sources\label{t-archival_samp}}
  \tablewidth{0pt}
  \tablehead{
    \colhead{Source}
    & \colhead{RA}
    & \colhead{DEC}
    & \colhead{$z$}
    & \colhead{$r$\tablenotemark{a}}
    & \colhead{OBSID}
    & \colhead{$\theta$\tablenotemark{b}}
    & \colhead{Counts\tablenotemark{c}}
    & \colhead{Exp. Time}
    & \colhead{$f_{X}$\tablenotemark{d}}
    & \colhead{Spec. Ref.\tablenotemark{e}}
    & \colhead{Phot. Ref.\tablenotemark{f}} \\
    & J2000 & J2000 & & & & &
    & ksec
    & $10^{-14}\ {\rm ergs\ cm^{-2}\ s^{-1}}$
    & & 
  }
  \startdata
  0002+0049 &  00 02 30.7 & +00 49 59.0 & 1.352 & 18.12 & 4861 & 0.58 & 188.0  & 5.7   & $ 10.52 \pm 1.07 $ & 1 & 1 \\
  0006-0015 &  00 06 54.1 & -00 15 33.4 & 1.725 & 18.20 & 4096 & 0.58 & 46.2   & 4.5   & $ 3.35 \pm 0.49 $ & 1 & 1 \\
  0022+0016 &  00 22 10.0 & +00 16 29.3 & 0.574 & 18.06 & 2252 & 7.12 & 343.4  & 71.2  & $ 2.82 \pm 0.48 $ & 1 & 1 \\
  0027+0026 &  00 27 52.4 & +00 26 15.7 & 0.205 & 18.18 & 4080 & 4.48 & 54.4   & 1.6   & $ 11.66 \pm 4.47 $ & 1 & 1 \\
  0031+0034 &  00 31 31.4 & +00 34 20.2 & 1.735 & 18.79 & 2101 & 1.20 & 73.9   & 6.7   & $ 3.10 \pm 0.51 $ & 1 & 1 \\
  0057+1450 &  00 57 01.1 & +14 50 03.0 & 0.623 & 18.81 & 865  & 1.15 & 4.6    & 4.6   & $ < 0.91        $ & 1 & 1 \\
  0059+0003 &  00 59 22.8 & +00 03 01.0 & 4.178 & 19.5  & 2179 & 0.58 & 9.0    & 2.7   & $ 1.21 \pm 0.38 $ & 2 & 5 \\
  0106+0048 &  01 06 19.2 & +00 48 22.0 & 4.437 & 19.1  & 2180 & 0.60 & 24.2   & 3.7   & $ 2.04 \pm 0.41 $ & 2 & 5 \\
  0113+1531 &  01 13 05.7 & +15 31 46.5 & 0.576 & 18.30 & 3219 & 2.09 & 1291.8 & 58.5  & $ 12.41 \pm 0.53 $ & 1 & 1 \\
  0113+1535 &  01 13 09.1 & +15 35 53.6 & 1.806 & 18.33 & 3219 & 6.29 & 528.3  & 58.5  & $ 6.20 \pm 0.29 $ & 1 & 1 \\
  0115+0020 &  01 15 37.7 & +00 20 28.7 & 1.275 & 18.72 & 3204 & 6.79 & 406.2  & 37.6  & $ 6.77 \pm 0.66 $ & 1 & 1 \\
  0133+0400 &  01 33 40.4 & +04 00 59.0 & 4.150 & 18.0  & 3152 & 0.59 & 39.4   & 6.1   & $ 2.03 \pm 0.33 $ & 3 & 3 \\
  0134+3307 &  01 34 21.5 & +33 07 56.6 & 4.530 & 18.9  & 3018 & 0.60 & 22.7   & 5.0   & $ 1.38 \pm 0.32 $ & 3 & 3 \\
  0148+0001 &  01 48 12.2 & +00 01 53.3 & 1.704 & 17.67 & 4098 & 0.57 & 42.3   & 3.7   & $ 3.57 \pm 0.55 $ & 1 & 1 \\
  0148-0002 &  01 48 21.0 & -00 02 25.8 & 0.930 & 18.39 & 4098 & 5.41 & 94.7   & 3.7   & $ 7.96 \pm 1.53 $ & 1 & 1 \\
  0152+0105 &  01 52 58.7 & +01 05 07.4 & 0.647 & 19.22 & 1448 & 4.13 & 104.6  & 7.9   & $ 6.86 \pm 1.30 $ & 1 & 1 \\
  0153+0052 &  01 53 09.1 & +00 52 50.1 & 1.161 & 18.81 & 3580 & 7.26 & 58.1   & 19.9  & $ 2.27 \pm 0.37 $ & 1 & 1 \\
  0156+0053 &  01 56 50.3 & +00 53 08.5 & 1.652 & 18.65 & 4100 & 0.59 & 95.2   & 5.6   & $ 4.58 \pm 0.49 $ & 1 & 1 \\
  0159+0023 &  01 59 50.2 & +00 23 40.8 & 0.162 & 15.97 & 4104 & 0.58 & 5708.2 & 9.7   & $ 189.5 \pm 8.19 $ & 1 & 1 \\
  0201-0919 &  02 01 18.7 & -09 19 35.8 & 0.660 & 17.61 & 3772 & 7.83 & 379.2  & 19.7  & $ 11.42 \pm 1.17 $ & 1 & 1 \\
  0208+0022 &  02 08 45.5 & +00 22 36.1 & 1.885 & 17.08 & 4099 & 0.58 & 58.2   & 3.5   & $ 4.69 \pm 0.65 $ & 1 & 1 \\
  0209+0517 &  02 09 44.7 & +05 17 14.0 & 4.140 & 17.8  & 3153 & 0.60 & 30.2   & 5.8   & $ 1.66 \pm 0.31 $ & 3 & 3 \\
  0232-0731 &  02 32 17.7 & -07 31 19.9 & 1.163 & 19.19 & 3030 & 10.1 & 0.0    & 4.2   & $ < 1.85 $ & 1 & 1 \\
  0241-0811 &  02 41 05.8 & -08 11 53.2 & 0.978 & 19.87 & 385  & 1.69 & 5.8    & 2.3   & $ < 1.56 $ & 1 & 1 \\
  0241+0023 &  02 41 10.0 & +00 23 01.4 & 0.790 & 20.47 & 4011 & 9.33 & 15.8   & 5.0   & $ 2.35 \pm 0.57 $ & 1 & 1 \\
  0244-0134 &  02 44 01.9 & -01 34 03.0 & 4.053 & 18.4  & 875  & 0.64 & 17.6   & 7.4   & $ 0.55 \pm 0.15 $ & 2 & 6 \\
  0248+1802 &  02 48 54.3 & +18 02 49.9 & 4.430 & 18.4  & 876  & 0.66 & 18.1   & 1.7   & $ 3.55 \pm 0.83 $ & 2 & 5 \\
  0259+0048 &  02 59 59.7 & +00 48 13.6 & 0.892 & 19.44 & 4145 & 0.54 & 40.6   & 4.7   & $ 3.13 \pm 0.47 $ & 1 & 1 \\
  0311-1722 &  03 11 15.2 & -17 22 47.3 & 4.000 & 18.0  & 3154 & 0.57 & 8.9    & 6.1   & $ < 1.07 $ & 3 & 3 \\
  0314-0111 &  03 14 27.5 & -01 11 52.3 & 0.387 & 18.05 & 4084 & 4.48 & 114.8  & 1.9   & $ 23.9 \pm 5.29 $ & 1 & 1 \\
  0403-1703 &  04 03 56.6 & -17 03 24.1 & 4.236 & 18.7  & 2182 & 0.59 & 13.9   & 3.8   & $ 1.08 \pm 0.29 $ & 2 & 6 \\
  0419-5716 &  04 19 50.9 & -57 16 13.1 & 4.460 & 18.7  & 4066 & 0.58 & 9.3    & 4.0   & $ 0.71 \pm 0.24 $ & 3 & 3 \\
  0755+2203 &  07 55 02.1 & +22 03 46.9 & 0.399 & 18.99 & 647, 3767 & 3.85, 7.26 & 457.0 & 156.2 & $ 5.83 \pm 0.41 $ & 1 & 1 \\
  0755+4058 &  07 55 35.6 & +40 58 03.0 & 2.417 & 18.85 & 3032 & 9.48 & 15.6   & 7.3   & $ 1.37 \pm 0.36 $ & 1 & 1 \\
  0755+4111 &  07 55 40.0 & +41 11 19.1 & 0.967 & 17.86 & 3032 & 10.0 & 9.6   & 7.3   & $ 1.13 \pm 0.35 $ & 1 & 1 \\
  0755+4056 &  07 55 45.6 & +40 56 43.6 & 2.348 & 19.17 & 3032 & 9.06 & 19.8   & 7.3   & $ 1.49 \pm 0.37 $ & 1 & 1 \\
  0819+3649 &  08 19 51.4 & +36 49 50.8 & 0.736 & 19.19 & 4119 & 11.5 & 86.6   & 7.3   & $ 11.22 \pm 2.71 $ & 1 & 1 \\
  0832+5243 &  08 32 06.0 & +52 43 59.3 & 1.572 & 19.47 & 1643 & 3.80 & 7.6    & 9.1   & $ < 0.56 $ & 1 & 1 \\
  0845+3431 &  08 45 26.6 & +34 31 02.0 & 2.046 & 19.88 & 818  & 10.1 & 3.4    & 4.5   & $ < 2.13 $ & 1 & 1 \\
  0849+4457 &  08 49 05.1 & +44 57 14.8 & 1.259 & 20.00 & 927, 1708 & 3.26, 3.26 & 938.2 & 186.5 & $ 2.37 \pm 0.11 $ & 1 & 1 \\
  0849+4500 &  08 49 43.7 & +45 00 24.3 & 1.592 & 18.39 & 927, 1708 & 10.4, 10.4 & 1149.6 & 186.5 & $ 3.87 \pm 0.13 $ & 1 & 1 \\
  0910+5427 &  09 10 29.0 & +54 27 19.0 & 0.525 & 18.76 & 2227 & 7.57 & 3327.4 & 105.7 & $ 19.95 \pm 0.74 $ & 1 & 1 \\
  0912+0547 &  09 12 10.3 & +05 47 42.1 & 3.240 & 18.06 & 419, 1629 & 10.9, 11.0 & 53.9 & 38.6 & $ 0.88 \pm 0.14 $ & 1 & 1 \\
  0918+5139 &  09 18 28.6 & +51 39 32.1 & 0.185 & 17.46 & 533  & 4.51 & 147.5  & 11.3  & $ 5.05 \pm 3.14 $ & 1 & 1 \\
  0918+0647 &  09 18 47.5 & +06 47 04.7 & 0.821 & 18.80 & 3563 & 11.1 & 78.0   & 4.9   & $ 5.62 \pm 1.63 $ & 1 & 1 \\
  0933+5515 &  09 33 59.3 & +55 15 50.8 & 1.863 & 19.08 & 805  & 1.27 & 779.7  & 40.8  & $ 4.57 \pm 0.24 $ & 1 & 1 \\
  0941+5948 &  09 41 33.7 & +59 48 11.3 & 0.967 & 16.38 & 3035 & 2.83 & 398.1  & 4.2   & $ 25.16 \pm 2.86 $ & 1 & 1 \\
  0950+5619 &  09 50 24.0 & +56 19 46.7 & 1.912 & 20.53 & 4151 & 6.76 & 29.0   & 8.9   & $ 1.87 \pm 0.34 $ & 1 & 1 \\
  0951+5940 &  09 51 30.2 & +59 40 37.1 & 1.056 & 18.74 & 3036 & 5.39 & 45.9   & 5.1   & $ 4.48 \pm 0.66 $ & 1 & 1 \\
  0951+5944 &  09 51 51.6 & +59 44 30.0 & 2.338 & 19.79 & 3036 & 1.12 & 28.5   & 5.1   & $ 1.26 \pm 0.28 $ & 1 & 1 \\
  0952+5152 &  09 52 40.2 & +51 52 50.0 & 0.553 & 18.47 & 3195 & 2.42 & 1487.6 & 26.9  & $ 12.81 \pm 0.94 $ & 1 & 1 \\
  0952+5151 &  09 52 43.0 & +51 51 21.1 & 0.861 & 17.34 & 3195 & 3.25 & 1688.2 & 26.9  & $ 14.79 \pm 0.76 $ & 1 & 1 \\
  0955+5935 &  09 55 05.6 & +59 35 17.6 & 0.912 & 18.91 & 3156 & 4.84 & 38.3   & 5.7   & $ 3.50 \pm 0.56 $ & 1 & 1 \\
  0955+5940 &  09 55 11.3 & +59 40 32.2 & 4.340 & 18.58 & 3156 & 0.60 & 17.8   & 5.7   & $ 0.90 \pm 0.21 $ & 1 & 1 \\
  0955+4116 &  09 55 42.1 & +41 16 55.3 & 3.420 & 19.36 & 5294 & 7.45 & 31.2   & 17.3  & $ 1.01 \pm 0.20 $ & 1 & 1 \\
  0955+4109 &  09 55 48.1 & +41 09 55.3 & 2.307 & 18.74 & 5294 & 2.67 & 71.9   & 17.3  & $ 2.47 \pm 0.30 $ & 1 & 1 \\
  0956+4110 &  09 56 40.4 & +41 10 43.5 & 1.887 & 20.49 & 5294 & 7.23 & 12.8   & 17.3  & $ 0.58 \pm 0.15 $ & 1 & 1 \\
  0958+0734 &  09 58 20.5 & +07 34 36.1 & 1.885 & 18.44 & 2990 & 9.38 & 296.9  & 14.1  & $ 14.01 \pm 1.66 $ & 1 & 1 \\
  0958+0747 &  09 58 22.6 & +07 47 47.7 & 3.218 & 20.07 & 2990 & 8.48 & 10.6   & 14.1  & $ 0.72 \pm 0.20 $ & 1 & 1 \\
  0958+0745 &  09 58 36.6 & +07 45 56.3 & 1.487 & 19.17 & 2990 & 6.23 & 192.1  & 14.1  & $ 4.70 \pm 0.63 $ & 1 & 1 \\
  1002+5542 &  10 02 05.4 & +55 42 57.9 & 1.151 & 18.03 & 2038 & 2.84 & 161.6  & 26.6  & $ 3.44 \pm 0.62 $ & 1 & 1 \\
  1003+4736 &  10 03 52.8 & +47 36 53.4 & 2.934 & 19.72 & 4152 & 0.59 & 64.5   & 13.7  & $ 1.24 \pm 0.16 $ & 1 & 1 \\
  1013-0052 &  10 13 14.9 & -00 52 33.6 & 0.275 & 17.78 & 4085 & 4.49 & 192.3  & 2.0   & $ 32.75 \pm 5.50 $ & 1 & 1 \\
  1019+4737 &  10 19 02.0 & +47 37 14.6 & 2.944 & 19.19 & 4153 & 0.58 & 28.7   & 8.0   & $ 0.98 \pm 0.19 $ & 1 & 1 \\
  1023+0415 &  10 23 50.9 & +04 15 42.0 & 1.809 & 19.40 & 1651, 909 & 6.66, 6.66 & 240.7 & 211.7 & $ 2.19 \pm 0.14 $ & 1 & 1 \\
  1030+0524 &  10 30 31.6 & +05 24 54.9 & 1.182 & 17.74 & 3357 & 0.88 & 11.2   & 8.0   & $ 0.57 \pm 0.15 $ & 1 & 1 \\
  1032+5738 &  10 32 27.9 & +57 38 22.5 & 1.968 & 20.58 & 3345, 3344 & 8.17, 8.18 & 589.7 & 77.0 & $ 3.79 \pm 0.16 $ & 1 & 1 \\
  1032+5800 &  10 32 36.2 & +58 00 34.0 & 0.686 & 19.83 & 3343 & 7.90 & 1.7    & 37.0  & $ < 0.25 $ & 1 & 1 \\
  1036-0343 &  10 36 23.8 & -03 43 20.0 & 4.509 & 18.5  & 877  & 0.65 & 15.7   & 3.4   & $ 1.29 \pm 0.33 $ & 2 & 6 \\
  1038+4727 &  10 38 08.7 & +47 27 34.9 & 1.047 & 18.56 & 4154 & 3.96 & 15.6   & 9.8   & $ 0.65 \pm 0.17 $ & 1 & 1 \\
  1042+0100 &  10 42 30.7 & +01 00 01.6 & 1.400 & 18.40 & 4086 & 4.87 & 21.2   & 1.7   & $ 4.05 \pm 1.27 $ & 1 & 1 \\
  1044+5921 &  10 44 54.9 & +59 21 34.1 & 1.291 & 19.03 & 5030 & 7.31 & 265.5  & 65.7  & $ 2.51 \pm 0.27 $ & 1 & 1 \\
  1049+5750 &  10 49 21.5 & +57 50 36.6 & 1.106 & 18.81 & 1673 & 8.73 & 23.9   & 4.9   & $ 2.92 \pm 0.62 $ & 1 & 1 \\
  1050+5702 &  10 50 15.6 & +57 02 55.7 & 3.273 & 20.17 & 1679, 1680 & 10.9, 8.19 & 24.0 & 9.4 & $ 1.47 \pm 0.33 $ & 1 & 1 \\
  1050+5738 &  10 50 50.1 & +57 38 20.0 & 1.281 & 19.09 & 1678 & 3.98 & 32.8   & 4.7   & $ 3.49 \pm 0.62 $ & 1 & 1 \\
  1052+5724 &  10 52 39.6 & +57 24 31.4 & 1.111 & 17.79 & 1683 & 2.96 & 90.8   & 4.7   & $ 8.81 \pm 1.64 $ & 1 & 1 \\
  1053+5735 &  10 53 16.8 & +57 35 50.8 & 1.204 & 19.08 & 1683, 1684 & 9.42, 7.40 & 161.8 & 9.4 & $ 10.6 \pm 0.96 $ & 1 & 1 \\
  1054+5740 &  10 54 04.1 & +57 40 19.8 & 1.100 & 18.04 & 1688 & 5.25 & 24.7   & 4.7   & $ 2.61 \pm 0.54 $ & 1 & 1 \\
  1054+5720 &  10 54 22.6 & +57 20 31.0 & 2.972 & 19.85 & 1687 & 7.13 & 3.1    & 4.7   & $ < 1.99 $ & 1 & 1 \\
  1055+5704 &  10 55 18.1 & +57 04 23.6 & 0.695 & 18.73 & 1686, 1691 & 8.96, 8.35 & 86.7 & 9.4 & $ 6.43 \pm 1.06 $ & 1 & 1 \\
  1056+5722 &  10 56 44.5 & +57 22 33.5 & 0.286 & 18.90 & 1693 & 4.96 & 26.2   & 5.7   & $ 2.56 \pm 0.52 $ & 1 & 1 \\
  1057+4555 &  10 57 56.4 & +45 55 52.0 & 4.100 & 17.48 & 878  & 0.65 & 34.8   & 2.8   & $ 3.12 \pm 0.53 $ & 2 & 1 \\
  1109+0900 &  11 09 05.3 & +09 00 48.7 & 1.674 & 19.42 & 3252 & 7.34 & 22.1   & 10.0  & $ 1.40 \pm 0.31 $ & 1 & 1 \\
  1111+5532 &  11 11 32.1 & +55 32 40.3 & 1.004 & 18.44 & 2025 & 7.89 & 333.8  & 59.4  & $ 3.70 \pm 0.37 $ & 1 & 1 \\
  1114+5315 &  11 14 52.8 & +53 15 31.7 & 1.213 & 19.02 & 3253, 3321 & 4.01, 8.41 & 39.8 & 13.6 & $ 4.60 \pm 0.46 $ & 1 & 1 \\
  1115+5309 &  11 15 20.7 & +53 09 22.1 & 0.877 & 18.05 & 3321 & 1.18 & 2.6    & 4.8   & $ < 0.55 $ & 1 & 1 \\
  1129-0137 &  11 29 43.9 & -01 37 52.3 & 1.294 & 18.15 & 2082 & 5.66 & 128.7  & 4.8   & $ 7.96 \pm 1.14 $ & 1 & 1 \\
  1129-0150 &  11 29 51.2 & -01 50 37.3 & 1.784 & 20.24 & 2082 & 7.89 & 23.7   & 4.8   & $ 2.86 \pm 0.56 $ & 1 & 1 \\
  1136+0159 &  11 36 21.2 & +01 59 27.9 & 0.766 & 19.24 & 4833 & 3.15 & 85.1   & 5.9   & $ 5.28 \pm 0.98 $ & 1 & 1 \\
  1136+0158 &  11 36 31.9 & +01 58 01.1 & 1.470 & 17.85 & 4833 & 0.57 & 9.8   & 5.9   & $ 0.74 \pm 0.21 $ & 1 & 1 \\
  1136+0207 &  11 36 33.1 & +02 07 47.7 & 0.239 & 18.07 & 4833 & 9.31 & 65.4   & 5.9   & $ 5.48 \pm 2.06 $ & 1 & 1 \\
  1202-0129 &  12 02 26.8 & -01 29 15.3 & 0.150 & 17.13 & 4108 & 0.59 & 2361.2 & 9.4   & $ 54.36 \pm 3.27 $ & 1 & 1 \\
  1204+0150 &  12 04 36.6 & +01 50 25.6 & 1.927 & 18.63 & 3234 & 5.49 & 148.6  & 30.0  & $ 1.85 \pm 0.19 $ & 1 & 1 \\
  1208+0016 &  12 08 29.6 & +00 16 42.7 & 1.063 & 18.97 & 2083 & 5.86 & 34.8   & 4.6   & $ 3.53 \pm 0.61 $ & 1 & 1 \\
  1213+0252 &  12 13 43.0 & +02 52 48.9 & 0.641 & 19.30 & 4110 & 4.55 & 57.9   & 10.0  & $ 1.69 \pm 0.47 $ & 1 & 1 \\
  1214+0055 &  12 14 15.2 & +00 55 11.5 & 0.395 & 18.35 & 4087 & 4.48 & 137.3  & 2.0   & $ 21.42 \pm 3.75 $ & 1 & 1 \\
  1215-0034 &  12 15 40.5 & -00 34 33.8 & 0.757 & 19.46 & 4201 & 3.45 & 551.2  & 44.5  & $ 7.16 \pm 0.56 $ & 1 & 1 \\
  1218+0546 &  12 18 36.1 & +05 46 28.1 & 0.795 & 18.89 & 3322 & 0.98 & 72.5   & 4.6   & $ 3.53 \pm 0.60 $ & 1 & 1 \\
  1220-0025 &  12 20 04.4 & -00 25 39.1 & 0.421 & 18.96 & 3141 & 0.58 & 2380.4 & 19.7  & $ 30.45 \pm 1.26 $ & 1 & 1 \\
  1223+1034 &  12 23 07.5 & +10 34 48.2 & 2.747 & 18.59 & 3232 & 4.65 & 302.6  & 30.1  & $ 3.02 \pm 0.20 $ & 1 & 1 \\
  1226-0011 &  12 26 52.0 & -00 11 59.6 & 1.175 & 17.88 & 4865 & 0.58 & 137.7  & 4.9   & $ 8.07 \pm 1.05 $ & 1 & 1 \\
  1228+4413 &  12 28 18.0 & +44 13 02.0 & 0.662 & 18.05 & 2031 & 5.81 & 3.0    & 26.6  & $ < 0.27 $ & 1 & 1 \\
  1228+4411 &  12 28 53.7 & +44 11 52.9 & 1.276 & 18.79 & 2031 & 9.09 & 96.6   & 26.6  & $ 2.71 \pm 0.30 $ & 1 & 1 \\
  1230+0302 &  12 30 05.8 & +03 02 04.2 & 1.604 & 18.91 & 4040 & 3.01 & 34.1   & 3.5   & $ 3.07 \pm 0.56 $ & 1 & 1 \\
  1230+0305 &  12 30 25.9 & +03 05 35.4 & 1.055 & 19.45 & 4040 & 3.17 & 35.9   & 3.5   & $ 4.72 \pm 0.80 $ & 1 & 1 \\
  1230+0306 &  12 30 27.4 & +03 06 27.5 & 0.628 & 18.65 & 4040 & 4.05 & 105.6  & 3.5   & $ 14.99 \pm 2.32 $ & 1 & 1 \\
  1230+0308 &  12 30 39.9 & +03 08 57.3 & 1.843 & 19.50 & 4040 & 8.02 & 12.7   & 3.5   & $ 1.73 \pm 0.54 $ & 1 & 1 \\
  1230+0305 &  12 30 54.7 & +03 05 37.2 & 0.427 & 19.19 & 4040 & 9.76 & 16.8   & 3.5   & $ 3.05 \pm 0.78 $ & 1 & 1 \\
  1236+6215 &  12 36 22.9 & +62 15 26.6 & 2.587 & 20.44 & 580, 2423, 2344, 3409,  & 6.83, 1.67, 5.38, 1.79, & 3741.3 & 1961.0 & $ 0.90 \pm 0.02 $ & 1 & 1 \\
            &             &             &       &       & 967, 966, 3389, 957,    & 5.46, 5.45, 4.11, 1.73, & & & & \\
            &             &             &       &       & 3408, 2233, 2232, 2386, & 4.12, 1.67, 1.69, 5.38, & & & & \\
            &             &             &       &       & 3388, 2421, 2234, 3293, & 4.12, 1.64, 1.64, 4.12, & & & & \\
            &             &             &       &       & 3294, 3390, 3391, 1671  & 1.79, 1.79, 1.79, 5.37  & & & & \\
  1237+6203 &  12 37 16.0 & +62 03 23.4 & 2.068 & 19.86 & 580, 2344, 967, 966,    & 9.23, 9.77, 9.59, 9.62, & 198.9 & 431.9 & $ 0.30 \pm 0.03 $ & 1 & 1 \\
            &             &             &       &       & 2386, 1671              & 9.78, 9.78              & & & & \\
  1242+0249 &  12 42 55.3 & +02 49 57.0 & 1.458 & 19.21 & 323, 3926 & 8.17, 6.56 & 657.2 & 127.1 & $ 3.250 \pm 0.17 $ & 1 & 1 \\
  1245-0027 &  12 45 41.0 & -00 27 44.9 & 1.693 & 18.58 & 4018 & 7.82 & 88.7   & 4.9   & $ 11.24 \pm 1.21 $ & 1 & 1 \\
  1255+5652 &  12 55 35.1 & +56 52 39.6 & 1.803 & 19.13 & 1031 & 6.00 & 339.3  & 39.3  & $ 1.79 \pm 0.18 $ & 1 & 1 \\
  1255+5650 &  12 55 36.2 & +56 50 00.1 & 1.373 & 19.83 & 1031 & 6.32 & 0.0    & 39.3  & $ < 0.14 $ & 1 & 1 \\
  1258-0143 &  12 58 49.8 & -01 43 03.3 & 0.967 & 17.06 & 4178 & 4.58 & 2024.9 & 27.3  & $ 21.09 \pm 1.00 $ & 1 & 1 \\
  1259+0102 &  12 59 43.6 & +01 02 55.0 & 0.394 & 18.34 & 4088 & 4.48 & 4.0    & 1.9   & $ < 2.47 $ & 1 & 1 \\
  1311+0031 &  13 11 08.5 & +00 31 51.7 & 0.429 & 17.92 & 4089 & 4.48 & 14.9   & 1.7   & $ 2.86 \pm 0.76 $ & 1 & 1 \\
  1317+3531 &  13 17 43.2 & +35 31 31.1 & 4.360 & 19.1  & 879  & 0.64 & 5.7    & 2.8   & $ < 1.78 $ & 2 & 5 \\
  1344-0000 &  13 44 25.9 & -00 00 56.2 & 1.095 & 18.58 & 2251 & 0.39 & 26.2   & 9.6   & $ 0.82 \pm 0.16 $ & 1 & 1 \\
  1411+5217 &  14 11 04.1 & +52 17 55.6 & 2.882 & 19.07 & 2254 & 6.15 & 247.3  & 90.9  & $ 1.48 \pm 0.11 $ & 1 & 1 \\
  1411+5205 &  14 11 04.9 & +52 05 16.8 & 1.083 & 18.97 & 2254 & 7.19 & 41.1   & 90.9  & $ 1.20 \pm 0.68 $ & 1 & 1 \\
  1417+4456 &  14 17 00.8 & +44 56 06.4 & 0.113 & 16.32 & 541  & 8.26 & 5313.3 & 31.2  & $ 92.89 \pm 4.08 $ & 1 & 1 \\
  1419+4709 &  14 19 51.9 & +47 09 01.4 & 2.288 & 17.37 & 3076 & 0.59 & 174.3  & 7.7   & $ 6.77 \pm 0.53 $ & 1 & 1 \\
  1424+4214 &  14 24 14.1 & +42 14 00.1 & 1.608 & 19.03 & 3077 & 5.91 & 133.1  & 5.9   & $ 6.81 \pm 0.79 $ & 1 & 1 \\
  1424+4210 &  14 24 36.0 & +42 10 30.5 & 2.217 & 17.51 & 3077 & 0.58 & 137.0  & 5.9   & $ 5.35 \pm 0.57 $ & 1 & 1 \\
  1432-0059 &  14 32 44.4 & -00 59 15.2 & 1.026 & 17.26 & 907  & 7.56 & 1715.9 & 21.4  & $ 43.53 \pm 2.13 $ & 1 & 1 \\
  1433+0227 &  14 33 35.3 & +02 27 18.3 & 2.072 & 19.94 & 3959 & 3.85 & 17.0   & 3.5   & $ 2.57 \pm 0.63 $ & 1 & 1 \\
  1434+0227 &  14 34 07.5 & +02 27 04.6 & 1.710 & 19.41 & 3959 & 4.24 & 28.4   & 3.5   & $ 2.91 \pm 0.56 $ & 1 & 1 \\
  1438+0341 &  14 38 42.0 & +03 41 10.4 & 1.737 & 18.27 & 3290 & 5.97 & 296.9  & 57.6  & $ 3.11 \pm 0.20 $ & 1 & 1 \\
  1438+0335 &  14 38 59.1 & +03 35 47.5 & 0.733 & 18.43 & 3290 & 8.23 & 498.2  & 57.6  & $ 7.25 \pm 0.50 $ & 1 & 1 \\
  1442+0110 &  14 42 31.7 & +01 10 55.3 & 4.560 & 20.90 & 3960 & 0.58 & 43.9   & 11.0  & $ 1.37 \pm 0.20 $ & 4 & 1 \\
  1443+5856 &  14 43 40.8 & +58 56 53.2 & 4.260 & 18.28 & 3160 & 0.58 & 16.3   & 5.8   & $ 0.71 \pm 0.19 $ & 1 & 1 \\
  1445+0129 &  14 45 54.8 & +01 29 03.3 & 1.845 & 20.00 & 2112 & 2.40 & 38.9   & 5.9   & $ 2.22 \pm 0.34 $ & 1 & 1 \\
  1448+4738 &  14 48 53.4 & +47 38 21.3 & 2.894 & 19.39 & 4155 & 0.58 & 19.0   & 6.9   & $ 0.93 \pm 0.21 $ & 1 & 1 \\
  1448+0015 &  14 48 56.7 & +00 15 10.3 & 0.832 & 18.80 & 4092 & 7.59 & 19.2   & 2.1   & $ 4.64 \pm 1.08 $ & 1 & 1 \\
  1449+0024 &  14 49 13.5 & +00 24 06.9 & 0.440 & 19.13 & 4092 & 3.10 & 48.0   & 2.1   & $ 7.82 \pm 1.18 $ & 1 & 1 \\
  1452+4304 &  14 52 15.6 & +43 04 48.7 & 0.296 & 18.89 & 1048, 2424 & 3.14, 3.11 & 785.1 & 47.2 & $ 5.010 \pm 0.49 $ & 1 & 1 \\
  1452+4308 &  14 52 40.9 & +43 08 14.4 & 1.704 & 19.41 & 1048 & 8.36 & 139.8  & 17.7  & $ 4.46 \pm 0.38 $ & 1 & 1 \\
  1511+5659 &  15 11 26.5 & +56 59 34.8 & 1.031 & 17.55 & 3334 & 9.81 & 3.9    & 4.9   & $ < 4.30 $ & 1 & 1 \\
  1515+5521 &  15 15 04.9 & +55 21 07.3 & 1.844 & 20.40 & 3006 & 9.83 & 18.2   & 9.6   & $ 1.02 \pm 0.39 $ & 1 & 1 \\
  1539+4313 &  15 39 47.6 & +43 13 41.6 & 0.347 & 18.75 & 2993 & 1.30 & 870.1  & 14.8  & $ 15.96 \pm 1.28 $ & 1 & 1 \\
  1543+5405 &  15 43 16.4 & +54 05 26.1 & 0.245 & 18.11 & 822  & 8.39 & 132.3  & 4.5   & $ 14.35 \pm 3.03 $ & 1 & 1 \\
  1545+4846 &  15 45 30.2 & +48 46 09.1 & 0.399 & 16.44 & 3339 & 8.93 & 245.8  & 4.9   & $ 34.29 \pm 5.79 $ & 1 & 1 \\
  1605-0109 &  16 05 17.8 & -01 09 55.5 & 1.572 & 19.14 & 2086 & 5.42 & 80.4   & 4.6   & $ 7.08 \pm 0.91 $ & 1 & 1 \\
  1618+3456 &  16 18 34.0 & +34 56 25.6 & 1.922 & 18.73 & 3341 & 5.29 & 17.4   & 4.9   & $ 1.83 \pm 0.46 $ & 1 & 1 \\
  1640+4644 &  16 40 25.0 & +46 44 49.1 & 0.537 & 18.38 & 896  & 1.17 & 1049.2 & 42.3  & $ 5.17 \pm 0.41 $ & 1 & 1 \\
  1641+4649 &  16 41 10.6 & +46 49 11.9 & 0.695 & 19.21 & 896  & 9.98 & 366.1  & 42.3  & $ 4.83 \pm 0.48 $ & 1 & 1 \\
  1641+4000 &  16 41 54.2 & +40 00 33.1 & 1.002 & 17.81 & 3575 & 1.29 & 311.8  & 46.5  & $ 3.09 \pm 0.38 $ & 1 & 1 \\
  1657+3524 &  16 57 13.2 & +35 24 39.4 & 2.328 & 19.37 & 3662 & 8.88 & 102.8  & 49.6  & $ 1.16 \pm 0.14 $ & 1 & 1 \\
  1701+6412 &  17 01 00.6 & +64 12 09.0 & 2.735 & 16.00 & 547  & 4.98 & 364.7  & 49.5  & $ 3.69 \pm 0.21 $ & 1 & 1 \\
  1702+3405 &  17 02 24.5 & +34 05 39.0 & 2.038 & 18.93 & 4179 & 5.25 & 97.8   & 57.0  & $ 0.65 \pm 0.06 $ & 1 & 1 \\
  1703+6045 &  17 03 55.8 & +60 45 11.7 & 0.284 & 18.77 & 435  & 4.97 & 250.3  & 9.1   & $ 12.0 \pm 2.11 $ & 1 & 1 \\
  1708+6154 &  17 08 17.9 & +61 54 48.6 & 1.414 & 17.84 & 4864 & 0.58 & 239.4  & 4.1   & $ 16.4 \pm 1.83 $ & 1 & 1 \\
  1719+2732 &  17 19 27.3 & +27 32 46.8 & 1.446 & 18.72 & 3245 & 10.1 & 92.7   & 10.0  & $ 7.27 \pm 1.29 $ & 1 & 1 \\
  1720+2638 &  17 20 26.5 & +26 38 16.0 & 1.141 & 19.13 & 3224, 4361 & 4.43, 5.43 & 170.6 & 49.5 & $ 2.66 \pm 0.27 $ & 1 & 1 \\
  1735+5355 &  17 35 51.9 & +53 55 15.7 & 0.955 & 17.91 & 4863 & 0.58 & 268.2  & 5.4   & $ 14.86 \pm 1.76 $ & 1 & 1 \\
  1737+5828 &  17 37 16.6 & +58 28 39.5 & 1.775 & 19.05 & 3038 & 3.28 & 21.2   & 4.6   & $ 3.71 \pm 0.83 $ & 1 & 1 \\
  1738+5837 &  17 38 36.2 & +58 37 48.6 & 1.279 & 17.71 & 4860 & 0.58 & 10.1   & 3.9   & $ 0.76 \pm 0.26 $ & 1 & 1 \\
  2215-1611 &  22 15 27.1 & -16 11 33.0 & 3.990 & 18.1  & 2185 & 0.59 & 16.6   & 3.2   & $ 1.35 \pm 0.37 $ & 2 & 6 \\
  2238-0921 &  22 38 19.8 & -09 21 06.0 & 3.259 & 18.04 & 2411 & 6.07 & 1.3      & 5.9   & $ < 5.97 $ & 1 & 1 \\
  2238-0937 &  22 38 54.7 & -09 37 36.2 & 1.472 & 19.14 & 2411 & 12.7 & 4.0    & 5.8   & $ < 17.5 $ & 1 & 1 \\
  2239-0933 &  22 39 17.3 & -09 33 40.9 & 1.817 & 19.28 & 2414 & 11.3 & 9.9    & 5.7   & $ < 14.6 $ & 1 & 1 \\
  2249-0808 &  22 49 03.3 & -08 08 41.7 & 0.457 & 19.42 & 583  & 8.04 & 304.5  & 11.7  & $ 13.6 \pm 1.97 $ & 1 & 1 \\
  2337+0025 &  23 37 18.1 & +00 25 50.7 & 2.053 & 19.28 & 3248 & 7.59 & 20.6   & 9.2   & $ 1.85 \pm 0.40 $ & 1 & 1 \\
  2337+0022 &  23 37 22.0 & +00 22 38.9 & 1.376 & 19.27 & 3248 & 4.50 & 8.0    & 9.2   & $ 0.97 \pm 0.33 $ & 1 & 1 \\
  2337+0026 &  23 37 39.1 & +00 26 56.2 & 1.703 & 18.85 & 3248 & 7.44 & 61.7   & 9.2   & $ 4.13 \pm 0.58 $ & 1 & 1 \\
  2348+0107 &  23 48 40.1 & +01 07 53.5 & 0.718 & 18.50 & 861  & 10.8 & 779.4  & 74.2  & $ 7.08 \pm 0.46 $ & 1 & 1
  \enddata

  \tablecomments{For sources with multiple observation IDs, the
    off-axis angles are reported for each. However, the reported counts
    and exposure time are summed over the observation IDs.}

  \tablenotetext{a}{$r$-band apparent magnitude.}
  \tablenotetext{b}{Off-axis angle, in arcmin.}
  \tablenotetext{c}{Number of observed background-subtracted source
    counts in the range 0.3--7.0 keV.}
  \tablenotetext{d}{Unabsorbed 2--10 keV flux, assuming a power-law. A
    photon index of $\Gamma_X = 1.9$ was assumed for those sources with $<
    50$ counts. A ``$<$''denotes an upper limit.}
  \tablenotetext{e}{Reference for the observed frame optical spectrum.}
  \tablenotetext{f}{Reference for the $r$-band magnitude.}
  \tablerefs{(1) SDSS; (2) \citet{const02}; (3) \citet{peroux01}; (4) \citet{and01}; 
    (5) \citet{kenn05}; (6) \citet{storr96}}

\end{deluxetable}

\clearpage

\begin{deluxetable}{lcccccc}
  \tabletypesize{\scriptsize}
  \tablecaption{Continuum and Fe Emission Fitting Windows\label{t-contwin}}
  \tablewidth{0pt}
  \tablehead{
    \colhead{} & \colhead{1} & \colhead{2} & \colhead{3} & \colhead{4} & \colhead{5} & \colhead{6}
  }
  \startdata
  UV      & $\lambda\lambda 1350$--$1365$ & $\lambda\lambda 1427$--$1500$ & $\lambda\lambda 1760$--$1860$ 
          & $\lambda\lambda 1950$--$2300$ & $\lambda\lambda 2470$--$2755$ & $\lambda\lambda 2855$--$3010$ \\
  Optical & $\lambda\lambda 3535$--$3700$ & $\lambda\lambda 4100$--$4200$ & $\lambda\lambda 4400$--$4700$ 
          & $\lambda\lambda 5100$--$6200$ & $\lambda\lambda 6800$--$7534$ & \nodata \\
  \enddata
\end{deluxetable}

\clearpage

\begin{deluxetable}{lcccccccc}
\tabletypesize{\scriptsize}
\tablecaption{X-ray and UV Parameters\label{t-xuvpar_samp}}
\tablewidth{0pt}
\tablehead{
\colhead{Source} 
& \colhead{Redshift}
& Gal. $N_H$\tablenotemark{a}
& $n_0$\tablenotemark{b}
& \colhead{$\Gamma_X$\tablenotemark{c}} 
& \colhead{$\log \nu L_{\nu} (2500\AA)$} 
& \colhead{$\log \nu L_{\nu} (2\ \rm{keV})$} 
& \colhead{$\alpha_{\rm ox}$} 
& \colhead{$\alpha_{UV}$\tablenotemark{d}} \\
&
& $10^{20}\ {\rm cm^{-2}}$
& $10^{-6}\ {\rm cm^{-2}\ s^{-1}\ keV^{-1}}$
& 
& ${\rm ergs\ s^{-1}}$
& ${\rm ergs\ s^{-1}}$
&
&
}
\startdata
0002+0049 & 1.353 & 2.82 & $ 41.87 \pm 3.08$ & $ 2.20 \pm 0.11$ & 45.87 & 44.88 & 1.38 & 1.33  \\
0006-0015 & 1.725 & 3.16 & $ 13.26^{+ 2.04}_{- 1.85} $ & $ 1.81^{+ 0.22}_{- 0.22} $ & 46.12 & 44.61 & 1.58 & 0.74  \\
0022+0016 & 0.575 & 2.70 & $ 11.40 \pm 0.88$ & $ 2.30 \pm 0.20$ & 44.97 & 43.35 & 1.62 & 0.52  \\
0027+0026 & 0.205 & 3.01 & $ 47.63 \pm 7.30$ & $ 1.68 \pm 0.27$ & 43.90 & 43.04 & 1.33 & $ \ldots $  \\
0031+0034 & 1.735 & 2.41 & $ 12.24 \pm 1.49$ & $ 2.21 \pm 0.18$ & 45.98 & 44.62 & 1.52 & 0.15  \\
0050-0053\tablenotemark{e} & 4.332 & 2.70 & $ 2.66^{+ 0.56}_{- 0.50} $ & $ 2.02^{+ 0.33}_{- 0.32} $ & 46.54 & 44.84 & 1.65 & 0.98  \\
0057+1450 & 0.624 & 4.35 & $ < 3.69 $ & $ \ldots $ & 45.11 & $ < 42.99 $ & $ > 1.81 $ & 0.64  \\
0059+0003 & 4.178 & 3.01 & $ 4.65^{+ 1.62}_{- 1.33} $ & $ 0.91^{+ 0.44}_{- 0.44} $ & 46.58 & 45.05 & 1.59 & 1.12  \\
0106+0048 & 4.437 & 3.16 & $ 7.85^{+ 1.70}_{- 1.49} $ & $ 1.95^{+ 0.31}_{- 0.30} $ & 46.83 & 45.33 & 1.57 & 0.60  \\
0113+1531 & 0.576 & 4.40 & $ 50.18 \pm 3.63$ & $ 1.91 \pm 0.08$ & 45.05 & 44.04 & 1.39 & 0.53  \\
0113+1535 & 1.807 & 4.38 & $ 24.49 \pm 1.37$ & $ 2.29 \pm 0.10$ & 46.17 & 44.98 & 1.46 & 0.53  \\
0115+0020 & 1.276 & 3.34 & $ 26.95 \pm 1.77$ & $ 1.92 \pm 0.11$ & 45.55 & 44.61 & 1.36 & 0.64  \\
0133+0400 & 4.150 & 3.01 & $ 7.84^{+ 1.34}_{- 1.21} $ & $ 2.16^{+ 0.28}_{- 0.27} $ & 46.98 & 45.27 & 1.66 & 1.00  \\
0134+3307 & 4.530 & 4.67 & $ 5.29^{+ 1.31}_{- 1.14} $ & $ 2.24^{+ 0.42}_{- 0.40} $ & 47.11 & 45.18 & 1.74 & 0.65  \\
0148+0001 & 1.705 & 2.88 & $ 14.12^{+ 2.29}_{- 2.07} $ & $ 2.39^{+ 0.24}_{- 0.24} $ & 46.29 & 44.62 & 1.64 & 0.16  \\
0148-0002 & 0.930 & 2.75 & $ 31.91 \pm 3.49$ & $ 2.31 \pm 0.19$ & 45.40 & 44.34 & 1.40 & 0.54  \\
0152+0105 & 0.647 & 2.80 & $ 27.70 \pm 2.75$ & $ 2.50 \pm 0.18$ & 44.90 & 43.85 & 1.40 & 0.18  \\
0153+0052 & 1.161 & 2.69 & $ 9.06 \pm 1.44$ & $ 2.19 \pm 0.26$ & 45.49 & 44.05 & 1.56 & 0.59  \\
0156+0053 & 1.652 & 2.69 & $ 18.15 \pm 2.12$ & $ 1.36 \pm 0.15$ & 46.01 & 44.63 & 1.53 & 1.38  \\
0159+0023 & 0.163 & 2.13 & $ 775.4 \pm 12.4$ & $ 2.39 \pm 0.03$ & 44.90 & 43.86 & 1.40 & $ \ldots $  \\
0201-0919 & 0.661 & 2.08 & $ 46.05 \pm 2.87$ & $ 2.16 \pm 0.12$ & 45.40 & 44.12 & 1.49 & 0.90  \\
0208+0022 & 1.885 & 2.78 & $ 18.49 \pm 2.65$ & $ 1.56 \pm 0.20$ & 46.69 & 44.79 & 1.73 & 0.49  \\
0209+0517 & 4.140 & 4.57 & $ 6.39^{+ 1.27}_{- 1.12} $ & $ 2.72^{+ 0.33}_{- 0.31} $ & 47.20 & 45.18 & 1.78 & 0.68  \\
0232-0731 & 1.164 & 3.30 & $ < 7.39 $ & $ \ldots $ & 45.24 & $ < 43.95 $ & $ > 1.50 $ & 1.41  \\
0241-0811 & 0.979 & 2.90 & $ < 6.24 $ & $ \ldots $ & 44.87 & $ < 43.70 $ & $ > 1.45 $ & 0.63  \\
0241+0023 & 0.790 & 3.37 & $ 9.46^{+ 2.45}_{- 2.12} $ & $ 2.01^{+ 0.48}_{- 0.44} $ & 44.55 & 43.64 & 1.35 & 0.78  \\
0244-0134 & 4.053 & 3.52 & $ 2.14^{+ 0.62}_{- 0.53} $ & $ 1.52^{+ 0.41}_{- 0.39} $ & 47.22 & 44.68 & 1.97 & 1.06  \\
0248+1802 & 4.430 & 9.02 & $ 13.64^{+ 3.43}_{- 2.94} $ & $ 1.97^{+ 0.39}_{- 0.37} $ & 46.97 & 45.57 & 1.54 & 0.34  \\
0259+0048 & 0.892 & 7.22 & $ 12.56^{+ 2.18}_{- 1.94} $ & $ 1.02^{+ -1.0}_{- 1.02} $ & 45.11 & 43.90 & 1.46 & 0.58  \\
0311-1722 & 4.000 & 3.81 & $ < 4.14 $ & $ \ldots $ & 47.21 & $ < 44.95 $ & $ > 1.87 $ & 1.90  \\
0314-0111 & 0.387 & 5.78 & $ 97.12 \pm 9.62$ & $ 2.75 \pm 0.17$ & 44.88 & 43.79 & 1.42 & $ \ldots $  \\
0403-1703 & 4.236 & 2.29 & $ 4.16^{+ 1.20}_{- 1.01} $ & $ 1.39^{+ 0.37}_{- 0.37} $ & 46.62 & 45.01 & 1.62 & 1.26  \\
0419-5716 & 4.460 & 1.67 & $ 2.73^{+ 1.01}_{- 0.82} $ & $ 2.13^{+ 0.51}_{- 0.49} $ & 47.20 & 44.88 & 1.89 & 2.18  \\
0755+2203 & 0.400 & 5.60 & $ 23.69 \pm 2.84$ & $ 1.38 \pm 0.10$ & 44.55 & 43.43 & 1.43 & $ \ldots $  \\
0755+4058 & 2.417 & 4.78 & $ 5.36^{+ 1.52}_{- 1.31} $ & $ 3.22^{+ 0.57}_{- 0.55} $ & 46.16 & 44.56 & 1.62 & 0.27  \\
0755+4111 & 0.967 & 4.98 & $ 4.53^{+ 1.51}_{- 1.27} $ & $ 2.09^{+ 0.61}_{- 0.55} $ & 45.78 & 43.54 & 1.86 & 0.20  \\
0755+4056 & 2.348 & 4.73 & $ 5.86^{+ 1.55}_{- 1.34} $ & $ 2.75^{+ 0.56}_{- 0.53} $ & 46.03 & 44.57 & 1.56 & 0.45  \\
0819+3649 & 0.736 & 4.82 & $ 45.20 \pm 5.24$ & $ 3.08 \pm 0.26$ & 45.00 & 44.18 & 1.32 & 0.38  \\
0832+5243 & 1.573 & 3.87 & $ < 2.23 $ & $ \ldots $ & 45.58 & $ < 43.74 $ & $ > 1.71 $ & 0.30  \\
0845+3431 & 2.046 & 3.41 & $ < 8.40 $ & $ \ldots $ & 45.60 & $ < 44.59 $ & $ > 1.39 $ & 0.62  \\
0849+4457 & 1.259 & 2.75 & $ 9.43 \pm 0.37$ & $ 2.07 \pm 0.07$ & 45.08 & 44.15 & 1.36 & 0.38  \\
0849+4500 & 1.592 & 2.70 & $ 15.33 \pm 0.57$ & $ 2.10 \pm 0.06$ & 45.94 & 44.61 & 1.51 & 0.53  \\
0910+5427 & 0.526 & 2.03 & $ 80.75 \pm 1.73$ & $ 2.11 \pm 0.04$ & 44.85 & 44.13 & 1.28 & -0.3  \\
0912+0547 & 3.241 & 3.65 & $ 3.40 \pm 0.56$ & $ 1.94 \pm 0.31$ & 46.75 & 44.67 & 1.80 & 0.67  \\
0914+5613\tablenotemark{e} & 4.035 & 2.89 & $ 0.69^{+ 0.24}_{- 0.20} $ & $ 1.04^{+ 0.55}_{- 0.52} $ & 46.14 & 44.18 & 1.75 & 1.43  \\
0918+5139 & 0.185 & 1.36 & $ 20.66 \pm 9.26$ & $ 0.92 \pm 0.33$ & 44.15 & 42.75 & 1.54 & $ \ldots $  \\
0918+0647 & 0.821 & 3.65 & $ 22.58 \pm 3.32$ & $ 2.81 \pm 0.28$ & 45.21 & 44.03 & 1.45 & 0.68  \\
0933+5515 & 1.864 & 2.00 & $ 18.05 \pm 1.18$ & $ 2.08 \pm 0.11$ & 45.83 & 44.85 & 1.37 & 0.81  \\
0941+5948 & 0.968 & 2.19 & $ 100.8 \pm 5.90$ & $ 2.44 \pm 0.12$ & 46.28 & 44.89 & 1.54 & 0.48  \\
0950+5619 & 1.913 & 1.19 & $ 7.36^{+ 1.40}_{- 1.25} $ & $ 2.87^{+ 0.38}_{- 0.36} $ & 45.20 & 44.46 & 1.28 & 0.68  \\
0951+5940 & 1.057 & 1.67 & $ 17.92^{+ 2.76}_{- 2.51} $ & $ 1.80^{+ 0.25}_{- 0.24} $ & 45.43 & 44.23 & 1.46 & 0.55  \\
0951+5944 & 2.339 & 1.51 & $ 4.95^{+ 1.17}_{- 1.01} $ & $ 2.01^{+ 0.38}_{- 0.36} $ & 45.60 & 44.49 & 1.43 & 0.21  \\
0952+5152 & 0.554 & 0.88 & $ 51.81 \pm 1.52$ & $ 2.29 \pm 0.06$ & 44.97 & 43.97 & 1.38 & 0.60  \\
0952+5151 & 0.862 & 0.86 & $ 59.38 \pm 1.62$ & $ 2.16 \pm 0.05$ & 45.59 & 44.53 & 1.41 & 0.88  \\
0955+5935 & 0.912 & 1.50 & $ 14.05^{+ 2.37}_{- 2.13} $ & $ 2.20^{+ 0.30}_{- 0.30} $ & 45.08 & 43.97 & 1.42 & 0.69  \\
0955+5940 & 4.340 & 1.33 & $ 3.47^{+ 0.89}_{- 0.76} $ & $ 1.87^{+ 0.34}_{- 0.33} $ & 46.76 & 44.96 & 1.69 & 0.47  \\
0955+4116 & 3.420 & 0.64 & $ 3.93^{+ 0.83}_{- 0.73} $ & $ 2.15^{+ 0.38}_{- 0.36} $ & 46.16 & 44.78 & 1.53 & 0.23  \\
0955+4109 & 2.308 & 0.59 & $ 9.68 \pm 1.36$ & $ 2.13 \pm 0.22$ & 46.09 & 44.82 & 1.49 & 0.25  \\
0956+4110 & 1.887 & 0.72 & $ 2.28^{+ 0.63}_{- 0.54} $ & $ 2.50^{+ 0.53}_{- 0.50} $ & 45.28 & 43.94 & 1.51 & -0.5  \\
0958+0734 & 1.885 & 2.95 & $ 59.53 \pm 11.9$ & $ 2.22 \pm 0.26$ & 46.10 & 45.40 & 1.27 & 0.72  \\
0958+0747 & 3.219 & 2.97 & $ 2.77^{+ 0.82}_{- 0.70} $ & $ 2.56^{+ 0.57}_{- 0.54} $ & 45.82 & 44.56 & 1.48 & 0.44  \\
0958+0745 & 1.488 & 3.05 & $ 18.66 \pm 1.47$ & $ 2.20 \pm 0.19$ & 45.46 & 44.64 & 1.31 & 0.72  \\
1002+5542 & 1.151 & 0.84 & $ 13.75 \pm 2.00$ & $ 1.27 \pm 0.15$ & 45.76 & 44.19 & 1.60 & 1.04  \\
1003+4736 & 2.934 & 0.93 & $ 4.83 \pm 0.64$ & $ 1.53 \pm 0.17$ & 45.88 & 44.60 & 1.49 & 0.52  \\
1013-0052 & 0.276 & 3.49 & $ 133.5 \pm 10.1$ & $ 2.74 \pm 0.12$ & 44.55 & 43.56 & 1.38 & $ \ldots $  \\
1019+4737 & 2.945 & 0.99 & $ 3.82^{+ 0.78}_{- 0.69} $ & $ 2.31^{+ 0.28}_{- 0.27} $ & 46.04 & 44.61 & 1.55 & 0.53  \\
1023+0415 & 1.809 & 2.89 & $ 8.66 \pm 0.68$ & $ 2.05 \pm 0.11$ & 45.65 & 44.50 & 1.45 & 0.55  \\
1030+0524 & 1.183 & 2.72 & $ 2.27^{+ 0.67}_{- 0.56} $ & $ 1.21^{+ 0.40}_{- 0.39} $ & 45.85 & 43.46 & 1.92 & 0.48  \\
1032+5738 & 1.969 & 0.59 & $ 14.95 \pm 0.81$ & $ 1.73 \pm 0.07$ & 45.15 & 44.77 & 1.15 & 0.73  \\
1032+5800 & 0.687 & 0.61 & $ < 1.00 $ & $ \ldots $ & 44.55 & $ < 42.52 $ & $ > 1.78 $ & 0.57  \\
1036-0343 & 4.509 & 4.77 & $ 4.96^{+ 1.38}_{- 1.18} $ & $ 2.67^{+ 0.50}_{- 0.45} $ & 47.09 & 45.15 & 1.74 & 0.62  \\
1038+4727 & 1.047 & 1.47 & $ 2.61^{+ 0.74}_{- 0.64} $ & $ 0.92^{+ 0.40}_{- 0.40} $ & 45.55 & 43.39 & 1.83 & 0.69  \\
1042+0100 & 1.401 & 3.95 & $ 16.10^{+ 3.88}_{- 3.36} $ & $ 2.25^{+ 0.36}_{- 0.35} $ & 45.85 & 44.51 & 1.51 & 0.61  \\
1044+5921 & 1.292 & 0.70 & $ 10.00 \pm 0.80$ & $ 1.85 \pm 0.11$ & 45.45 & 44.19 & 1.48 & 0.41  \\
1049+5750 & 1.106 & 0.60 & $ 11.66^{+ 2.65}_{- 2.31} $ & $ 1.79^{+ 0.40}_{- 0.38} $ & 45.42 & 44.10 & 1.51 & 0.72  \\
1050+5702 & 3.273 & 0.59 & $ 5.70^{+ 1.38}_{- 1.20} $ & $ 2.45^{+ 0.49}_{- 0.46} $ & 45.92 & 44.89 & 1.40 & 0.81  \\
1050+5738 & 1.281 & 0.59 & $ 13.89^{+ 2.61}_{- 2.32} $ & $ 2.68^{+ 0.36}_{- 0.35} $ & 45.38 & 44.33 & 1.41 & 0.53  \\
1052+5724 & 1.112 & 0.60 & $ 35.22 \pm 4.58$ & $ 1.80 \pm 0.18$ & 45.76 & 44.58 & 1.45 & 0.72  \\
1053+5735 & 1.205 & 0.59 & $ 42.26 \pm 4.19$ & $ 1.86 \pm 0.15$ & 45.29 & 44.74 & 1.21 & 1.37  \\
1054+5740 & 1.101 & 0.59 & $ 10.45^{+ 2.30}_{- 2.01} $ & $ 2.12^{+ 0.37}_{- 0.36} $ & 45.72 & 44.04 & 1.64 & 0.54  \\
1054+5720 & 2.972 & 0.59 & $ < 7.75 $ & $ \ldots $ & 45.91 & $ < 44.93$ & $ > 1.38$ & 0.62  \\
1055+5704 & 0.696 & 0.60 & $ 25.93 \pm 3.29$ & $ 2.12 \pm 0.20$ & 44.92 & 43.93 & 1.38 & 0.87  \\
1056+5722 & 0.286 & 0.60 & $ 10.45^{+ 2.24}_{- 1.97} $ & $ 0.28^{+ 0.32}_{- 0.33} $ & 44.01 & 42.66 & 1.52 & $ \ldots $  \\
1057+4555 & 4.100 & 1.16 & $ 12.05^{+ 2.17}_{- 1.94} $ & $ 2.08^{+ 0.27}_{- 0.27} $ & 47.67 & 45.44 & 1.86 & 2.14  \\
1109+0900 & 1.674 & 2.79 & $ 5.56^{+ 1.29}_{- 1.13} $ & $ 1.67^{+ 0.38}_{- 0.36} $ & 45.60 & 44.20 & 1.54 & 0.85  \\
1111+5532 & 1.004 & 0.78 & $ 14.82 \pm 0.91$ & $ 3.23 \pm 0.10$ & 45.50 & 44.10 & 1.54 & 0.71  \\
1114+5315 & 1.213 & 0.96 & $ 18.34 \pm 2.19$ & $ 1.65 \pm 0.16$ & 45.29 & 44.38 & 1.35 & 2.11  \\
1115+5309 & 0.877 & 0.97 & $ < 2.21 $ & $ \ldots $ & 45.53 & $ < 43.13 $ & $ > 1.92 $ & 0.56  \\
1129-0137 & 1.295 & 3.58 & $ 31.69 \pm 3.31$ & $ 1.54 \pm 0.15$ & 45.71 & 44.67 & 1.40 & 0.75  \\
1129-0150 & 1.785 & 3.56 & $ 11.30^{+ 2.37}_{- 2.09} $ & $ 2.11^{+ 0.37}_{- 0.35} $ & 45.15 & 44.57 & 1.22 & 0.97  \\
1136+0159 & 0.766 & 2.50 & $ 21.23 \pm 2.44$ & $ 2.03 \pm 0.19$ & 44.94 & 43.96 & 1.38 & 0.73  \\
1136+0158 & 1.471 & 2.61 & $ 2.95^{+ 0.89}_{- 0.75} $ & $ 1.45^{+ 0.39}_{- 0.40} $ & 46.10 & 43.80 & 1.88 & 0.57  \\
1136+0207 & 0.239 & 2.61 & $ 22.37 \pm 4.05$ & $ 1.63 \pm 0.26$ & 44.04 & 42.87 & 1.45 & $ \ldots $  \\
1202-0129 & 0.150 & 2.22 & $ 222.5 \pm 5.86$ & $ 3.02 \pm 0.04$ & 44.05 & 43.09 & 1.37 & $ \ldots $  \\
1204+0150 & 1.927 & 1.88 & $ 7.29 \pm 0.63$ & $ 2.10 \pm 0.15$ & 45.93 & 44.50 & 1.55 & 0.29  \\
1208+0016 & 1.063 & 1.99 & $ 14.11^{+ 2.56}_{- 2.29} $ & $ 1.25^{+ 0.28}_{- 0.27} $ & 45.21 & 44.14 & 1.41 & 0.92  \\
1213+0252 & 0.641 & 1.74 & $ 6.82 \pm 0.99$ & $ 2.10 \pm 0.25$ & 44.64 & 43.27 & 1.53 & 0.98  \\
1214+0055 & 0.396 & 1.95 & $ 87.02 \pm 7.77$ & $ 2.44 \pm 0.13$ & 44.67 & 43.82 & 1.33 & $ \ldots $  \\
1215-0034 & 0.758 & 2.07 & $ 28.80 \pm 1.72$ & $ 2.08 \pm 0.11$ & 44.69 & 44.07 & 1.24 & 0.86  \\
1218+0546 & 0.795 & 1.57 & $ 14.21 \pm 2.11$ & $ 1.28 \pm 0.20$ & 44.88 & 43.86 & 1.39 & 1.73  \\
1220-0025 & 0.421 & 2.02 & $ 123.6 \pm 2.93$ & $ 1.45 \pm 0.03$ & 44.36 & 44.18 & 1.07 & $ \ldots $  \\
1223+1034 & 2.747 & 2.13 & $ 11.80 \pm 0.81$ & $ 1.59 \pm 0.11$ & 46.35 & 44.95 & 1.54 & 0.27  \\
1226-0011 & 1.175 & 1.93 & $ 32.20 \pm 2.91$ & $ 2.43 \pm 0.13$ & 45.79 & 44.62 & 1.45 & 1.28  \\
1228+4413 & 0.662 & 1.45 & $ < 1.07 $ & $ \ldots $ & 45.29 & $ < 42.51 $ & $ > 2.07 $ & 0.59  \\
1228+4411 & 1.277 & 1.34 & $ 10.81 \pm 1.18$ & $ 2.12 \pm 0.18$ & 45.61 & 44.21 & 1.54 & 0.12  \\
1230+0302 & 1.605 & 1.80 & $ 11.27^{+ 2.17}_{- 1.93} $ & $ 1.61^{+ 0.29}_{- 0.28} $ & 45.72 & 44.48 & 1.48 & 0.40  \\
1230+0305 & 1.056 & 1.83 & $ 18.88^{+ 3.37}_{- 3.02} $ & $ 2.37^{+ 0.32}_{- 0.31} $ & 45.08 & 44.26 & 1.32 & 0.46  \\
1230+0306 & 0.628 & 1.81 & $ 59.90 \pm 6.60$ & $ 2.03 \pm 0.17$ & 44.99 & 44.20 & 1.30 & 0.33  \\
1230+0308 & 1.843 & 1.81 & $ 6.85^{+ 2.35}_{- 1.92} $ & $ 2.61^{+ 0.59}_{- 0.57} $ & 45.63 & 44.39 & 1.48 & 0.29  \\
1230+0305 & 0.428 & 1.83 & $ 12.38^{+ 3.41}_{- 2.92} $ & $ 1.73^{+ 0.52}_{- 0.48} $ & 44.28 & 43.13 & 1.44 & $ \ldots $  \\
1236+6215 & 2.587 & 1.52 & $ 1.19 \pm 0.12$ & $ 1.79 \pm 0.19$ & 45.53 & 43.95 & 1.61 & 0.55  \\
1237+6203 & 2.068 & 1.44 & $ 3.54 \pm 0.08$ & $ 1.85 \pm 0.03$ & 45.66 & 44.21 & 1.56 & 1.02  \\
1242+0249 & 1.459 & 1.92 & $ 12.89 \pm 0.58$ & $ 2.32 \pm 0.08$ & 45.58 & 44.47 & 1.43 & 0.06  \\
1245-0027 & 1.693 & 1.73 & $ 44.48 \pm 6.08$ & $ 1.80 \pm 0.19$ & 45.96 & 45.10 & 1.33 & 0.72  \\
1255+5652 & 1.804 & 1.25 & $ 7.06 \pm 0.48$ & $ 2.44 \pm 0.11$ & 45.74 & 44.46 & 1.49 & 0.34  \\
1255+5650 & 1.374 & 1.25 & $ < 0.56 $ & $ \ldots $ & 45.23 & $ < 43.00 $ & $ > 1.85 $ & 0.28  \\
1258-0143 & 0.967 & 1.54 & $ 84.49 \pm 2.06$ & $ 2.30 \pm 0.05$ & 46.03 & 44.81 & 1.47 & 0.56  \\
1259+0102 & 0.395 & 1.62 & $ < 10.05 $ & $ \ldots $ & 44.48 & $ < 42.96 $ & $ > 1.58 $ & $ \ldots $  \\
1311+0031 & 0.429 & 1.84 & $ 11.60^{+ 3.33}_{- 2.80} $ & $ 2.58^{+ 0.46}_{- 0.44} $ & 44.82 & 43.11 & 1.66 & $ \ldots $  \\
1317+3531 & 4.360 & 0.99 & $ < 6.84 $ & $ \ldots $ & 46.54 & $ < 45.26 $ & $ > 1.49 $ & 0.90  \\
1321+0038\tablenotemark{e} & 4.716 & 1.88 & $ 1.49^{+ 0.36}_{- 0.31} $ & $ 2.50^{+ 0.38}_{- 0.37} $ & 46.60 & 44.67 & 1.74 & 2.04  \\
1344-0000 & 1.096 & 1.89 & $ 3.27^{+ 0.68}_{- 0.60} $ & $ 1.80^{+ 0.32}_{- 0.30} $ & 45.47 & 43.53 & 1.74 & 0.43  \\
1411+5217 & 2.883 & 1.33 & $ 5.79 \pm 0.50$ & $ 1.82 \pm 0.14$ & 46.16 & 44.75 & 1.54 & 0.57  \\
1411+5205 & 1.084 & 1.40 & $ 4.80 \pm 2.07 $ & $ 2.44 \pm 0.49 $ & 45.30 & 43.70 & 1.61 & 0.73  \\
1413+0000\tablenotemark{e} & 4.078 & 3.15 & $ 1.67^{+ 0.42}_{- 0.37} $ & $ 1.93^{+ 0.43}_{- 0.39} $ & 46.46 & 44.58 & 1.72 & 1.66  \\
1417+4456 & 0.114 & 1.13 & $ 380.7 \pm 5.53$ & $ 2.38 \pm 0.03$ & 44.11 & 43.21 & 1.35 & $ \ldots $  \\
1419+4709 & 2.288 & 1.56 & $ 26.56 \pm 2.06$ & $ 1.85 \pm 0.12$ & 46.83 & 45.19 & 1.63 & 1.25  \\
1424+4214 & 1.608 & 1.25 & $ 26.00 \pm 2.42$ & $ 1.97 \pm 0.15$ & 45.63 & 44.84 & 1.30 & -0.0  \\
1424+4210 & 2.218 & 1.25 & $ 23.20 \pm 2.15$ & $ 2.39 \pm 0.12$ & 46.55 & 45.21 & 1.51 & 0.35  \\
1432-0059 & 1.027 & 3.39 & $ 174.2 \pm 5.15$ & $ 2.06 \pm 0.05$ & 45.88 & 45.19 & 1.26 & 0.78  \\
1433+0227 & 2.072 & 2.75 & $ 9.59^{+ 2.54}_{- 2.16} $ & $ 2.36^{+ 0.46}_{- 0.44} $ & 45.54 & 44.66 & 1.34 & 0.75  \\
1434+0227 & 1.711 & 2.76 & $ 10.73^{+ 2.19}_{- 1.94} $ & $ 2.05^{+ 0.34}_{- 0.32} $ & 45.60 & 44.51 & 1.42 & 0.08  \\
1438+0341 & 1.737 & 2.62 & $ 12.71 \pm 0.93$ & $ 2.11 \pm 0.13$ & 46.15 & 44.63 & 1.59 & 1.22  \\
1438+0335 & 0.734 & 2.62 & $ 20.96 \pm 3.70$ & $ 1.70 \pm 0.17$ & 45.18 & 43.92 & 1.48 & 0.88  \\
1442+0110 & 4.560 & 3.36 & $ 5.25^{+ 0.82}_{- 0.74} $ & $ 1.98^{+ 0.23}_{- 0.23} $ & 46.38 & 45.19 & 1.46 & 0.71  \\
1443+5856 & 4.260 & 1.56 & $ 2.74^{+ 0.80}_{- 0.67} $ & $ 2.29^{+ 0.42}_{- 0.40} $ & 46.97 & 44.84 & 1.82 & 0.29  \\
1444-0123\tablenotemark{e} & 4.179 & 4.03 & $ 1.59^{+ 0.53}_{- 0.44} $ & $ 2.95^{+ 0.77}_{- 0.66} $ & 46.63 & 44.58 & 1.79 & 1.60  \\
1445+0129 & 1.846 & 3.48 & $ 8.75^{+ 1.42}_{- 1.28} $ & $ 2.45^{+ 0.29}_{- 0.28} $ & 45.61 & 44.50 & 1.43 & 0.17  \\
1448+4738 & 2.894 & 2.05 & $ 3.61^{+ 0.86}_{- 0.74} $ & $ 1.72^{+ 0.32}_{- 0.31} $ & 46.29 & 44.57 & 1.66 & 1.12  \\
1448+0015 & 0.832 & 3.58 & $ 18.64^{+ 4.73}_{- 4.05} $ & $ 2.24^{+ 0.48}_{- 0.45} $ & 45.17 & 44.00 & 1.45 & 0.60  \\
1449+0024 & 0.441 & 3.58 & $ 31.72^{+ 5.06}_{- 4.58} $ & $ 2.50^{+ 0.27}_{- 0.27} $ & 44.55 & 43.57 & 1.38 & $ \ldots $  \\
1452+4304 & 0.296 & 1.69 & $ 20.43 \pm 0.99$ & $ 1.97 \pm 0.07$ & 44.07 & 42.97 & 1.42 & $ \ldots $  \\
1452+4308 & 1.704 & 1.64 & $ 17.64 \pm 1.90$ & $ 1.78 \pm 0.15$ & 45.63 & 44.70 & 1.36 & 0.20  \\
1453+3352\tablenotemark{e} & 4.191 & 1.22 & $ 2.39^{+ 0.56}_{- 0.49} $ & $ 1.38^{+ 0.35}_{- 0.34} $ & 46.23 & 44.76 & 1.56 & 2.05  \\
1511+5659 & 1.031 & 1.54 & $ < 17.19 $ & $ \ldots $ & 45.80 & $ < 44.19 $ & $ > 1.62 $ & 0.61  \\
1515+5521 & 1.844 & 1.44 & $ 4.02^{+ 1.63}_{- 1.44} $ & $ 4.56^{+ 0.90}_{- 0.90} $ & 45.30 & 44.16 & 1.44 & 0.33  \\
1539+4313 & 0.348 & 2.03 & $ 64.94 \pm 2.34$ & $ 1.96 \pm 0.06$ & 44.27 & 43.63 & 1.24 & $ \ldots $  \\
1543+5405 & 0.245 & 1.31 & $ 58.57 \pm 5.41$ & $ 2.08 \pm 0.15$ & 44.28 & 43.22 & 1.41 & $ \ldots $  \\
1545+4846 & 0.400 & 1.61 & $ 139.3 \pm 10.9$ & $ 2.23 \pm 0.13$ & 45.59 & 44.06 & 1.59 & $ \ldots $  \\
1605-0109 & 1.573 & 8.88 & $ 28.07 \pm 3.18$ & $ 2.03 \pm 0.18$ & 45.85 & 44.85 & 1.38 & 0.62  \\
1618+3456 & 1.922 & 1.46 & $ 7.21^{+ 1.96}_{- 1.66} $ & $ 1.76^{+ 0.44}_{- 0.42} $ & 46.10 & 44.46 & 1.63 & 1.60  \\
1640+4644 & 0.537 & 1.74 & $ 20.91 \pm 0.73$ & $ 2.13 \pm 0.06$ & 45.07 & 43.56 & 1.58 & 0.36  \\
1641+4649 & 0.695 & 1.77 & $ 19.47 \pm 1.22$ & $ 2.11 \pm 0.11$ & 44.79 & 43.81 & 1.38 & 0.67  \\
1641+4000 & 1.003 & 1.02 & $ 12.38 \pm 0.95$ & $ 1.74 \pm 0.12$ & 45.72 & 44.02 & 1.65 & 0.63  \\
1657+3524 & 2.329 & 1.75 & $ 4.55 \pm 0.68$ & $ 1.69 \pm 0.24$ & 45.85 & 44.41 & 1.55 & 0.45  \\
1701+6412 & 2.736 & 2.59 & $ 14.40 \pm 0.93$ & $ 2.04 \pm 0.10$ & 47.41 & 45.15 & 1.86 & 0.37  \\
1702+3405 & 2.038 & 2.04 & $ 2.55 \pm 0.29$ & $ 1.37 \pm 0.16$ & 46.01 & 43.97 & 1.78 & 1.11  \\
1703+6045 & 0.285 & 2.32 & $ 49.18 \pm 3.43$ & $ 2.18 \pm 0.13$ & 44.04 & 43.27 & 1.30 & $ \ldots $  \\
1708+6154 & 1.415 & 2.49 & $ 65.18 \pm 4.69$ & $ 2.00 \pm 0.14$ & 46.03 & 45.11 & 1.35 & 1.15  \\
1719+2732 & 1.447 & 3.68 & $ 28.89 \pm 3.64$ & $ 2.22 \pm 0.21$ & 45.72 & 44.80 & 1.35 & 1.11  \\
1720+2638 & 1.141 & 3.86 & $ 10.63 \pm 0.94$ & $ 2.40 \pm 0.15$ & 45.33 & 44.10 & 1.47 & 0.20  \\
1735+5355 & 0.956 & 3.39 & $ 59.55 \pm 3.96$ & $ 1.98 \pm 0.12$ & 45.62 & 44.65 & 1.37 & 1.01  \\
1737+5828 & 1.776 & 3.51 & $ 14.65^{+ 3.51}_{- 3.03} $ & $ 2.39^{+ 0.40}_{- 0.38} $ & 45.80 & 44.68 & 1.43 & 0.79  \\
1738+5837 & 1.279 & 3.59 & $ 3.04^{+ 1.16}_{- 0.94} $ & $ 2.23^{+ 0.55}_{- 0.53} $ & 45.97 & 43.66 & 1.89 & 1.41  \\
2215-1611 & 3.990 & 2.65 & $ 5.23^{+ 1.53}_{- 1.29} $ & $ 1.30^{+ 0.38}_{- 0.38} $ & 46.81 & 45.05 & 1.68 & 0.80  \\
2238-0921 & 3.259 & 4.64 & $ < 23.19 $ & $ \ldots $ & 46.70 & $ < 45.50 $ & $ > 1.46 $ & 0.31  \\
2238-0937 & 1.472 & 4.78 & $ < 69.50 $ & $ \ldots $ & 45.58 & $ < 45.17 $ & $ > 1.16 $ & 1.07  \\
2239-0933 & 1.818 & 4.63 & $ < 57.74 $ & $ \ldots $ & 45.84 & $ < 45.30 $ & $ > 1.21 $ & 0.63  \\
2249-0808 & 0.457 & 3.46 & $ 55.44 \pm 4.29$ & $ 1.81 \pm 0.11$ & 44.11 & 43.86 & 1.09 & $ \ldots $  \\
2337+0025 & 2.054 & 3.81 & $ 6.89^{+ 1.59}_{- 1.39} $ & $ 2.29^{+ 0.47}_{- 0.43} $ & 45.87 & 44.50 & 1.53 & 0.18  \\
2337+0022 & 1.376 & 3.30 & $ 3.87^{+ 1.46}_{- 1.18} $ & $ 2.93^{+ 0.70}_{- 0.65} $ & 45.48 & 43.85 & 1.63 & 0.99  \\
2337+0026 & 1.703 & 3.80 & $ 21.37 \pm 3.03$ & $ 2.67 \pm 0.28$ & 45.84 & 44.90 & 1.36 & 0.15  \\
2348+0107 & 0.718 & 3.98 & $ 28.51 \pm 1.28$ & $ 2.05 \pm 0.08$ & 45.19 & 44.01 & 1.45 & 0.32  \\
2357+0043\tablenotemark{e} & 4.362 & 3.33 & $ 2.00^{+ 0.51}_{- 0.44} $ & $ 1.58^{+ 0.40}_{- 0.38} $ & 46.54 & 44.72 & 1.70 & 1.74  \\
\enddata

\tablecomments{Quoted errors are at $68\%\ (1\sigma)$ confidence.}

\tablenotetext{a}{Galactic $N_H$, inferred from COLDEN \citep{colden}.}
\tablenotetext{b}{When fitting $n_0$ for the sources with $< 50$ counts, we fix $\Gamma_X = 1.9$.}
\tablenotetext{c}{The photon index could not be estimated for those sources with
  upper limits.}
\tablenotetext{d}{The UV spectral slope could not be estimated for
  sources with $z < 0.5$.}
\tablenotetext{e}{One of seven new \emph{Chandra} observations.}

\end{deluxetable}

\clearpage

\begin{deluxetable}{cccccccccc}
  \tabletypesize{\scriptsize}
  \tablecaption{Behavior of Kendall's Generalized Partial $\tau$ From Simulation\label{t-tau}}
  \tablewidth{0pt}
  \tablehead{
    & & \multicolumn{4}{c}{$H_0 : L_X \propto L_{UV}^{0.65}$} & 
    \multicolumn{4}{c}{$H_1 : L_X \propto L_{UV}^{0.4} e^{-t(z) / 5.5}$} \\
    \colhead{$\rho$\tablenotemark{a}}
    & \colhead{$\tau_{lz}$\tablenotemark{b}}
    & \colhead{$\tau_{xz,l}$\tablenotemark{c}}
    & \colhead{${\rm pow}(\tau_{xz,l}|H_0)$\tablenotemark{d}}
    & \colhead{$\tau_{\alpha z,l}$\tablenotemark{e}}
    & \colhead{${\rm pow}(\tau_{\alpha z,l}|H_0)$\tablenotemark{f}}
    & \colhead{$\tau_{xz,l}$\tablenotemark{c}}
    & \colhead{${\rm pow}(\tau_{xz,l}|H_1)$\tablenotemark{d}}
    & \colhead{$\tau_{\alpha z,l}$\tablenotemark{e}}
    & \colhead{${\rm pow}(\tau_{\alpha z,l} | H_1)$\tablenotemark{f}}
  }
  \startdata
  0.9 & 0.715 & 0.153 & 0.988 & 0.105 & 0.845 & 0.290 & 1.000 &  0.011 & 0.061 \\
  0.6 & 0.412 & 0.101 & 0.696 & 0.074 & 0.390 & 0.338 & 1.000 & -0.126 & 0.935 \\
  0.3 & 0.197 & 0.048 & 0.196 & 0.038 & 0.126 & 0.329 & 1.000 & -0.204 & 1.000 \\
  0.0 & 0.000 & 0.000 & 0.050 & 0.000 & 0.050 & 0.299 & 1.000 & -0.258 & 1.000
  \enddata
  
  \tablenotetext{a}{The correlation between $l_{UV}$ and $\log z$ used
    for the simulation.}
  \tablenotetext{b}{The average value of Kendall's $\tau$ between
    $L_{UV}$ and $z$.}
  \tablenotetext{c}{The average value of Kendall's generalized partial
    $\tau$ between $L_X$ and $z$, controlling for $L_{UV}$. The left
    value of $\tau_{xz,l}$ corresponds to when the null hypothesis,
    $H_0$, is true, and the right value corresponds to when the
    alternative hypothesis, $H_1$, is true.}
  \tablenotetext{d}{The power of the test when $\tau_{xz,l}$ is used
    and the null hypothesis $H_0$ is true (left) or the alternative
    hypothesis $H_1$ is true (right).}
  \tablenotetext{e}{Same as $\tau_{xz,l}$, but when using $\alpha_{\rm
      ox}$ instead of $L_X$.}
  \tablenotetext{f}{Same as ${\rm pow}(\tau_{xz,l}|\cdot)$, but when
      using $\tau_{az,l}$ instead of $\tau_{xz,l}$.}

\end{deluxetable}

\clearpage

\end{document}